\documentclass[structabstract]{aa}  
\usepackage{graphicx,aalongtable}
\usepackage{txfonts}
\begin{document}
   \title{Beryllium abundances and star formation in the halo and in the 
thick disk\thanks{Based on observations made with
  ESO VLT, at Paranal Observatory, under programs 076.B-0133 and
  077.B-0507, and on data obtained from the ESO/ST-ECF Science Archive
  Facility and the UVES Paranal Observatory Project 266.D-5655.}}
\author{R. Smiljanic \inst{1,2}
        \and
        L. Pasquini\inst{2}
         \and
         P. Bonifacio\inst{3,4,5}
        \and
         D. Galli\inst{6}
        \and
         R.~G. Gratton\inst{7}
         \and
         S. Randich\inst{6}
         \and
         B. Wolff\inst{2}
}
\institute{Universidade de S\~ao Paulo, IAG, Dpt. de Astronomia, Rua
     do Mat\~ao 1226, S\~ao Paulo-SP 05508-090, Brazil\\ 
\email{rodolfo@astro.iag.usp.br} 
\and
European Southern Observatory, Karl-Schwarzschild-Str. 2, 85748
        Garching bei M\"unchen, Germany 
\and
CIFIST Marie Curie Excellence Team
\and
GEPI, Observatoire de Paris, CNRS, Universit\'e Paris Diderot, Place Jules Janssen, 92190 Meudon, France 
\and
INAF - Osservatorio Astronomico di Trieste, Via G.B. Tiepolo 11, 34143 Trieste, Italy
\and
INAF - Osservatorio Astrofisico di Arcetri, Largo E. Fermi 5, 50125 Firenze, Italy
\and
INAF - Osservatorio Astronomico di Padova, Vicolo dell'Osservatorio 5, 35122 Padova, Italy
}

\date{Received ; accepted }

 
  \abstract
{Beryllium is a pure product of cosmic ray spallation. This implies a
relatively simple evolution in time of the beryllium abundance
and suggests its use as a time-like observable.}
{Our goal is to derive abundances of Be in a sample of 90 stars, the largest 
sample of halo and thick disk stars analyzed to date.
We study the evolution of Be in the early Galaxy and 
its dependence on kinematic and orbital parameters, and investigate its use as 
a cosmochronometer. Abundances of Be, Fe, and $\alpha$-elements of 73 stars are employed to 
study the formation of the halo and the thick disk of the Galaxy.}
   {Beryllium abundances are determined from high-resolution, high signal-to-noise 
UVES spectra with spectrum synthesis. Atmospheric parameters and abundances of 
$\alpha$-elements are adopted from the literature. Lithium abundances are used to eliminate 
mixed stars from the sample. The properties of halo and thick disk stars are investigated 
in diagrams of log(Be/H) vs.\@ [$\alpha$/H], log(Be/H) vs.\@ [Fe/H], and [$\alpha$/Fe] 
vs.\@ log(Be/H) and with orbital and kinematic parameters.}
  {We present our observational results in various diagrams. ({\it i}) In a 
log(Be/H) vs.\@ [Fe/H] diagram we find a marginal statistical detection 
of a real scatter, above what is expected from measurement errors, with 
a larger scatter among halo stars. The detection of the scatter is 
further supported by the existence of pairs of stars with identical
atmospheric parameters and different Be abundances. ({\it ii}) In a 
log(Be/H) vs.\@ [$\alpha$/Fe] diagram, the halo stars separate into 
two components; one is consistent with predictions of evolutionary models, 
while the other has too high $\alpha$ and Be abundances and is chemically 
indistinguishable from thick disk stars. This suggests that the halo 
is not a single uniform population where a clear age-metallicity 
relation can be defined. ({\it iii}) In diagrams of 
R$_{\rm min}$ vs.\@ [$\alpha$/Fe] and log(Be/H), the thick disk stars 
show a possible decrease in [$\alpha$/Fe] with R$_{\rm min}$, whereas no 
dependence of Be with R$_{\rm min}$ is seen. This anticorrelation suggests that 
the star formation rate was lower in the outer regions of the thick disk, pointing 
towards an inside-out formation. The lack of correlation for Be indicates that it is 
insensitive to the local conditions of star formation.
  } 
{}

   \keywords{stars: abundances -- stars: late-type -- Galaxy: halo -- Galaxy: thick disk}

   \titlerunning{Be and star formation in the halo and thick disk}

   \maketitle
%

\section{Introduction}\label{sec:intr}

 The nucleosynthetic origin of beryllium is different from most chemical 
elements. Beryllium is neither a product of stellar nucleosynthesis nor expected 
to be created by the standard homogeneous primordial nucleosynthesis in a 
detectable amount (Thomas et al.\@ \cite{Tho93}). Its single long-lived isotope,
 $^{9}$Be, is a pure product of cosmic-ray spallation of heavy (mostly CNO) nuclei 
in the interstellar medium (Reeves et al.\@ \cite{RFH70}; Meneguzzi et al.\@ 
\cite{MAR71}).

 Early theoretical models of Be production in the Galaxy assumed the cosmic-ray 
composition to be similar to the composition of the interstellar medium (ISM). In 
this scenario, Be is produced by accelerated protons and $\alpha$-particles 
colliding with CNO nuclei of the ISM (Meneguzzi \& Reeves \cite{MeRe75}; 
Vangioni-Flam et al. \cite{Van90}; Prantzos et al. \cite{Pra93}), resulting in a 
quadratic dependence of the Be abundance with metallicity. However, 
observations of Be in metal-poor stars (Rebolo et al.\@ \cite{RABM88}; Gilmore et 
al.\@ \cite{GGEN92}; Molaro et al.\@ \cite{MBCP97}; Boesgaard et al.\@ \cite{BDKR99}) 
find a slope equal or close to one between log(Be/H) and [Fe/H], and just slightly 
higher for log(Be/H) and [O/H]\footnote{[A/B] = log [N(A)/N(B)]$_{\rm \star}$ - log 
[N(A)/N(B)]$_{\rm\odot}$}. Such slopes argue that Be behaves as a primary element 
and its production mechanism is independent of ISM metallicity. Thus, the dominant 
production mechanism is now thought to be the collision of cosmic-rays 
composed of accelerated CNO nuclei with protons and $\alpha$-particles of the ISM 
(Duncan et al.\@ \cite{DLL92}; Cass\'e et al.\@ \cite{CLV95}; Vangioni-Flam et al.\@ 
\cite{VaF98}).

On the other hand, in light of some controversy about the behavior of O in 
metal-poor stars, Fields \& Olive (\cite{FO99}) argue that a secondary behavior 
cannot be excluded. King (\cite{Ki01,Ki02}), however, reassessing data of 
different oxygen indicators, as well as data for other $\alpha$-elements, 
show that a metal-poor primary mechanism is necessary to explain 
the observations. 

As a primary element and considering cosmic-rays to be globally transported 
across the Galaxy, one may expect the Be abundance to be rather homogeneous 
at a given time in the early Galaxy. It should have a smaller scatter 
than the products of stellar nucleosynthesis (Suzuki et al.\@ \cite{SYK99}; Suzuki \& 
Yoshii \cite{SY01}). Thus, Be would show a good correlation with time and could be 
employed as a cosmochronometer for the early stages of the Galaxy (Beers et al.\@ 
\cite{ipBSY00}; Suzuki \& Yoshii \cite{SY01}).

 Pasquini et al.\@ (\cite{Pas04,Pas07}) have tested this suggestion deriving Be abundances 
in turn-off stars in the globular clusters NGC 6397 and NGC 6752. The Be ages derived 
from a model of the evolution of Be with time (Valle et al.\@ \cite{Va02}) are in 
excellent agreement with those derived from theoretical isochrones. Moreover, the Be 
abundances of these globular cluster stars are similar to the abundances of field 
stars with the same metallicity. These results strongly suggest that the stellar Be abundances 
are independent of the environment where the star was formed and support the use of 
Be as a cosmochronometer. 

 Using abundances determined by Boesgaard et al.\@ (\cite{BDKR99}), Pasquini et al.\@ (\cite{Pas05}) 
extended the use of Be as a time scale to a sample 
of 20 halo and thick disk stars and investigated the evolution of the star formation 
rate in the early-Galaxy. Stars belonging to the two different kinematic components 
identified by Gratton et al.\@ (\cite{G03}) seem to separate in a log(Be/H)  vs.\@ [O/Fe]
diagram. Such separation is interpreted as indicating the formation of the two components 
to occur under different conditions and time scales.

 Our aim in this work is to better understand the evolution of Be in the early Galaxy 
and its dependence on different parameters, in particular on the stellar population. We 
analyze an unprecedentedly large sample of halo and thick disk stars, and further investigate 
the use of Be as a cosmochronometer and its role as a discriminator of different stellar 
populations in the Galaxy. The sample and the observational data are described in Sect.\@ 
\ref{sec:data}. The details on the adopted atmospheric parameters are given in Sect.\@ 
\ref{sec:par}. The abundance determination and its uncertainties are discussed in 
Sect.\@ \ref{sec:abun} while a comparison with previous results of the literature is presented 
in Sect.\@ \ref{sec:comp}. Lithium abundances are used to clean the sample from mixed 
stars in Sect.\@ \ref{sec:mix}. The Galactic evolution of Be is discussed in Sect.\@ 
\ref{sec:evo}, the use of Be as a chronometer
is discussed in Sect.\@ \ref{sec:chrono}, while a summary is given in Sect.\@ \ref{sec:con}.


\section{Sample selection and observational data}\label{sec:data}

 The sample stars were selected from the compilation by Venn et al.\@ (\cite{Venn04}) 
of several abundance and kinematic analyses of Galactic 
stars available in the literature. Using the available kinematic data, Venn et al.\@ 
(\cite{Venn04}) calculated the probabilities that each star belongs to the thin disk, 
the thick disk, or the halo. A total of 90 stars were selected for this work: 9 of 
them have higher probability of being thin disk stars, 30 of being thick disk stars, 
49 of being halo stars; and 2 have 50\% probabilities of being halo or thick disk stars. 
 We simply assume a star belongs to a certain kinematic group when the probability 
of belonging to that group is 
higher than the probability of being in the other two groups. One of our aims 
is to compare stars of different populations but of similar abundances, 
so we tried to maximize the metallicity overlap between the two sub-samples. The halo 
stars range from [Fe/H] = $-$2.48 to $-$0.50 and the thick disk stars from [Fe/H] = 
$-$1.70 to $-$0.50, although strongly concentrated in [Fe/H] $\geq$ $-$1.00. The sample 
stars are listed in Table \ref{tab:data}, together with V band magnitudes, parallaxes, 
absolute magnitudes, bolometric corrections, bolometric magnitudes, luminosities, 
and information on multiplicity. Details on these are given in the following sections. 

 Spectra for all stars were obtained using UVES, the \emph{Ultraviolet and Visual 
Echelle Spectrograph} (Dekker et al.\@ \cite{De00}) fed by UT2 of the VLT. UVES is 
a cross-dispersed echelle spectrograph able to obtain spectra from the atmospheric 
cut-off at 300 nm to $\sim$ 1100 nm.

 For 55 stars, new spectra were obtained in service mode during two 
observing periods between October 2005 and September
 2006. UVES was operated in dichroic mode with cross dispersers \#1,
 \#3, and \#4 resulting in spectral coverage of $\lambda\lambda$ 300-390
 nm in the blue arm and $\lambda\lambda$ 420-680 nm and
 $\lambda\lambda$ 660-1100 nm in the red arm, with some gaps. The spectra have
 a resolving power of R $\sim$ 35\,000 and S/N between 25 and 150 in the Be region.
 The reduction was conducted with the UVES pipeline within MIDAS. While
 the blue arm of the spectra was always found to have enough quality
 for the analysis, the same was not true for the red arm that, in some
 cases, presented residual fringing. For them, a new reduction of the 
raw frames was conducted using the most recent
 release of the UVES Common Pipeline Library (CPL) recipes within
 ESORex, the ESO Recipe Execution Tool.

 Reduced spectra for 4 stars were downloaded from the UVES Paranal Observatory 
 Project (POP) Library - a library of high-resolution spectra of stars
 across the HR diagram (Bagnulo et al.\@ \cite{Ba03}). In this library the stars were
 observed with both dichroic \#1 and \#2 covering almost all the
 interval between 300 and 1000 nm. The spectra have R $\sim$ 80\,000 and a 
 typical S/N ratio varying from 60 to 400 in the Be region.

\setcounter{table}{3}
\begin{table*}
\caption{Methods adopted by the reference papers cited
  in the text to calculate the atmospheric parameters.}
\label{tab:lit}
\centering
\begin{tabular}{cccccc}
\noalign{\smallskip}
\hline\hline
\noalign{\smallskip}
Reference & Temp.  & Surface gravity & Microt. & Model atm. & LTE/NLTE? \\
\hline
F00 & Exc. Eq. of \ion{Fe}{i} lines & Ion. Eq. \ion{Fe}{i} and
\ion{Fe}{ii} & \ion{Fe}{i} lines & Kurucz + overshooting & LTE \\
    & log (EW/$\lambda$) $\leq$ $-$4.80 & log (EW/$\lambda$) $\leq$
$-$4.80 & & & \\
SB02 & Exc. Eq. of \ion{Fe}{i} lines  & Average of Ion. Eq. of &
\ion{Fe}{i} lines & Kurucz  & LTE \\
     & log (EW/$\lambda$) $\leq$ $-$5.15  & \ion{Fe}{i}/\ion{Fe}{ii}
and \ion{Ti}{i}/\ion{Ti}{ii} & agrees with Ed93 relation  & Castelli
et al. (\cite{Cas97}) & \\ 
NS97 & Exc. Eq. of \ion{Fe}{i} lines & Ion. Eq. \ion{Fe}{i} and
\ion{Fe}{ii} & \ion{Fe}{i} lines & (NEW)MARCS (Ed93) & LTE \\
     & log (EW/$\lambda$) $\leq$ $-$5.15 &  & with $\chi \geq$ 4 eV &
& \\
Ed93 & Str\"omgren photometry & Balmer discontinuity &  Empiric
relation &  (NEW)MARCS &  LTE \\
     & calib. of $(b-y)$ and $\beta$ & index, c$_{1}$ & dependent
on T$_{\rm eff}$ and log g &  & \\
\noalign{\smallskip}
\hline
\end{tabular}
\end{table*}

 For the remaining 31 stars, archive raw data in the $\lambda\lambda$
 300-390 nm region were retrieved from the ESO/ST-ECF science archive
 facility and employed in the analysis. For some stars, in particular
 those that do not have previous determinations of lithium abundance
 in the literature, raw data in the $\lambda\lambda$
 420-680 nm region were also retrieved. The spectra were reduced
 using the UVES pipeline within MIDAS and ESORex. The resolving
 power of these spectra varies between 35\,000 and 50\,000 and the S/N
 ratio varies between 45 and 170 in the Be region. The log book of 
the observations is given in Table \ref{tab:log} in the Appendix.


\section{Atmospheric parameters}\label{sec:par}

\subsection{Source of the parameters}

 All the selected stars have been targets of previous abundance analyses. In 
this work, we decided to adopt the atmospheric parameters, effective temperature 
(T$_{\rm eff}$), surface gravity (log g), microturbulence velocity ($\xi$), and 
metallicity ([Fe/H]), determined by previous works. The adopted parameters are 
given in Table \ref{tab:par}. We adopted as main reference the same high-resolution 
analyses quoted in the compilation by Venn et al.\@ (\cite{Venn04}). Our sample 
contains 59 stars analyzed by Fulbright (\cite{F00}, F00 hereafter), 15 by Nissen 
\& Schuster (\cite{NS97}, NS97 hereafter), 7 by Edvardsson et al.\@ (\cite{Ed93}, 
Ed93 hereafter), 4 by Stephens \& Boesgaard (\cite{SB02}, SB02 hereafter), 2 by 
Prochaska et al.\@ (\cite{Pr00}, Pr00 hereafter), 1 by Gratton \& Sneden 
(\cite{GS88}, GS88 hereafter), 1 by Gratton \& Sneden (\cite{GS91}, GS91 hereafter), 
and 1 by McWilliam et al.\@ (\cite{McW95}, McW95 hereafter). 

 Even though different methods were adopted to determine the parameters, we do not 
expect this choice to introduce any systematic effect in our analysis. The likely 
effect of adopting atmospheric parameters with possibly different scales is an 
increase in the scatter of the abundances. We note that Venn et al.\@ (\cite{Venn04}) 
argue that a superficial analysis of their whole sample did not show a major 
inconsistency or zero point difference between the results. We refer the reader to 
the original papers for detailed descriptions of the methods and comparisons between 
the results with other literature determinations. We note, however, that all the papers 
claim reasonable agreement with previous determinations. Moreover, as the parameters 
of most of the stars analyzed here (59 out of 90) come from a single reference, F00, we 
expect any possible difference in the scale of the parameters to have only a minor effect 
on the abundances.

 For the completeness of the discussion, we list some information in Table \ref{tab:lit}  
on the methods adopted by each relevant\footnote{We do not include in Table 
\ref{tab:lit} details on the analyses by Gratton \& Sneden (\cite{GS88,GS91}) and 
by McWilliam et al.\@ (\cite{McW95}), which contribute only 1 star each, and 
on the one by Prochaska et al.\@ (\cite{Pr00}), with only 2 stars, since they will 
have no influence at all on a possibly larger scatter of the results.} work 
as the reference for the atmospheric parameters. As is clear from Table \ref{tab:lit}, F00, 
NS97, and SB02 adopt very similar methods based on spectroscopy to determine the 
atmospheric parameters. On the other hand, the method adopted by Ed93 is completely 
diverse, based only on photometry. In NS97 the same model atmospheres used by 
Edvardsson et al. are adopted. In addition, their work is a differential one with 
respect to two bright stars analyzed by Ed93. They therefore obtain parameters 
essentially on the same scale as the ones derived by Ed93, as demonstrated by their 
Fig. 03, which shows excellent agreement between their T$_{\rm eff}$ with temperatures 
derived using $(b-y)$ calibrations. 

As Be abundances are calculated from lines of the ionized species, log g 
is the most relevant parameter for the analysis. Our sample stars are 
relatively bright and nearby, so we used Hipparcos parallaxes 
(\cite{ESA}) to estimate gravities. Apart from eight stars that show a 
significant difference (larger than 0.28 dex), the agreement is excellent. 
A linear fit to the points yields a line that is statistically indistinguishable from 
the x = y line, with a r.m.s. of 0.20 dex when all the points are considered, 
or of 0.14 dex when the discrepant stars are excluded. These eight stars 
have uncertain parallaxes $\sigma_{\pi}$/$\pi$ $\geq$ 0.20 or are binary stars.

Thus, only a comparison between the results of F00 and Ed93 is necessary to 
ensure the consistency between the atmospheric parameters of most of the 
sample stars. Such a comparison is shown below.

\subsection{Comparison between Edvardsson et al. and Fulbright}\label{ed93f00}

 The comparison between the results of F00 and Ed93 was conducted to assure the 
detailed understanding and proper identification of any possible systematic effect
introduced in the analysis.

\begin{figure*}
\begin{centering}
\includegraphics[width=14cm]{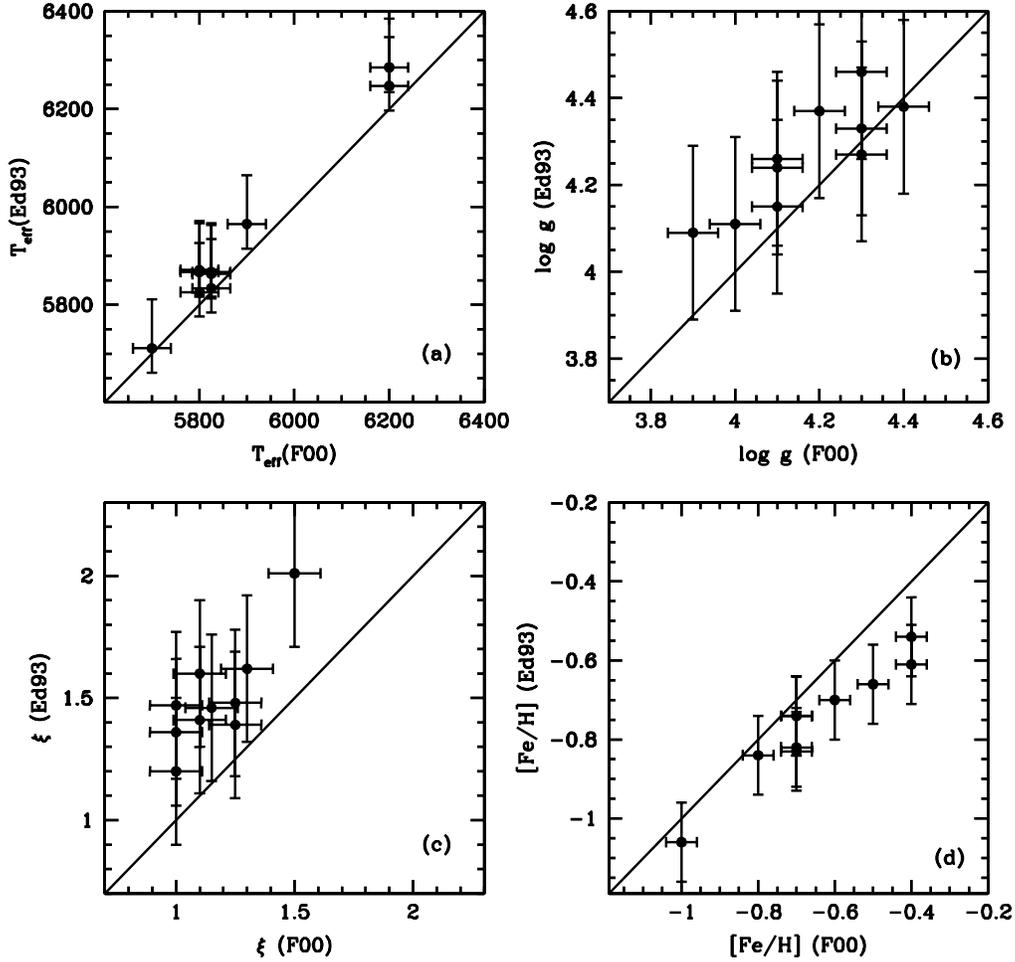}
\caption{Comparison of the adopted atmospheric parameters for
 stars shared by F00 and ED93. The error bars 
with the values listed in the original papers are also shown.}
\label{fig:f00ed93}
\end{centering}
\end{figure*}

 Ten stars are shared between the two analyses. Of these, seven are included in 
our sample. The atmospheric parameters of all the ten stars derived by the two 
papers are compared in Fig.\@ \ref{fig:f00ed93}. The effective temperatures show 
excellent agreement, with a mean difference of 46 K and a maximum difference of 85 K. 
Although the values derived by F00 are systematically lower than the ones derived 
by Ed93, the difference is small, well within the uncertainties of the calculations. 
The difference in surface gravity is also small, with a mean of 0.11 dex and a maximum 
of 0.19 dex, and might be related to the different adopted temperature scales. The 
differences in gravity are close to the uncertainty value. Finally, the metallicity 
([Fe/H]) values show a mean difference of 0.10 dex and maximum of 0.21 dex. This 
difference seems to be systematic, because the values of F00 are always slightly higher than 
the ones derived by Ed93. 

The comparison shows that the differences between F00 and Ed93 are not 
large, but can be ascribed mostly to the uncertainties in each analysis, 
except for the microturbulence (Fig.\@ \ref{fig:f00ed93}). This shows that no 
strong systematic effect will be present on our abundances and argues in favor 
of our assumption that the scales of the different papers are indeed similar, and 
the use of the combined set of parameters introduces only a small and acceptable 
dispersion on the abundances. However, we note 
that the observed difference in [Fe/H] might introduce some dispersion on the data points 
when analyzing plots like log(Be/H) vs.\@ [Fe/H] or [Be/Fe] vs. [Fe/H]. This point has 
to be kept in mind and will be recalled when discussing our results.

\subsection{Uncertainties of the atmospheric parameters}

 In Table \ref{tab:uncer} we list the uncertainties in the atmospheric parameters 
as calculated by each of the four papers that contribute more than two stars 
to our sample.

\begin{table}
\caption{Uncertainties in the adopted atmospheric parameters.}
\label{tab:uncer}
\centering
\begin{tabular}{ccccc}
\noalign{\smallskip}
\hline\hline
\noalign{\smallskip}
Ref. &  T$_{\rm eff}$ (K) & log g & $\xi$ (km s$^{-1}$) & [Fe/H] \\
\hline
F00 & $\pm$40 & $\pm$0.06 & $\pm$0.11 & $\pm$0.04 \\
SB02 & $\pm$75 & $\pm$0.35 & $\pm$0.30 & $\pm$0.10 \\
NS97 & $\pm$50 & $\pm$0.10 & -- & $\pm$0.05 \\
Ed93 & $^{+100}_{-50}$ & $\pm$0.20 & $\pm$0.30 & $\pm$0.10 \\
\hline
adopted & $\pm$100 & $\pm$ 0.15 & $\pm$ 0.30 & $\pm$
0.15\\
\noalign{\smallskip}
\hline
\end{tabular}
\end{table}

 As seen in Table \ref{tab:uncer}, the adopted works claim a rather low uncertainty 
for their T$_{\rm eff}$ values, ranging from 40 to 100 K. As discussed before, 
for the ten stars in common, the mean difference between the temperatures derived 
by Ed93 and F00 is 46 K with maximum of 85 K. These values argue in favor of 
the quoted uncertainties. Nevertheless, in this case we decided to adopt a rather 
conservative value of 100 K as representative of the associated uncertainty. 
This value equals the higher of the values suggested by Ed93 and might be a good 
estimate of the effect of mixing the temperature scales of the different 
papers together. However, as we will see in the following sections the uncertainty in 
T$_{\rm eff}$ has little influence on the derived beryllium abundance.

 For the log g values, the adopted papers claim uncertainties ranging form 
0.06 to 0.35 dex. The comparison between Ed93 and F00 shows a mean difference 
of 0.11 dex with maximum of 0.19 dex. Moreover, the comparison with log g values 
estimated using parallax show an r.m.s. of 0.14 dex. In particular, this 
comparison seems to show that the log g values are rather well constrained, in 
spite of the larger uncertainties quoted by Ed93 and SB02. Therefore, in this case 
we adopted a value of 0.15 dex as representative of the uncertainty in log g.

 The quoted uncertainty on the microturbulence velocity ($\xi$) varies from 
0.11 to 0.30 km s$^{-1}$ in the adopted papers. The comparison between Ed93 and F00, 
however, shows larger differences with a mean of 0.31 km s$^{-1}$ and maximum of 
0.54 km s$^{-1}$. In this case, however, we simply decided to adopt the higher 
value of the quoted uncertainties, 0.30 km s$^{-1}$, as representative. Again, 
as we see below, this choice and the large differences in $\xi$ have no 
significant impact on the beryllium abundances.

The adopted papers list values for the metallicity ([Fe/H]) ranging 
from 0.04 to 0.10 dex for its uncertainty. The comparison between 
Ed93 and F00, however, shows a mean difference of 0.10 dex with maximum of 
0.21 dex. Mainly because of this comparison, we decided to adopt a somewhat 
higher value as representative of the uncertainty than the ones quoted 
in the adopted papers, 0.15 dex. As for the other parameters, except for 
log g, its influence on the final beryllium abundances is weak. The 
adopted uncertainty values are listed in Table \ref{tab:uncer}.

\section{Abundances determination}\label{sec:abun}

\subsection{Synthetic spectra}

 Abundances of beryllium and, for some stars, of lithium were calculated 
using synthetic spectra. In this work, we adopted the codes for calculating 
synthetic spectrum described by Barbuy et al.\@ (\cite{Bar03}) and Coelho 
et al.\@ (\cite{Co05}). We use the grids of model atmospheres calculated by 
the ATLAS9 program (Castelli \& Kurucz \cite{CK03}), without overshooting. 
These models assume local thermodynamic equilibrium, plane-parallel geometry, 
and hydrostatic equilibrium. Codes for interpolating among the grids were adopted.

 On the spectra of late-type stars, only the \ion{Be}{ii} $^{2}$S - $^{2}$P$_{0}$ 
resonance lines at 3131.065 \AA\@ and 3130.420 \AA\@ are strong enough to be 
useful for an abundance analysis. These lines are near the atmospheric cut-off 
at 3000 \AA\@ in a region of low detector sensitivity. This near-UV region is 
extremely crowded, full of atomic and molecular lines, some of them still lacking 
proper identification. The determination of Be abundances, thus, needs to be done 
with spectrum synthesis taking all the blending nearby features into account. Our 
database of molecular lines include the OH (A$^{2}\Sigma$-X$^{2}\Pi$), NH 
(A$^{3}\Pi$-X$^{3}\Sigma$), and CN (B$^{2}\Pi$-X$^{2}\Sigma$) band systems as 
implemented by Castilho et al.\@ (\cite{Ca99}) and the CH band systems 
(A$^{2}\Delta$-X$^{2}\Pi$), (B$^{2}\Sigma$-X$^{2}\Pi$), and (C$^{2}\Sigma$-X$^{2}\Pi$) 
as implemented by Mel\'endez et al.\@ (\cite{Me03}), all of them affecting this 
spectral region.

\subsection{Beryllium}

\begin{figure}
\begin{centering}
\includegraphics[width=9cm,height=11cm]{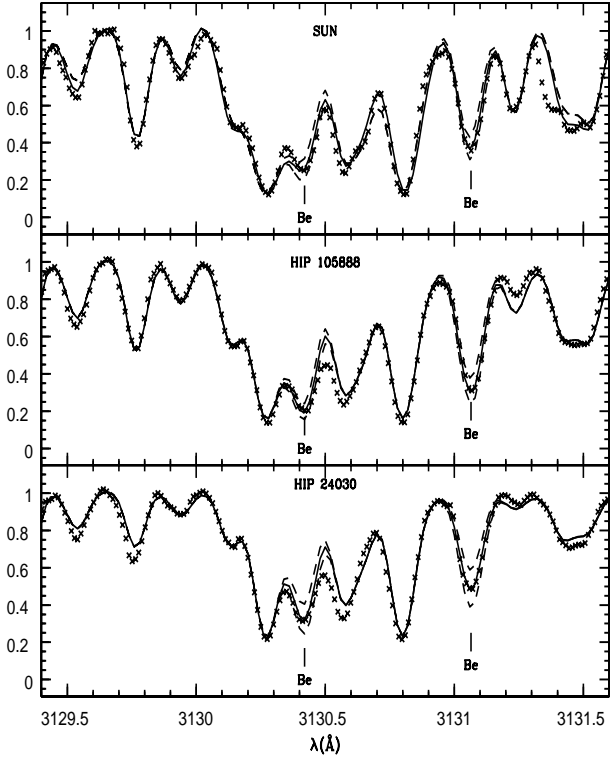}
\caption{Fit to the region of the beryllium lines in the solar
  spectrum and in the stars HIP 105888 and HIP 24030. The crosses represent 
the observed spectrum, the solid line the best synthetic fit, and 
the dashed lines represent changes of $\pm$0.20 dex in the Be abundance of the 
best fit. A solar abundance of A(Be) = 1.10 was determined.} 
\label{fig:sun}
\end{centering}
\end{figure}
\begin{figure}
\begin{centering}
\includegraphics[width=9cm,height=11cm]{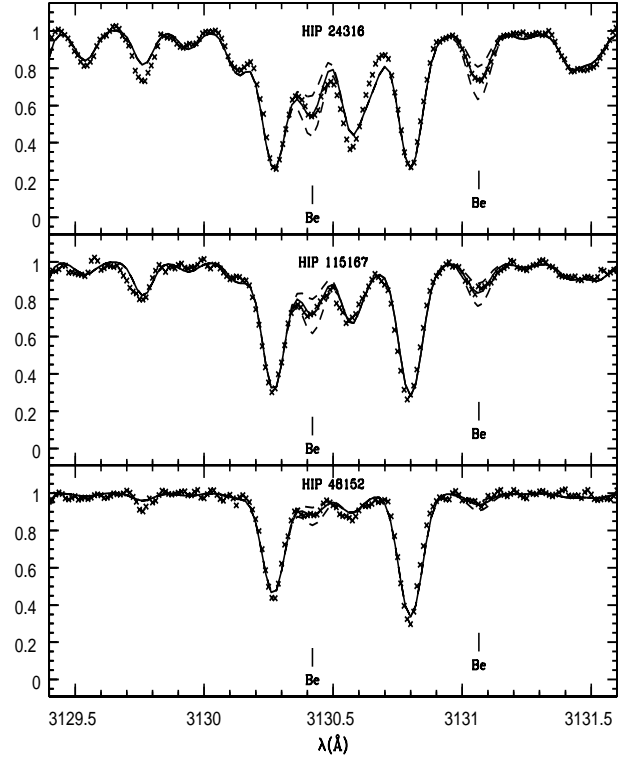}
\caption{Fit to the region of the beryllium lines in the stars HIP 24316, 
HIP 115167, and HIP 48152. The lines are as in Fig.\@ \ref{fig:sun}.}
\label{fig:be01}
\end{centering}
\end{figure}

 The line list of atomic lines compiled by Primas et al.\@
 (\cite{Pr97}) was adopted in this work. This same line list was
 adopted in many analyses of Be abundances in literature (Garc\'{\i}a
 P\'erez \& Primas \cite{GPP06}, Primas et al.\@
 \cite{Pr97,Pr00a,Pr00b}, Randich et al.\@ \cite{Ran02,Ran07}). The
 adopted log $gf$ of the Be lines are those commonly used in the
 literature, $-$0.168 and $-$0.468 for 3131.065 \AA\@ and 3130.420
 \AA, respectively. In this line list, the blending line affecting
 the blue wing of the Be 3131 line is assumed to be an \ion{Fe}{i} line
 in 3131.043 \AA, with log $gf$ = $-$2.517 and $\chi$ = 2.85 eV. The
 parameters of this line were constrained using several stars of
 different parameters and metallicities (see Primas et al.\@
 \cite{Pr97} for details). The proper identification of this line,
 however, remains controversial in the literature. Castilho et al.\@
 (\cite{Ca99}), for example, adopt an \ion{Fe}{i} line at 3130.995 \AA\, 
in the analysis of Be in red giants, with log $gf$ = $-$3.30 and
 $\chi$ = 3.00 eV. Other possibilities explored in the literature
 include an \ion{Mn}{i} line at 3131.037 \AA\@ adopted by Garcia Lopez
 et al.\@ (\cite{GL95a}) and an \ion{Mn}{ii} line at 3131.015 adopted
 by Boesgaard \& King (\cite{BK02}). In this work, we opted for the
 \ion{Fe}{i} included by Primas et al. (\cite{Pr97}), in particular to
 maintain the consistency with all the papers that adopted the same
 atomic line list.

There is currently a controversy about a possible missing 
continuum opacity source affecting the near-UV region, hence Be
abundances (Asplund \cite[and references therein]{As04}). The code
used to calculate the synthetic spectra does not include bound-free
opacities due to metals (Smiljanic \& Barbuy \cite{SB07}). However, we
are mostly interested in relative abundances between stars of similar
composition, so the bulk of the analysis is not affected by this
uncertainty. The slope between Be and Fe or $\alpha$ could be
affected. With respect to this, we note that our most metal-rich stars
have [Fe/H] = $-$0.50; for metal-poor objects, the missing opacity
effect is expected to be negligible. 

 In fact, Balachandran \& Bell (\cite{BB98}) argue that the
  \ion{Fe}{i} bound-free opacity should be increased by a factor
  1.6. This is equivalent to an increase of 0.20 dex in Fe
  abundance. We ran a test using the Kurucz suite of codes to test the
  influence of this increase in Fe abundance in the calculation of Be 
abundance for a model with [Fe/H] = $-$0.50, appropriate for 
star HIP 31639. The difference of 0.022 dex is negligible. We therefore 
expect that this issue does not really affect our analysis and 
conclusions in any significant way.  

 With the line list described above, we fitted the Be lines region in the solar 
UVES\footnote{The spectrum is available for download at the ESO website: 
www.eso.org/observing/dfo/quality/UVES/pipeline/solar\_spectrum
.html} spectrum, adopting the parameters T$_{\rm eff}$ = 5777 K, log g = 4.44, 
and $\xi$ = 1.00 km/s, and obtained A(Be) = 1.10 (Fig.\@ \ref{fig:sun}). This 
abundance is in excellent agreement with the one found by Chmielewski et 
al.\@ (\cite{Ch75}), A(Be) = 1.15, usually adopted as the reference photospheric 
solar abundance. We also note excellent agreement with the abundance derived 
by Randich et al.\@ (\cite{Ran02}), A(Be) = 1.11, using this same atomic line 
list but a different molecular line list and by fitting the Kurucz Solar Flux 
Atlas (Kurucz et al.\@ \cite{K84}). 

 From the 90 stars in the sample, abundances of Be were determined for 83: 
76 detections and 7 upper limits. The 7 stars for which abundances could not be 
calculated were affected by different problems, some with profiles affected by cosmic 
rays, and others were found to be spectroscopic binaries. These stars will not be 
considered further in the analysis. Examples of the fits used to derive the Be 
abundance are given in Figs.\@ \ref{fig:sun} and \ref{fig:be01}. 

 For most of the stars, both beryllium lines are well-fitted with the same abundance. 
In some stars, however, the lines are best fitted with slightly different values. In 
most cases, the difference amounts to $\sim$ 0.04 dex, a value that can be completely 
attributed to the uncertainty of the fitting procedure itself
and/or to the determination 
of the continuum. In one extreme case, however, the difference amounts to 0.24 dex 
(HIP 31639). This discrepancy is likely due to an inefficient treatment of the blending 
features caused, for example, by abundance ratios that are significantly different from solar of 
the blending lines directly affecting the Be lines. For these discrepant cases, we consider the 
abundance derived solely from the 3131 \AA\@ line to be more reliable, since its profile is 
visually less affected by blends. In Table \ref{tab:par} we list both the beryllium 
abundance determined solely from the 3131 \AA\@ line and the average of the two lines.

 We tested the dependence on atmospheric models by also using a version of the Kurucz model 
atmospheres with enhanced abundances of $\alpha$-elements, [$\alpha$/Fe] = +0.40 dex. 
Metal-poor stars are known to have enhanced abundances of $\alpha$ elements. Some of these 
elements are particularly important as electron donors and, as such, may influence some 
continuum opacity sources (bound-free and free-free absorption of H$^{-}$, for example). 
We found, however, this effect to be small. In the following discussion, we adopt the beryllium 
abundance derived solely from the 3131 line with normal model atmosphere as the reference value.

\subsection{Lithium}

 In our analysis, we expect to recover the initial beryllium abundance the star 
had at the time of its formation, since this is the one produced by the interaction 
between the Galactic cosmic-rays (GCR) and the ISM. It is important then to identify 
stars where significant mixing of the photospheric material with deep hotter regions 
may have altered the initial Be abundance. The abundances of Li will help in this.

 Most stars in our sample have been analyzed for Li abundances before. Our interest 
in Li in this work is only related to the information it might give on the possible 
effects of convective mixing, not on the absolute value of the Li abundance. Thus, 
we decided to adopt Li values from the literature whenever it was available. As 
main reference for lithium abundances (32 stars), we adopted the work by Charbonnel 
\& Primas (\cite{CP05}), who analyzed a large number of previous literature results 
in detail. Whenever possible, we adopted their so-called \emph{most consistent values} 
including NLTE corrections. As a second preferred reference (24 stars), we adopted the 
results of the survey of Li abundances by Chen et al. (\cite{Ch01}). These Li abundances 
also include corrections for NLTE effects. For stars not included in these papers, the Li 
abundance was obtained from a number of other papers: for seven stars, abundances were taken 
from Boesgaard et al. (\cite{Bo05}); for four stars, abundances were taken from F00; for 
three stars, from Romano et al. (\cite{Ro99}); while Favata et al. (\cite{Fav96}), 
Gratton et al. (\cite{Gr00}), Ryan \& Deliyannis (\cite{RyD95}), Spite et al. (\cite{Sp94}), 
and Takeda \& Kawanomoto (\cite{TK05}) each contributed with one star. 

\begin{figure}
\begin{centering}
\includegraphics[width=7cm]{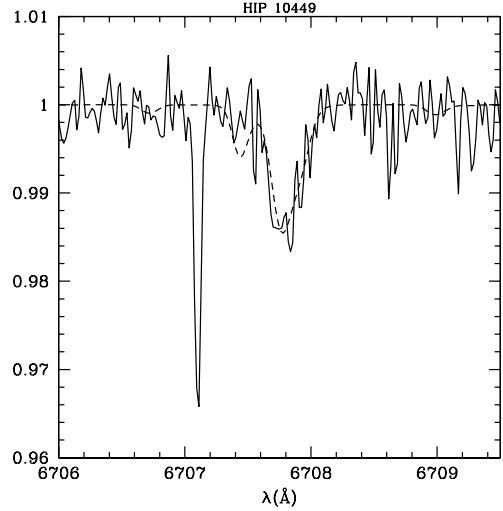}
\caption{Fit to the \ion{Li}{i} doublet at 6707.8 \AA\@ in star HIP 10449.
The solid line is the observed spectrum and the dashed line the synthetic one.} 
\label{fig:li}
\end{centering}
\end{figure}

 Lithium abundances were not available in the literature for 15 stars of our sample. For 
these stars, we present new abundances derived using spectral synthesis. We adopted the 
atomic data for the \ion{Li}{i} doublet and neighboring lines listed by Andersen et al.\@
 (\cite{And84}). The Li abundance of star HIP 92781 could not be calculated because of 
a damaged profile. The lithium abundances are listed in table \ref{tab:par} along with a flag 
indicating the sources. An example of a synthetic fit to the \ion{Li}{i} doublet is 
shown in Fig. \ref{fig:li}.

 Since Li and Be are destroyed at different depths, even stars depleted in Li are 
expected to have preserved their original Be abundance (as is likely in the case of 
the Sun and of cool main sequence stars, cfr. e.g. Randich \cite{Ran07}). Whatever 
the Li reference adopted, non-mixed metal-poor dwarfs are expected to have a lithium 
abundance close to the primordial value of the Spite plateau (Spite \& Spite \cite{SS82}). 
The exact value of the plateau and possible dependencies with [Fe/H] and/or T$_{\rm eff}$ 
are still controversial, but for our purpose it is sufficient to assume a typical value of 
A(Li) = 2.20.

\subsection{Alpha elements}

 Abundances for the $\alpha$-elements were taken from the work of Venn et al.\@ 
(\cite{Venn04}). These are average abundances of Mg, Ca, and Ti determined in the 
same papers, when available, used as the source of the atmospheric parameters we adopt 
in this work. The abundances have not been homogenized since different works use 
different spectral lines to calculate the abundances. We refer the reader to Venn 
et al.\@ (\cite{Venn04}) and the original papers for more details. We list the 
abundances in Table \ref{tab:prob}. We plan to improve this analysis 
by evaluating CNO abundances for all the stars in a subsequent work (Smiljanic 
et al.\@ 2009, in preparation). 

\subsection{Uncertainties}

\setcounter{table}{6}
\begin{table*}
\caption{The uncertainties of the Be abundance introduced by the
  uncertainties of the atmospheric parameters.}
\centering 
\label{tab:sigma}
\begin{tabular}{cccccccc}
\noalign{\smallskip}
\hline\hline
\noalign{\smallskip}
Star & $\sigma_{\rm Teff}$ & $\sigma_{\rm log g}$ & $\sigma_{\rm \xi}$ &
 $\sigma_{\rm [Fe/H]}$ & $\sigma_{\rm cont.}$  & $\sigma_{\rm blend}$  & $\sigma_{\rm total}$ \\
\hline
 HIP 53070 & $\pm$ 0.03 & $\pm$ 0.06 & $\pm$ 0.01 & $\pm$ 0.00 & $\pm$ 0.05 & $\pm$ 0.08 &
 $\pm$ 0.12 \\
 HIP 104660 & $\pm$ 0.03 & $\pm$ 0.08 & $\pm$ 0.01 & $\pm$ 0.02 & $\pm$ 0.05 & $\pm$ 0.08 &
 $\pm$ 0.13 \\
 HD 159307 & $\pm$ 0.04 & $\pm$ 0.08 & $\pm$ 0.01 & $\pm$ 0.03 & $\pm$ 0.05 & $\pm$ 0.08 &
 $\pm$ 0.13 \\
\noalign{\smallskip}
\hline
\end{tabular}
\end{table*}
\begin{table*}
\caption{The uncertainties of the Li abundances, calculated in this work, introduced by the
  uncertainties of the atmospheric parameters. The total uncertainty 
also includes the effect of the continuum uncertainties ($\pm$ 0.05 dex) on the Li abundance.}
\centering 
\label{tab:sigma2}
\begin{tabular}{ccccccc}
\noalign{\smallskip}
\hline\hline
\noalign{\smallskip}
Star & $\sigma_{\rm Teff}$ & $\sigma_{\rm log g}$ & $\sigma_{\rm \xi}$ &
 $\sigma_{\rm [Fe/H]}$ & $\sigma_{\rm cont.}$  & $\sigma_{\rm total}$ \\
\hline
 HIP 74067 & $\pm$ 0.10 & $\pm$ 0.00 & $\pm$ 0.00 & $\pm$ 0.00 & $\pm$ 0.05 &
 $\pm$ 0.11 \\
\noalign{\smallskip}
\hline
\end{tabular}
\end{table*}

\subsubsection*{Beryllium}

 The Be abundance determination is affected by the uncertainty in the atmospheric 
parameters and uncertainty in the determination of the pseudo-continuum. To 
estimate the effect of the atmospheric parameters, we change each one by its 
own error, keeping the other ones with the original adopted values, and recalculate 
the abundances. Thus, we measure what effect the variation of one parameter has in 
the abundances. These effects are listed in Table \ref{tab:sigma}. The calculations 
were done for three stars representative of the range of parameters defined by our 
sample. The total uncertainty also includes the effect of the continuum 
uncertainties ($\pm$ 0.05 dex) and the influence of the unidentified blends 
($\pm$ 0.08 dex) on the Be abundance. We recall that we adopted larger uncertainties 
for T$_{\rm eff}$ and [Fe/H] when compared to the values adopted by F00 and Ed93. 
Thus, the uncertainties caused by these parameters in the abundances already consider 
some of the effects of putting the two samples together.

 The uncertainty due to the continuum was determined by estimating the sensitivity 
of the derived Be abundance on the continuum level adopted. This uncertainty is 
mostly related to the S/N of the spectrum. After a number of tests with different 
choices for the continuum level, we found this effect to introduce a typical 
uncertainty of $\pm$ 0.05 dex.

 Another factor that may influence the determination of Be abundances are the blends 
affecting the Be lines. In particular, we adopt the value given only by the 3131 
\AA\@ line as the abundance. In this case, the predicted line of \ion{Fe}{i} is the most 
important blend. The influence of this predicted line decreases with increasing 
temperature and/or decreasing metallicity. In a series of tests, we noticed that 
removing the Fe line would increase the beryllium abundance in a star with 
T$_{\rm eff}$ $\sim$ 5200 K and metallicity $\sim$ $-$0.90 by 0.15 dex. In a star with 
T$_{\rm eff}$ $\sim$ 6200 K and metallicity $\sim$ $-$0.70, no effect is seen at all. 
We adopt a value of $\pm$ 0.08 dex, the average between the maximum and minimum effect 
of neglecting blend, to represent this source of uncertainty and extend this value to 
all the sample stars. If our tentative identification of the line is wrong, we are 
wrongly modeling the effects of its variation with [Fe/H] and Teff. This effect, 
however, is not likely to be as strong as the complete neglect of the line.

 Assuming that the effects of the uncertainties from the parameters, the continuum, and 
the blends are independent, we may add them in quadrature to estimate the total uncertainty, 
which is also listed in Table \ref{tab:sigma}.

\subsubsection*{Lithium}

 In the case of lithium, the main uncertainties also come from the uncertainties in 
the atmospheric parameters and from the determination of the pseudo-continuum. These  
uncertainties were determined in a similar way to the one described above for Be and are 
listed in Table \ref{tab:sigma2}. While the main parameter affecting the abundance 
is log g for Be, for Li it is T$_{\rm eff}$. The uncertainties we derive are only valid for 
the Li abundances determined in this work, for the values adopted from the literature 
we refer the reader to the original papers.

\subsubsection*{Alpha elements}

 As in Venn et al.\@ (\cite{Venn04}), we do not conduct a detailed error analysis on 
the $\alpha$-elements since they were determined from a variety of indicators. We decided 
to adopt the same representative uncertainty as adopted by Venn et al.\@ (\cite{Venn04}), 
$\sigma_{[\alpha/Fe]}$ = $\pm$ 0.15 dex.

\section{Comparison with previous results}\label{sec:comp}

 A number of the stars analyzed in this work have beryllium abundances 
previously determined in the literature. A detailed comparison of our 
results with those obtained previously for each star is shown in 
Appendix \ref{app:comp} for the interested reader. As is clear from 
this comparison, there is also a difference in log g in most cases 
where our abundances do not agree with previous results. This shows that 
to establish a consistent and reliable gravity scale is important 
for properly determining the Be abundance. We have confidence in our 
results given the good agreement found with log g derived using the 
Hipparcos parallaxes.

In the following, we discuss both the spread in the atmospheric parameters between the 
results found in these papers (in particular the one of [Fe/H], that 
might have some influence in the spread of the relation between Be and Fe) 
and the spread of the Be abundances. The spread of Be abundances might 
show to what extent the absolute scale of the abundances is reliable.

\begin{figure}
\begin{centering}
\includegraphics[width=7cm]{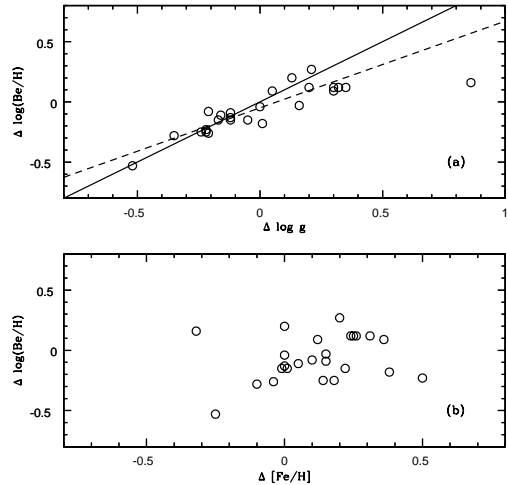}
\caption{(a) Diagram showing the difference between the Be abundances determined in this work 
and the ones from the literature, as a function of the difference in log g. The correlation 
coefficient, $\rho$ = 0.84, including the outlier, and $\rho$ = 0.90, excluding it, indicates the 
reality of the linear correlation apparent from the figure. The dashed line indicates a linear 
fit, excluding the outlier, and the solid line is the line x = y. (b) The difference between the 
Be abundances as a function of the difference in [Fe/H]. The correlation coefficient, $\rho$ = 0.24, 
indicates the lack of correlation as apparent from the figure.}
\label{fig:delta}
\end{centering}
\end{figure}
\begin{figure}
\begin{centering}
\includegraphics[width=7cm]{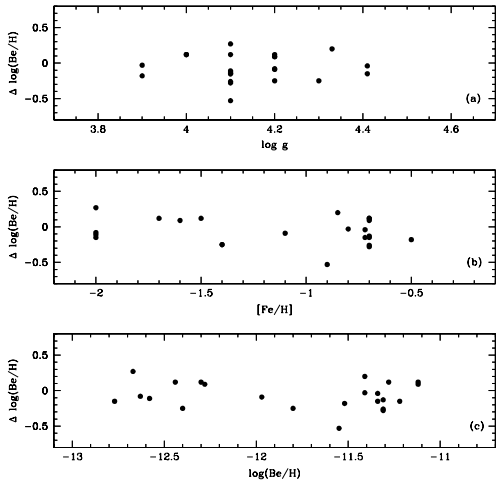}
\caption{(a) Diagram showing the difference between the Be abundances determined in this work 
and the ones from the literature, as a function of the log g values. The correlation 
coefficient, $\rho$ = 0.14, indicates the lack of correlation also apparent from the figure.
(b) The difference between the Be abundances as a function of the [Fe/H] values. The correlation 
coefficient, $\rho$ = $-$0.26, indicates the lack of correlation also apparent from the figure.
(c) The difference between the Be abundances as a function of the log(Be/H) values. The correlation 
coefficient, $\rho$ = $-$0.20, indicates the lack of correlation also apparent from the figure.}
\label{fig:deltabe}
\end{centering}
\end{figure}

\subsection{The spread of the parameters}

Although in the comparison shown in Appendix \ref{app:comp} we show results from all papers that 
previously determined Be abundances for our sample stars, in what follows we consider only 
papers published after 1997 (including). Early works most likely employed data and techniques of 
lower quality then modern analyses and therefore are prone to greater uncertainties. We also do 
not consider upper limits for this comparison, resulting in 25 possible comparisons. Multiple 
comparisons for a given star are possible.

The average difference between our log g values and the ones adopted in the literature is 
0.00 $\pm$ 0.28. If we exclude the comparison of our log g of HIP 114962, 4.30, with the 
one adopted by Molaro et al.\@ (\cite{MBCP97}), 3.44, the average is $-$0.03 $\pm$ 0.23. This 
shows that, although no systematic effect seems to be present, a wide spread of the log g 
values adopted in the literature exists. 

Regarding the metallicity, [Fe/H], the average difference between our values and the 
ones adopted in the literature is +0.12 $\pm$ 0.19. If again we exclude the comparison with 
of HIP 114962 by Molaro et al.\@ (\cite{MBCP97}), the average is +0.13 $\pm$ 0.17. This 
comparison seems to indicate a weak systematic effect in the sense of our adopted [Fe/H] values 
to be higher then the ones adopted in other analyses. It also indicates the existence of 
a spread of the [Fe/H] values adopted in the literature. This spread is a measure of the 
extent to which the determinations can be considered reliable. It will be important to 
keep this in mind when we discuss the relation between log(Be/H) and [Fe/H] in the 
sections to follow. We note that the the uncertainty we assume for the [Fe/H] values 
is similar to the magnitude of the spread found in this comparison.

\subsection{The spread on the Be abundances}

The average difference between our Be abundances and the ones determined in the literature 
is $-$0.06 $\pm$ 0.19. This indicates that, on average, our values are slightly lower 
than the ones previously derived in the literature. As shown in Fig.\@ \ref{fig:delta}, 
the difference in the Be abundance correlates very well with the difference in log g but 
not with the difference in [Fe/H]. This comparison shows the extent to which different 
analyses of Be can be trusted when compared to each other. It also shows that most of 
this spread is related to the difference in the adopted log g. Also, no correlation between 
the difference in the Be abundance and the own abundance, [Fe/H], and log g was found 
(Fig.\@ \ref{fig:deltabe}). Therefore, this difference does not seem to be due to some 
systematic effect of the analysis. As shown before, our log g values are in excellent 
agreement to the ones derived using Hipparcos parallaxes, giving confidence in the 
abundances we derived.


\section{Be depleted stars}\label{sec:mix}

 The light elements Li, Be, and B are very fragile and are destroyed
 when in regions inside stars with temperatures of $2.5 \times 10^{6}$
 K, $3.5 \times 10^{6}$ K, and $5 \times 10^{6}$ K,
 respectively. Main-sequence stars with M $\lesssim$ 1.5 M$_{\odot}$ 
have a surface convective zone that might reach regions hot
 enough to deplete the surface abundance of these elements. The depth
 of the convective zone is a function of effective temperature and
 metallicity; it is larger when the star is cooler and more metal-rich.

In this section we use Li and Be abundances to evaluate which stars
are affected by some kind of mixing event. The stars that do not
display the original Be abundances in their photospheres will be
excluded from further discussion. We also exclude star HIP 59490, 
the Be-rich star discussed in detail in Smiljanic et al. (\cite{Sm08}). 
Both Li and Be abundances of this star are not a result of the normal 
evolution of these elements in the Galaxy, but a likely result of a 
peculiar event tentatively identified as a hypernova.

Seven stars in our sample have only Be upper limits. All have 
also low Li abundances. Three of them are subgiants, HIP 17001, HIP 71458, 
and HIP 77946, and have probably diluted both Li and Be due to the 
deepening of the convective envelope, as expected from stellar 
evolution. One is HIP 55022 (HD 97916), a star first found to be 
Li depleted by Spite et al.\ (\cite{Sp84}). It is one of the 
so-called ``ultra-lithium-deficient'' stars (Ryan et al.\ \cite{Ry01}), 
a group of a few metal-poor stars that deviate from the Li plateau and 
only have Li upper limits. Ryan et al.\ (\cite{Ry01}) suggest they 
are formed by the same mechanism as forms field blue stragglers. 
Boesgaard (\cite{Boe07}) find these stars are also Be-depleted, 
a result we confirm for HIP 55022. The thick-disk dwarf HIP 59750 is 
more metal rich ([Fe/H] = $-$0.60) than the stars that usually define 
the Li plateau. At its temperature (6200 K), no Li (or Be) depletion is 
expected. Star HIP 36818 has a lower temperature (5672 K) and higher 
metallicity ([Fe/H] = $-$0.83) than plateau stars. Stars of similar 
temperature and metallicity also show Li depletion but normal Be. Both 
seem to be affected by stronger mixing than other similar stars. They 
might as well be metal-rich counterparts of the ``ultra-lithium-deficient'' 
stars mentioned above. Star HIP 19814 has a rather high Be upper limit, 
consistent with stars of similar temperature and metallicity where Be 
was detected. Thus, strictly speaking it cannot be considered to be Be-depleted. 
All these 7 stars will not be considered in the discussion presented in 
the following sections. We thus start with a sample of 75 stars
where Be was detected.

\begin{figure}
\begin{centering}
\includegraphics[width=7cm]{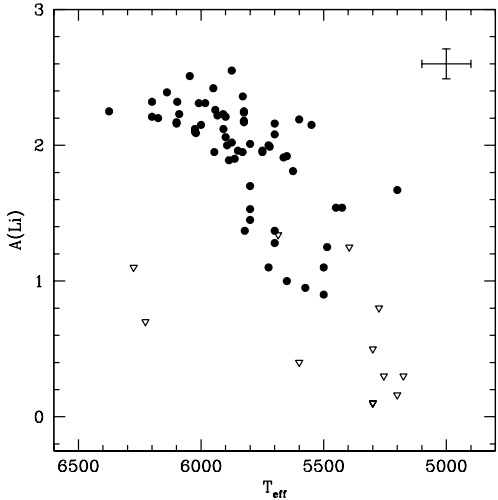}
\caption{Abundances of Li as a function of temperature for the sample
  stars. Detections are shown as filled circles and upper limits as
  upside down open triangles.}
\label{fig:litempall}
\end{centering}
\end{figure}

Spite \& Spite (\cite{SS82}) were the first to show halo stars with a
range of metallicity to have a constant Li abundance (the so-called
Spite plateau). Thick-disk, metal-poor dwarfs also show the same
feature (Molaro et al.\@ \cite{MBP97}). Although departures from the
plateau due to lithium production by various stellar sources seem to
start at [Fe/H] $\sim$ $-$1.00, a sudden large increase with stars
reaching A(Li) $\sim$ 3.0 is observed only for [Fe/H] $>$ $-$0.45, out
of the explored range (Travaglio et al.\@ \cite{Tr01}). Although
some discussion exists on whether the plateau is primordial in origin
or a result of some depletion mechanism (Bonifacio \& Molaro
\cite{BM97}; Piau et al.\@ \cite{PiBe06}; Bonifacio et
al. \cite{B07}; Korn et al.\@ \cite{Ko07}), we assume in this work
that plateau stars did not suffer a previous significant depletion of
Li (and Be) in their atmospheres.

In Fig. \ref{fig:litempall} we show the lithium abundances as a
function of T$_{\rm eff}$. As previously found in the literature,
stars with T$_{\rm eff}$ $\gtrsim$ 5800 K define the Li plateau, while
stars with  T$_{\rm eff}$ $\lesssim$ 5800 K start to show the effects
of Li depletion. There are only two stars with T$_{\rm eff}$ above
6000 K and A(Li) below the plateau. HIP 85963 is a thin-disk star with
[Fe/H] = $-$0.71, T$_{\rm eff}$ = 6227, and A(Li) $\leq$ 0.70, and HIP
107975 is a thin-disk star with [Fe/H] = $-$0.50, T$_{\rm eff}$ =
6275, and A(Li) $\leq$ 1.10. Both seem to be lithium-dip\footnote{
    The so-called lithium-dip is a strong decrease in Li abundances of
    main sequence stars in an interval of $\sim$ 300 K around $\sim$
    6700 K, an effect first noticed by Wallerstein et al.\ (\cite{Wal65}) 
in the Hyades and later confirmed by Boesgaard \& Tripicco
    (\cite{BT86}). A discussion of the physical mechanisms possibly
    associated to the lithium dip can be found in Charbonnel \& Talon
    (\cite[and references therein]{CT99}).} stars or to be evolving
from the lithium dip. Since they have most likely experienced strong
mixing, we do not consider them in the further analysis, reducing
our sample to 73 stars. Similar plots of A(Li) as a function of log g
and [Fe/H] do not reveal any other star showing strong mixing.

\begin{figure}
\begin{centering}
\includegraphics[width=7cm]{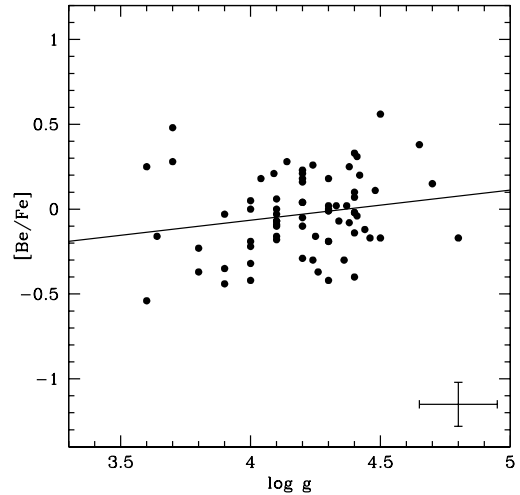}
\caption{Abundances of Be relative to Fe as a function of log g for the sample stars. 
A linear fit to the points is also shown.} 
\label{fig:belogg}
\end{centering}
\end{figure}

No trends or deviating stars are seen in plots of
[Be/Fe]\footnote{ We adopt the meteoritic solar abundance,
    A(Be)$_{\odot}$ = 1.41 (Lodders \cite{L03}) to calculate the
    [Be/Fe] ratio.} as a function of T$_{\rm eff}$ and [Fe/H]. 
We use [Be/Fe] in this case because log(Be/H) is expected to vary
  from star to star, while [Be/Fe] is expected to be roughly
  constant. A small trend with a correlation coefficient of $\rho$ =
0.19 is seen in a plot of [Be/Fe] as a function of log g (Fig.\@
\ref{fig:belogg}), suggesting a small decrease in Be abundance with
decreasing log g. Stars with log g $\geq$ 4.00 seem to form a plateau,
while some stars with log g $<$ 4.00 seem to have low [Be/Fe]. A
similar effect is not seen in the Li abundances, which argues against
a simple dilution or mixing effect. These stars span almost the whole
range of metallicity (from $-$0.50 to $-$1.80), which argues against a
chemical evolution effect. The trend might be related to 
the dependence of Be on gravity, suggesting that log g for the lower 
gravity stars is possibly not well constrained. Nevertheless, we 
see no strong reason to exclude these stars from the discussion.

Thus, only two stars are excluded from the sample because of 
possible mixing effects. The remaining 73 stars retain their
original Be abundances and are used in the following sections to
discuss the galactic evolution of this element and the use of Be as a
time scale tracing the different star formation history of the halo
and the thick disk.

\begin{figure}
\begin{centering}
\includegraphics[width=7cm]{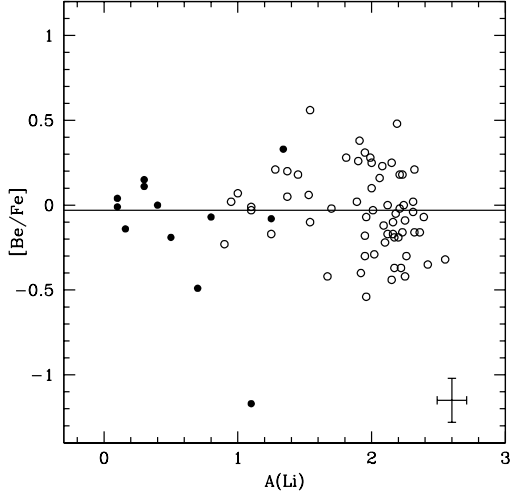}
\caption{The [Be/Fe] ratio as a function of the Li abundance. Stars
  with beryllium abundances but lithium upper limits are shown as
  filled circles and stars with both beryllium and lithium detections
  are shown as open circles. A linear fit to the points, without the
  excluded stars, is also shown. No correlation between the
  abundances of Li and Be is seen.}  
\label{fig:beli}
\end{centering}
\end{figure}

Finally, in Fig.\@ \ref{fig:beli} we plot the [Be/Fe] as a function of
the lithium abundances. No clear trend between Be and Li is seen. The
stars with lower Be abundance are not the ones with smaller Li
abundance. Actually, most of the stars with Li upper-limits have Be
abundances compatible with stars that show no signs of Li
depletion. We note that three stars for which we determined only
a Be upper limit have Li detection, however, they are cool,
  T$_{\rm eff}$ $<$ 5500 K and Li-depleted relative to the plateau.
This is probably a result of the much higher S/N obtained in the red
regions of the spectra. These stars are nevertheless very interesting
and should be further investigated.

\section{Galactic evolution of Be }\label{sec:evo}

Once we have cleaned our original sample by identifying the stars that
have suffered Be dilution or depletion, or for which we cannot
  determine Be abundances, we can investigate the evolution of Be in
the Galaxy. We are left with 73 stars, 6 from the thin disk, 27 from
the thick disk, and 39 from the halo. One star has a 50\% probability of
being either halo or thick disk. We keep this star and the thin disk
stars in the sample because according to different kinematic criteria
they receive a different classification (such as in the dissipative
and accretion components defined by Gratton et al.\@ \cite{G03} - see
below). In Fig. \ref{fig:logbefehall} we show a diagram of the [Fe/H]
vs.\@ log(Be/H) and in Fig. \ref{fig:logbealfahall} a diagram of
[$\alpha$/H] vs.\@ log(Be/H) for all the 73 stars. In both figures,
the points are distributed along linear relations, as previously found
in the literature. We fitted a straight line to the data, taking the errors 
on both axes into account. Following the discussion in
Sect. \ref{sec:comp} we adopted an error of 0.15 dex on [Fe/H] and
[$\alpha$/H]. The adopted error on the Be abundance was 0.13 dex. The
resulting fits are the following:

\begin{equation}
{\rm log(Be/H)} \; = \; (-10.38 \pm 0.08) \; + \; (1.24 \pm 0.07 ) \; {\rm [Fe/H]}
\end{equation}

\begin{equation}
{\rm log(Be/H)} \; = \; (-10.62 \pm 0.07) \; + \; (1.36 \pm 0.08) \; {\rm [\alpha/H].}
\end{equation}
\noindent 
For the fit with [Fe/H] the goodness-of-fit provides a probability
of 0.14, while for the one with [$\alpha$/H] this is 0.76.
Although both fits are formally acceptable, the difference
in goodness-of-fit  between the two is remarkable.
The probability is very sensitive to the error estimates. 
If we decrease our error estimate
on [Fe/H] and [$\alpha$/H] to 0.12 dex, the probability
of the two fits drops to 0.002 and 0.14 respectively.
Therefore, while the linear fit with  [$\alpha$/H] remains
acceptable, the one with  [Fe/H] would not, generally, be considered
acceptable.
Visual inspection of Figs.\@ \ref{fig:logbefehall}  and 
\ref{fig:logbealfahall} confirms the impression derived
from the statistical analysis. While in Fig. \ref{fig:logbealfahall} 
one can see a clear trend, albeit with an obvious scatter, one has the impression 
in Fig.\ref{fig:logbefehall} of being able to see two parallel linear relations.

The most important relation  to study the chemical evolution of Be is the 
one with [O/H], since oxygen gives the highest contribution to the spallation process 
responsible for forming Be. In the absence of oxygen abundances for our 
complete sample, we decided to use  the average abundance of $\alpha$-elements, as 
compiled by Venn et al.\@ (\cite{Venn04}). These are typically an average
of Mg and Ca; when available Ti also enters into the definition
of $\alpha $ by Venn et al.\@ (\cite{Venn04}).
These elements show behavior with metallicity similar to 
oxygen. It is therefore reasonable to use [$\alpha$/H]
as a proxy for [O/H]. No major inconsistency 
should be introduced by this choice.
That a single linear relation links Be and $\alpha$ elements,
even when considering stars extracted from different
Galactic populations (halo, thick disk, thin disk)
is a consequence of the uniqueness of the physical process
leading to Be formation.
A non-uniqueness of the relation between Be and Fe, if present,
could be a consequence of the different evolution 
of Fe in the different Galactic populations.

\begin{figure}
\begin{centering}
\includegraphics[width=7cm]{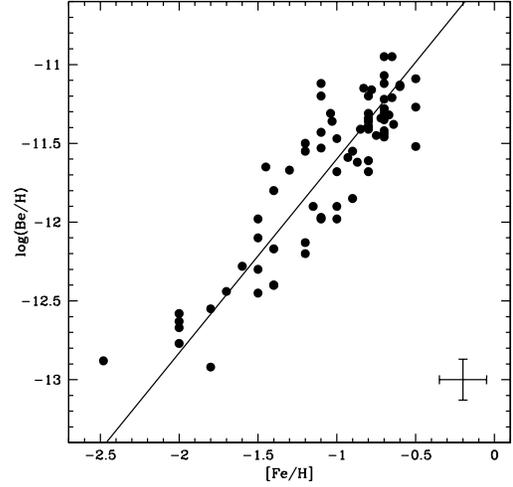}
\caption{Diagram of [Fe/H] vs.\@ log(Be/H) for all the 73 stars showing a linear fit to 
the points. A correlation coefficient of $\rho$ = 0.89 is found. An example of error bar 
is shown in the lower right corner.}
\label{fig:logbefehall}
\end{centering}
\end{figure}
\begin{figure}
\begin{centering}
\includegraphics[width=7cm]{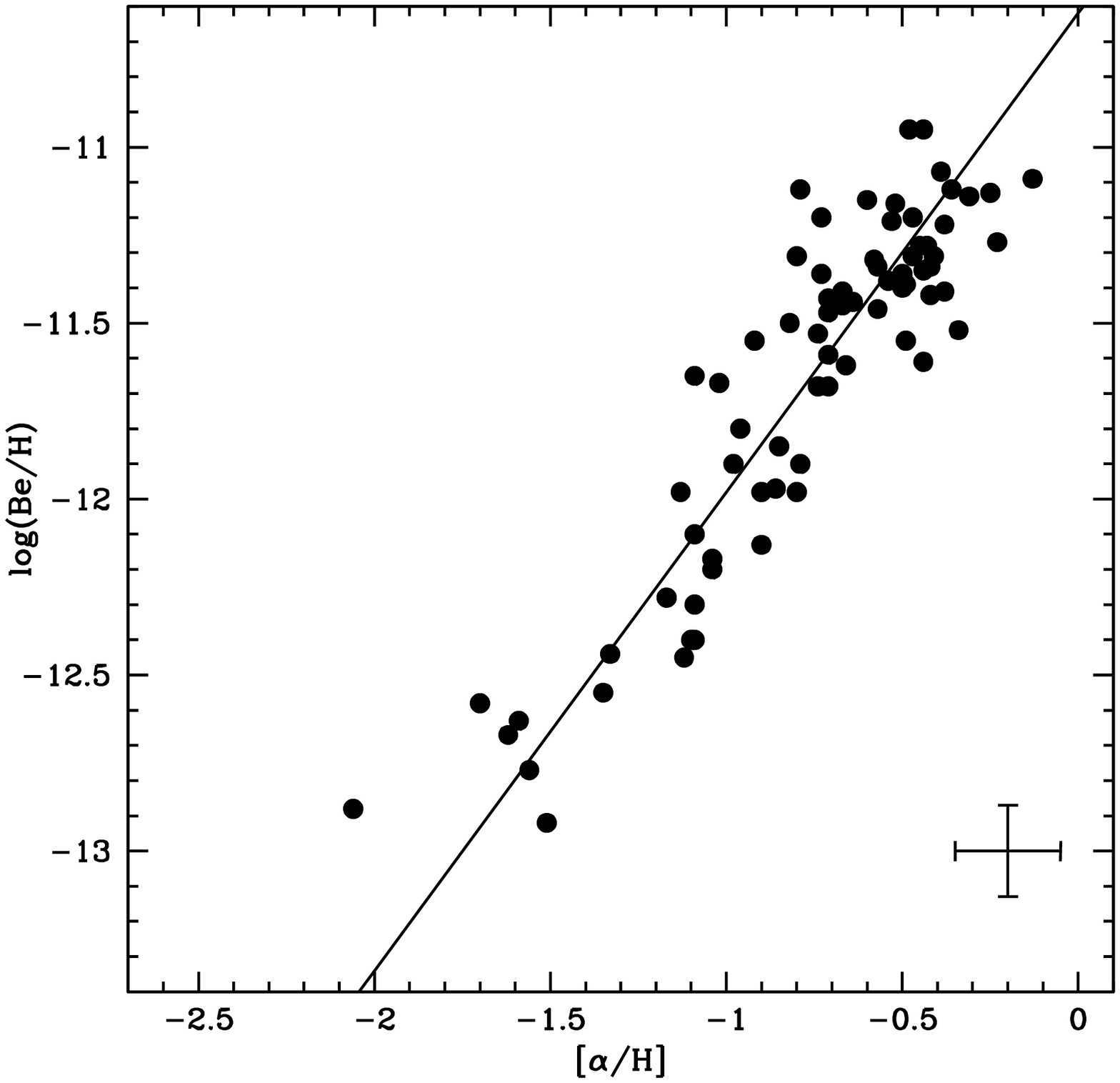}
\caption{Diagram of [$\alpha$/H] vs.\@ log(Be/H) for all the 73 stars
  showing a linear fit to the points. A correlation coefficient of
  $\rho$ = 0.91 is found. An example of error bar is shown in the
  lower right corner.}
\label{fig:logbealfahall}
\end{centering}
\end{figure}
\begin{table*}
\caption{A comparison of the linear relations between Be and Fe and O
  (or any other $\alpha$ indicators adopted) as found in the
  literature.}
\label{tab:lin}
\centering
\begin{tabular}{ll}
\noalign{\smallskip}
\hline\hline
\noalign{\smallskip}
log(Be/H) = (A $\pm$ $\sigma_{A}$) \@ + \@ (B $\pm$ $\sigma_{B}$) \@ [Fe/H] & Ref.  \\
\hline
log(Be/H) = ($-$10.38 $\pm$ 0.08) \@ + \@ (1.24 $\pm$ 0.07) \@ [Fe/H] & This work (all stars)\\
log(Be/H) = ($-$10.34 $\pm$ 0.09) \@ + \@ (1.27 $\pm$ 0.08) \@ [Fe/H] & This work (F00 stars)\\
log(Be/H) = ($-$10.40 $\pm$ 0.08) \@ + \@ (1.22 $\pm$ 0.07) \@ [Fe/H] & This work (halo stars)\\
log(Be/H) = ($-$10.38 $\pm$ 0.08) \@ + \@ (1.16 $\pm$ 0.07) \@ [Fe/H] & This work (thick disk stars)\\
log(Be/H) = ($-$10.22 $\pm$ 0.07) \@ + \@ (1.16 $\pm$ 0.04) \@ [Fe/H] & King (\cite{Ki01}) \\
log(Be/H) = ($-$10.59 $\pm$ 0.03) \@ + \@ (0.96 $\pm$ 0.04) \@ [Fe/H] & Boesgaard et al.\@ (\cite{BDKR99}) \\
log(Be/H) = ($-$10.19 $\pm$ 0.11) \@ + \@ (1.07 $\pm$ 0.08) \@ [Fe/H] & Molaro et al.\@ (\cite{MBCP97}) \\
log(Be/H) = ($-$9.76 $\pm$ 0.22) \@ + \@ (1.26 $\pm$ 0.11) \@ [Fe/H] & Boesgaard \& King (\cite{BK93}) \\
log(Be/H) = ($-$10.87 $\pm$ 0.51) \@ + \@ (0.77 $\pm$ 0.23) \@ [Fe/H] & Gilmore et al.\@ (\cite{GGEN92})$^{2}$ \\
\hline\hline
log(Be/H) = (A $\pm$ $\sigma_{A}$) \@ + \@ (B $\pm$ $\sigma_{B}$) \@ [$\alpha$/H] & Ref.  \\
\hline
log(Be/H) = ($-$10.62 $\pm$ 0.07) \@ + \@ (1.36 $\pm$ 0.08) \@ [$\alpha$/H] & This work (all stars)\\
log(Be/H) = ($-$10.71 $\pm$ 0.07) \@ + \@ (1.34 $\pm$ 0.08) \@ [$\alpha$/H] & This work (F00 stars)\\
log(Be/H) = ($-$10.62 $\pm$ 0.07) \@ + \@ (1.37 $\pm$ 0.08) \@ [$\alpha$/H] & This work (halo stars)\\
log(Be/H) = ($-$10.64 $\pm$ 0.07) \@ + \@ (1.31 $\pm$ 0.08) \@ [$\alpha$/H] & This work (thick disk stars)\\
log(Be/H) = ($-$10.87 $\pm$ 0.28) \@ + \@ (1.10 $\pm$ 0.18) \@ [Mg/H] & King (\cite{Ki02}) \\
log(Be/H) = ($-$10.33 $\pm$ 0.16) \@ + \@ (1.31 $\pm$ 0.18) \@ [Ca/H] & King (\cite{Ki02}) \\
log(Be/H) = ($-$10.61 $\pm$ 0.06) \@ + \@ (1.51 $\pm$ 0.05) \@ [O/H]$^{1}_{mean}$ & King (\cite{Ki01}) \\
log(Be/H) = ($-$10.69 $\pm$ 0.04) \@ + \@ (1.45 $\pm$ 0.04) \@ [O/H] & Boesgaard et al.\@ (\cite{BDKR99}) \\
log(Be/H) = ($-$10.62 $\pm$ 0.13) \@ + \@ (1.13 $\pm$ 0.11) \@ [O/H] & Molaro et al.\@ (\cite{MBCP97}) \\
log(Be/H) = $-$10.68 \@ + \@ 1.12 \@ [O/H] & Boesgaard \& King (\cite{BK93}) \\
log(Be/H) = ($-$11.19 $\pm$ 0.25) \@ + \@ (0.85 $\pm$ 0.15) \@ [O/H] & Gilmore et al.\@ (\cite{GGEN92})$^{2}$ \\
\noalign{\smallskip}
\hline
\end{tabular}
\\
(1) A mean oxygen abundance from different indicators, such as the molecular OH UV lines 
and the $\lambda$ 6300 [OI] forbidden line. \\ (2) The fits were recalculated in this work 
based on the original published data. \\
\end{table*}
In Table\@ \ref{tab:lin} we compare the linear relations we derived
with the ones previously obtained in the literature for metal-poor
stars. The relation with oxygen or $\alpha$-elements shows a sizable
scatter in the fitted slope; this is likely due to the difficulty in
the measurement of oxygen abundances in metal-poor stars. Our own
slope is close to the ones obtained by King (\cite{Ki02}) using [Ca/H]
and Boesgaard et al.\@ (\cite{BDKR99}) using [O/H]. The zero points
seem to agree better, the exception being the one determined by King
(\cite{Ki02}) using [Ca/H]. Our own value is the same as obtained by King
(\cite{Ki01}) and Molaro et al.\@ (\cite{MBCP97}) both using
[O/H]. The slope with the $\alpha$-elements abundance implies the need
for a primary production of Be. All the relations with Fe have a slope
very close to one. The zero points of the fits have a larger
scatter. The zero point of our fit is closer to the one obtained by
King (\cite{Ki01}), who analyzed results available in the
literature. 

The possible presence of a real scatter in the Be--Fe relationship
needs to be considered further. The spread is also present in earlier
data samples but, the limited size of the samples and the large errors
on Be abundance did not allow a robust assessment of its existence. In
particular, the presence of this scatter in samples of smaller size
might have motivated the claim of a change in the slope, as in Molaro
et al.\@ (\cite{MBCP97}) and Boesgaard et al.\@
(\cite{BDKR99}). Molaro et al.\@ (\cite{MBCP97}) were the first to
suggest a possible change in the slope of the relation of Be with Fe
at [Fe/H] $\sim$ $-$1.6, $-$1.1. No such change is apparent in our
data.

In the hypothesis that Be is a good cosmochronometer, there could be a
natural explanation for the scatter. At a given time in the
early-Galaxy, the Galactic halo has a higher metallicity than the
thick disk. Thus, a star formed at this time in the halo will have a
higher [Fe/H] than a star formed at the same time in the thick disk,
in spite of a similar Be abundance. We devote the next sections to a
detailed investigation of the presence of the scatter and to its
possible causes. This investigation will help to constrain the limits
of validity of Be as a cosmochronometer.

\begin{figure}
\centering
\includegraphics[width=7cm]{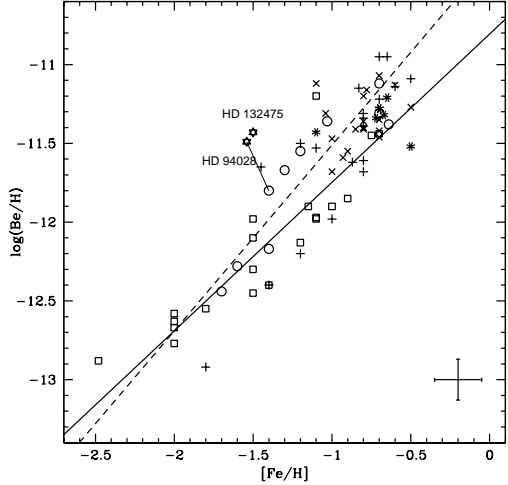}
\caption{\label{figbefe_plateau}
Our sample of 73 stars in the [Fe/H] - log(Be/H) plane split
in several sub samples.
Circles are thick-disk stars with A(Li) $\geq 2.0$,
squares halo stars with A(Li) $\geq 2.0$, $\times$ symbols
are thick-disk stars with A(Li) $< 2.0$ and
$+$ symbols are halo stars with A(Li) $< 2.0$.
Asterisks are thin-disk stars or stars with ambiguous
kinematic classification. 
The two star symbols are the Be-rich stars
HD 94028 and HD 132475 from Boesgaard \& Novicki (\cite{BN06}). 
HD 94028 (HIP 53070) is also included in our sample, where 
a smaller Be abundance was derived. A short line connects the two values.
The solid line is a least squares fit to the open square symbols
and the dashed line is a least squares fit to the open circles
symbols. A representative error bar is shown at the bottom right of
the plot.
}
\end{figure}

\subsection{Scatter in the Be-Fe relation}

\begin{figure}
\centering
\includegraphics[width=7cm]{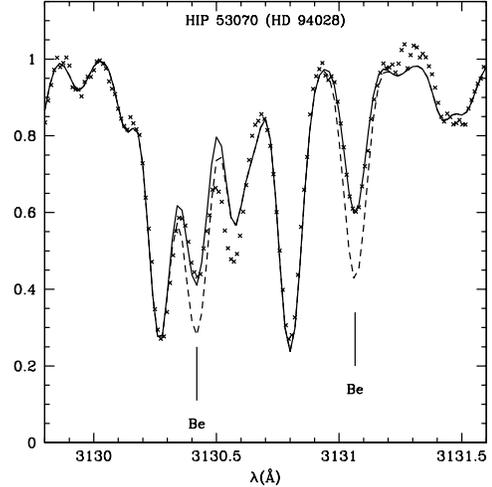}
\caption{\label{fig:53070} Fit to the region of the beryllium 
lines in star HIP 53070 (HD 94028). The crosses represent the observed 
spectrum, the solid line the best fit, and the dashed line a synthetic 
spectrum calculated with our adopted atmospheric parameters and the 
beryllium abundance determined by Boesgaard \& Novicki (\cite{BN06}), 
log(Be/H) = $-$11.49.}
\end{figure}

In Fig. \ref{figbefe_plateau}
we display our working sample of 73 stars, using
different symbols to identify 
several subsets. In addition in the plot the 
two stars, HD 94028 and HD 132475,
identified by
Boesgaard \& Novicki (\cite{BN06}) 
to  lie at about 4$\sigma$ above the mean Be--Fe trend
are also shown.
In this sense these two stars must be considered
Be-rich. However, we note that HD 94028 is also included in 
our sample. While Boesgaard \& Novicki (\cite{BN06}) find log(Be/H) = $-$11.49, 
we derived log(Be/H) = $-$11.80 and thus do not confirm the `Be rich status' 
of this star. Although the results basically agree within 1 $\sigma$
(we estimate $\pm$ 0.13 dex for our results and Boesgaard \& Novicki
$\pm$ 0.12 for theirs), the difference is associated to the different 
log g values adopted, log g = 4.20 by us and log g = 4.44 by Boesgaard \& Novicki. 
The log g derived using the Hipparcos parallax is log g = 4.27. As we point out 
in Sect.\@ \ref{sec:comp}, correct determination of log g is essential for 
deriving consistent Be abundances. In Fig.\ \ref{fig:53070} we show 
our fit to the spectrum of this star and a synthetic spectrum calculated 
with the larger abundance found by Boesgaard \& Novicki (but calculated with 
our adopted atmospheric parameters). Another, even more striking example of 
Be-rich star is HD 106038 (=HIP 59490,
Smiljanic et al. \cite{Sm08}), which we have removed from our analyzed sample.
Boesgaard \& Novicki (\cite{BN06}) suggests that large Be abundance 
arises from local 
enrichment phenomena, such as may happen in ``Super Bubbles'' in the
vicinity of a Supernova (see 
Parizot \cite{parizot}, Parizot \& Drury \cite{parizot_drury} and
references therein).
For the large Be enhancement  of HD 106038,
Smiljanic et al. (\cite{Sm08}) found such a scenario
insufficient and invoked a hypernova.
Such Be-rich stars seem to be quite rare, and at present it
seems unlikely that they contribute to a large
scatter in Be abundances. 
The Galactic orbits of all three stars have been 
computed by Caffau et al. (\cite{zolfo05}) and while
HD 132475 and  HD 106038 are halo stars, HD 94028 displays a 
disk-like kinematics. This suggests that the events
giving rise to Be-rich stars are indeed quite local and do not
depend on the stellar population to which the star belongs.

\begin{figure*}
\begin{centering}
\includegraphics[width=12cm]{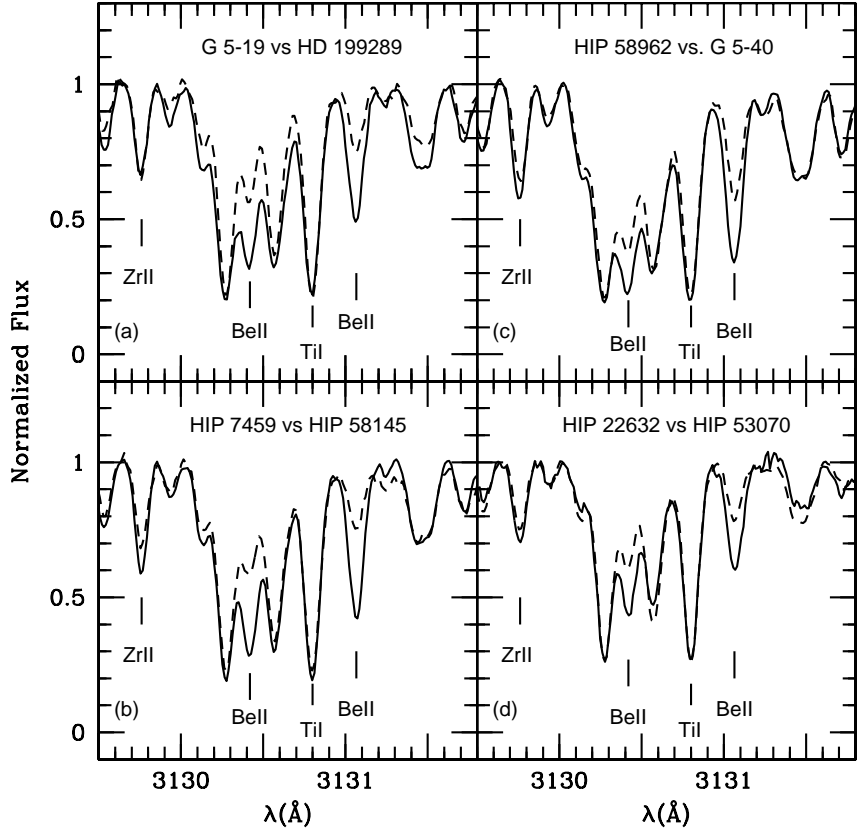}
\caption{Comparison in the Be region between spectra of stars with similar metallicity but different 
Be abundances. (a) G 5-19 with T$_{\rm eff}$/log g/[Fe/H]/log(Be/H) = 5942/4.24/$-$1.10/$-$11.98, shown 
as a dashed line, vs.\@ HD 199289 with T$_{\rm eff}$/log g/[Fe/H]/log(Be/H) = 5894/4.38/$-$1.03/$-$11.36, 
shown as a solid line. (b) HIP 7459 with T$_{\rm eff}$/log g/[Fe/H]/log(Be/H) = 5909/4.46/$-$1.15/$-$11.90, 
dashed line, vs.\@ HIP 58145 with T$_{\rm eff}$/log g/[Fe/H]/log(Be/H) = 5946/4.41/$-$1.04/$-$11.31, solid 
line. (c) HIP 58962 with T$_{\rm eff}$/log g/[Fe/H]/log(Be/H) = 5831/4.36/$-$0.80/$-$11.68, dashed line, 
vs.\@ G 5-40 with T$_{\rm eff}$/log g/[Fe/H]/log(Be/H) = 5863/4.24/$-$0.83/$-$11.15, solid line. (d) HIP 
22632 with T$_{\rm eff}$/log g/[Fe/H]/log(Be/H) = 5825/4.30/$-$1.40/$-$12.17, dashed line, vs.\@ HIP 53070 
with T$_{\rm eff}$/log g/[Fe/H]/log(Be/H) = 5900/4.20 /$-$1.40/$-$11.80, solid line.}
\label{fig:comp}
\end{centering}
\end{figure*}

Inspection of Fig. \ref{figbefe_plateau} gives
the visual impression that squares and circles (halo and thick disk)
define two distinct linear relations. At the same time it
is clear that the Li-depleted stars would blur this
picture, especially those belonging to the halo.
This is confirmed by fitting a straight line to the halo stars 
alone. The fit to all the halo stars (39 stars) 
has a probability of 0.04, marginally acceptable,
but suspicious. If we remove from the sample  
all the Li-depleted stars we are left with a sample 
of 20 stars, and the probability of the fit jumps up to 0.19.
This fit is shown in Fig. \ref{figbefe_plateau} corresponding to

\begin{equation}
{\rm log(Be/H)} \; = \; (-10.76 \pm 0.15) \; + \; (0.97 \pm 0.10 ) \; {\rm [Fe/H].}
\end{equation}

The sample of thick-disk stars, on the other hand, shows a reasonable fit
(0.69 probability) without the need to remove the Li-depleted stars.
The fits obtained by keeping all the thick-disk stars and by removing
the Li-depleted stars are very similar. For consistency in Fig. \ref{figbefe_plateau}
we give the one without Li-depleted stars, which corresponds to

\begin{equation}
{\rm log(Be/H)} \; = \; (-10.30 \pm 0.25) \; + \; (1.21 \pm 0.20 ) \; {\rm [Fe/H].}
\end{equation}

The choice of eliminating the halo Li-depleted 
stars from the fit is somewhat arbitrary and we
have no particular justification for it. Even more so, if
we note that many of the halo Li-depleted stars
have Be abundances that are {\em higher}
than the mean trend defined by the other stars.
It is nevertheless remarkable that making this 
choice the goodness-of-fit increases.

Our conclusion is that there is marginal evidence
of scatter in the Be-Fe relation, above what can be 
justified by observational errors. One possibility is
that two distinct Be-Fe relations exist:
one for the halo and one for the thick disk. 
Another possibility is that there is simply a dispersion
in Be abundances at any given metallicity.

\subsubsection{Stars with similar parameters}

We are analyzing a set of high quality data of stars that have, for the vast 
majority, parallaxes measured by Hipparcos and have been quite
extensively studied in the literature. Nevertheless, we are aware that
the uncertainties in an abundance analysis are often underestimated. 

We believe a meaningful test of the existence of real scatter  is to
identify stars with similar metallicity and similar atmospheric
parameters, hence almost identical spectra, but different Be
abundances. The difference in strength of the Be lines should be
obvious and largely independent  of our ability to model the spectra.

We identified 4 pairs of stars with similar atmospheric parameters and
metallicities. Their spectra are compared in Fig.\@ \ref{fig:comp}. In
this figure, all the pairs of spectra show very good agreement in
the behavior of the neighboring metal lines, but clearly differ in the
two Be lines. This difference is a definitive and convincing test that
the scatter in the abundances is a real feature. The stars being compared 
range from [Fe/H] = $-$0.80 to $-$1.40, showing that the
scatter is not concentrated in a single metallicity value, the stars
in each pair have very similar gravities, as well as similar Li
abundances. We cannot therefore invoke differences in gravity or in
dilution/depletion to explain the observed Be difference.

\begin{figure}
\begin{centering}
\includegraphics[width=7cm]{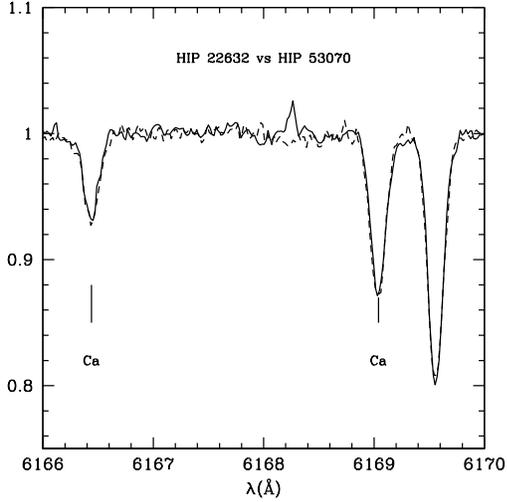}
\caption{Comparison between two calcium lines in the spectra of stars
  HIP 22632 (dashed line) and HIP 53070 (solid line). The Be region of
  these stars is shown in the last panel of the previous  Figure. The
  stars have similar  metallicity, similar atmospheric parameters, and
  similar abundances of $\alpha$-elements but different Be
  abundances.} 
\label{fig:cal}
\end{centering}
\end{figure}

The situation for $\alpha$ elements is less clear. In the above pairs
of stars, the [$\alpha$/H] ratio is not the same, so we could
expect stars with the same metallicity but different oxygen
abundance to have different Be abundances. It is more difficult to
identify stars in our sample with similar atmospheric parameters and
also the same [$\alpha$/H],  but different Be. Of the stars in Fig.\@
\ref{fig:comp}, HIP 22632 and HIP 53070 are the ones with the the most
similar [$\alpha$/H], $-$1.04 and $-$0.96 dex, respectively. In
Fig. \ref{fig:cal} we compare two calcium lines, $\lambda$ 6166.44
\AA\@ and $\lambda$ 6169.04 \AA\@ of these stars. As is clear from
this figure, there is no difference in the calcium abundances of these
stars. The difference in $\alpha$-element abundances of the other
stars in Fig.\@ \ref{fig:comp} are on the order of $-$0.15, which is
clodse to the 1$\sigma$ uncertainty of the abundances. The case
of HIP 22632 and HIP 53070 shows that stars with similar abundances of
$\alpha$-elements may have different abundances of Be.  

\subsubsection{Binaries}

We have finally investigated some other possible factors that may
influence the scatter, such as the use of data from different sources
or the possible presence of binaries. We collected information on the
multiplicity of our sample stars in the literature, which is listed in
Table \ref{tab:data}, where whenever possible we list the
number of components of the system (AB for two components, ABC for
three, and so on). When the number of stars in the system was not clear
we only list a flag \emph{yes} to indicate the star is a binary or
multiple system. The references consulted employed a variety of
methods, for example, radial velocities, photometry, or speckle
interferometry. A flag \emph{no} does not necessarily mean the star is
single, but that a reference was found where at least one method
failed to detect multiplicity. For some stars no reference reporting
an investigation on multiplicity was found, in this case no flag is
shown in the table. Again, no difference between the subsample of
possible binaries and the other stars is found. Binaries 
therefore do not seem to be responsible for the observed scatter.

\subsubsection{The F00 subsample}

 As discussed in Sect.\@ \ref{ed93f00} there is a systematic 
difference in temperature and [Fe/H] between the results 
of Edvardsson et al.\@ (\cite{Ed93}) and Fulbright (\cite{F00}), although within the
  uncertainties,. This difference in metallicity may introduce
  some scatter into the relations shown in Figs.\@ \ref{fig:logbefehall}
  and \ref{fig:logbealfahall}. In the final sample of 73 stars, 49
  have parameters calculated by F00, of which 26 are halo stars and 19
  are thick-disk stars. We repeated the analysis with this subsample
  of stars. The linear fits using only the Fullbright stars and rms
  are statistically identical to the ones obtained with the whole
  clean sample (also listed in Table \ref{tab:lin}). The equivalent
  of Figs.\@ \ref{fig:logbefehall} and \ref{fig:logbealfahall} for the
  F00 subsample are given in Appendix \ref{ap:f00}.

\subsection{log(Be/H) vs.\@ [Fe/Be]}

\begin{figure}
\begin{centering}
\includegraphics[width=7cm]{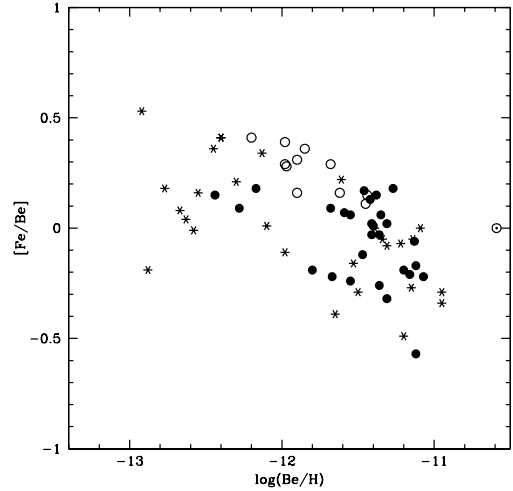}
\caption{Diagram of log(Be/H) vs.\@ [Fe/Be]. Thick disk stars are shown 
as filled circles, halo stars as 
starred symbols, and the subgroup of 'Fe-rich' 
halo stars discussed Sect. \ref{sec:chrono} is shown as open circles.}
\label{fig:febe}
\end{centering}
\end{figure}

The evolution of a given element in the Galaxy is usually analyzed 
in a plot of [Element/Fe] vs.\@ [Fe/H]. 
If Be is a better time scale than Fe, then 
changing Fe for Be in this kind of discussion would be then 
the natural follow-up. In Fig.\@ \ref{fig:febe} we plot 
[Fe/Be] as a function of log(Be/H). 

A decrease in [Fe/Be] with increasing log(Be/H) is seen. 
It is interesting to see that for log(Be/H) $> -11.5$ the 
majority of the stars have a sub-solar Fe/Be ratio.
In principle similar plots could be constructed also for
elements other that iron.

\section{Be as a chronometer\label{sec:chrono} }

The hypothesis that Be can be used as an age indicator relies 
on assumptions about the uniformity of the GCR in 
the early-Galaxy. This uniformity is necessary to assure the Be clock runs 
at a uniform pace within the relevant region of the Galaxy. 
In general, models of Galactic chemical 
evolution do not include a detailed treatment of GCR 
propagation and confinement. The model we use to compare 
with our observations (see Valle et al. \cite{Va02} and references 
therein) follows a multizone treatment of the Galaxy, where each 
component is characterized by different star formation rates with 
a radial dependence. The GCR, however, has neither radial nor 
vertical dependence. The interpretation of the results 
in light of these models may, therefore, be somewhat limited.

\begin{figure}
\begin{centering}
\includegraphics[width=7cm]{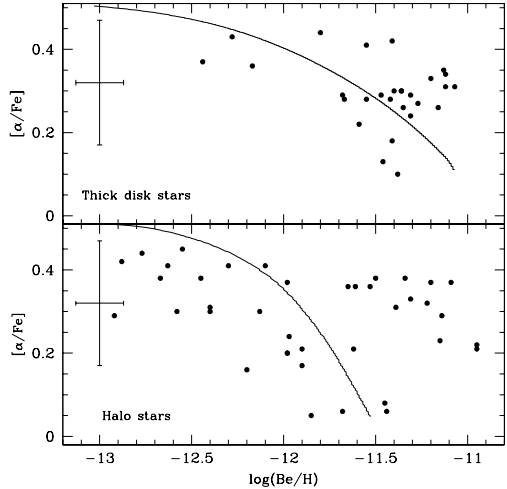}
\caption{Diagram of [$\alpha$/Fe] vs.\@ log(Be/H). The thick-disk 
stars are shown in the upper panel, while the halo stars are shown 
in the lower panel. The thick-disk stars diagram is characterized by 
a large scatter, while the halo stars diagram clearly divides into two 
sequences. The curves are the predictions of the models by Valle 
et al.\@ (\cite{Va02}).}
\label{fig:alfafebe}
\end{centering}
\end{figure}

\begin{figure}
\begin{centering}
\includegraphics[width=7cm]{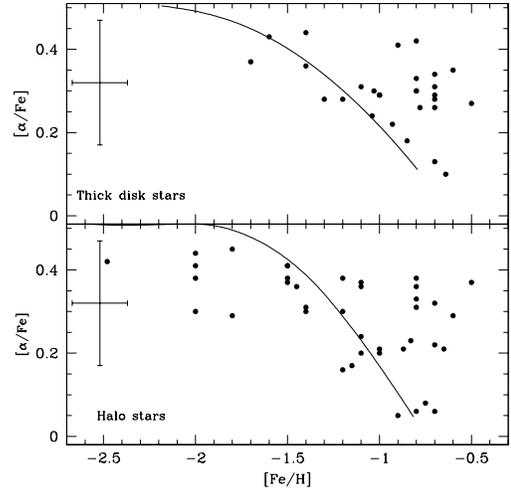}
\caption{Diagram of [$\alpha$/Fe] vs.\@ [Fe/H]. The thick-disk 
stars are shown in the upper panel, while the halo stars are 
shown in the lower panel. The curves are the predictions of the models by Valle 
et al.\@ (\cite{Va02})}
\label{fig:alfafefe}
\end{centering}
\end{figure}

Pasquini et al.\@ (\cite{Pas05}) began using the 
Be abundances as a time scale, and show that halo and 
thick-disk stars separate in a diagram of log(Be/H) vs.\@ [$\alpha$/Fe]. 
Their sample, however, is composed of only twenty stars 
analyzed by Boesgaard et al.\@ (\cite{BDKR99}). In a  diagram [O/Fe] vs.\@ log(Be/H),  the beryllium 
abundance in the abscissa can be considered as `increasing time' 
while the [O/Fe] ratio in the ordinate 
can be considered as the `star formation rate' (SFR). 
We use [$\alpha$/Fe] here rather than  [O/Fe]. 
In Fig.\@ \ref{fig:alfafebe} we show 
the [$\alpha$/Fe] vs.\@ log(Be/H) for our sample. 
The figure is divided into two panels; in the upper panel we plot the thick 
disk stars and in the lower panel we plot the halo stars. 
In the same figure the two models from 
Pasquini et al.\@ (\cite{Pas05}), derived from 
Valle et al. (\cite{Va02}), are also shown. 
In Fig.\@ \ref{fig:alfafefe} we show the 
classical [$\alpha$/Fe] vs.\@ [Fe/H] diagram, also divided with 
the thick-disk stars in the upper panel and the halo stars in the lower panel. 

From Fig.\@ \ref{fig:alfafebe} it appears that the model
is fairly successful in reproducing the general trend
of the thick-disk stars.
However the halo stars 
appear to split along a double sequence. 
The most striking is the group of stars 
with high [$\alpha$/Fe] and high log(Be/H),
which the model totally fails to reproduce.
From the  point of view of $\alpha$ elements and
beryllium abundance, this group of stars
is essentially indistinguishable from the thick-disk stars.

\begin{figure}
\begin{centering}
\includegraphics[width=7cm]{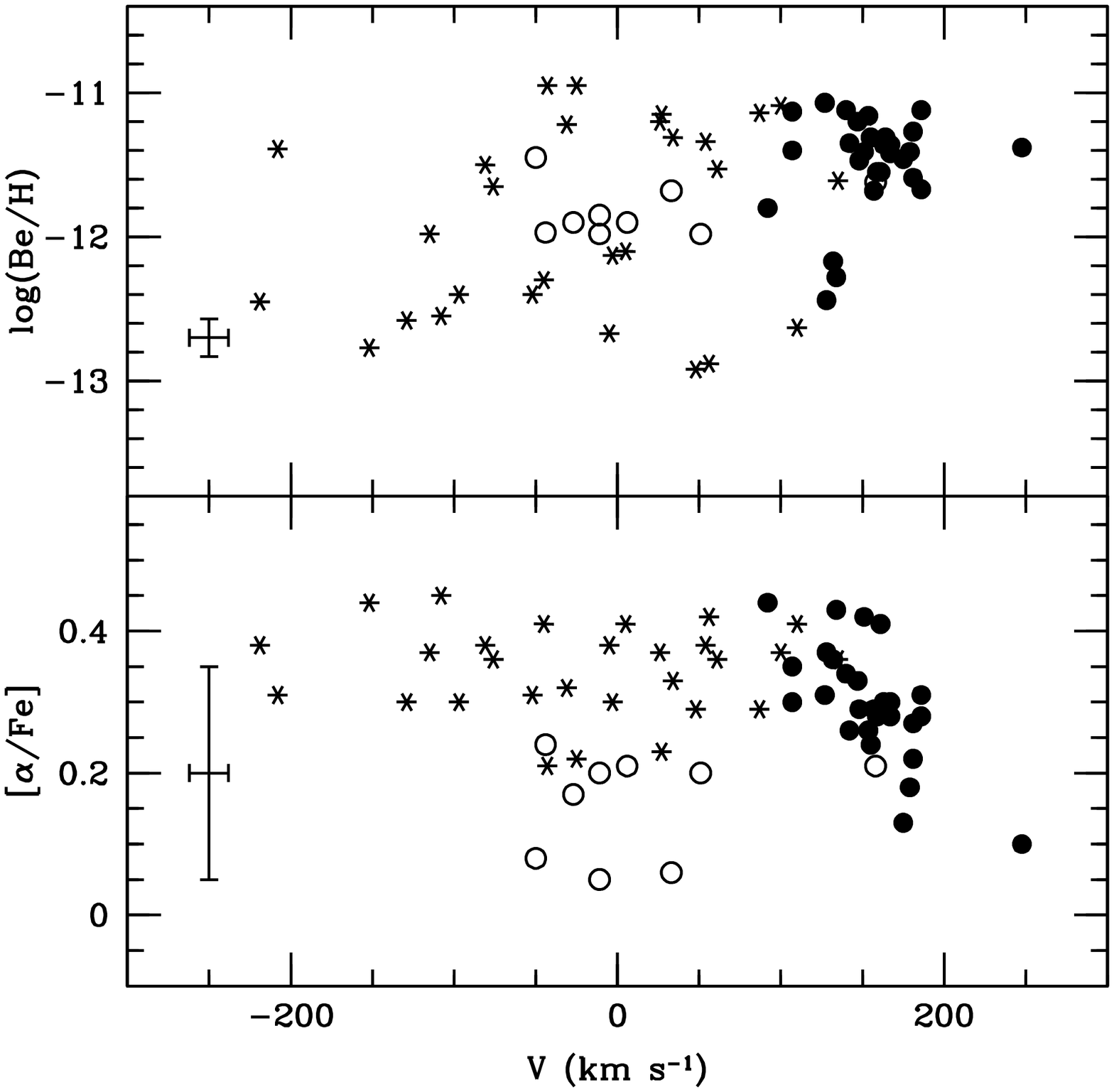}
\caption{Diagram of \@ log(Be/H) vs.\@ the V, the component of the
  space velocity of the star in the direction of the disk
  rotation. The thick-disk stars are shown as filled circles, halo
  stars are shown as starred symbols, and the subgroup of halo stars
  as open circles. A typical error of $\pm$ 12 Km s$^{-1}$ in V was 
adopted (see Gratton et al.\@ \cite{G03b}).}
\label{fig:plotv}
\end{centering}
\end{figure}

The first conclusion that may be drawn from this
plot is that the thick disk appears to be a fairly
homogeneous population, where the ``Be chronometer'' may
be used with some confidence. The halo instead appears
to be more complex. 
In the following sections we discuss the use
of Be as a chronometer in the two populations
separately. 
Galactic orbits for the majority of the sample stars were computed by 
Gratton et al.\@ (\cite{G03}). 
For the few remaining stars, new orbits were calculated following the 
same method. We refer to Gratton et al.\@ (\cite{G03}) 
for the details. The orbital parameters 
are listed in Table \ref{tab:prob}.
In Figs.\@ \ref{fig:plotv} and \ref{fig:plotrmin}
we show the correlation of chemical and dynamical
properties.
Alternative to the halo -- thick disk distinction,
we could have used the accretion -- dissipative component
distinction introduced by 
Gratton et al.\@ (\cite{G03,G03b}). We refer to the original papers for a detailed 
presentation of the kinematic criteria used for this division. 
We did this exercise, and the main conclusions are essentially the same.
The corresponding plots are provided in Appendix \ref{acc-diss}.

\subsection{The Be chronometer in the thick disk}

\begin{figure}
\begin{centering}
\includegraphics[width=7cm]{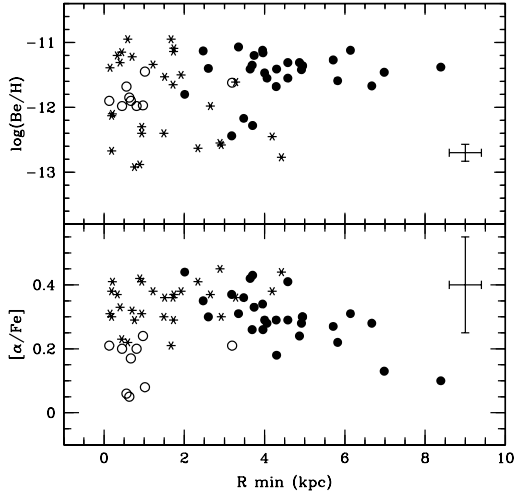}
\caption{Diagram of \@ log(Be/H) vs.\@ R$_{min}$, the perigalactic
  distance of the stellar orbit. The thick disk stars are shown as
  filled circles, halo stars are shown as starred symbols, and the
  subgroup of halo stars as open circles. A typical error of $\pm$ 
0.40 Kpc in R$_{\rm min}$ was adopted (see Gratton et al.\@ \cite{G03b}).} 
\label{fig:plotrmin}
\end{centering}
\end{figure}

Thick-disk stars show smaller scatter in all diagrams, and 
it seems possible to understand their formation in a relatively
simple way. The most interesting feature is, however, the 
anticorrelation of [$\alpha$/Fe] with R$_{\rm min}$ in our sample 
(Fig.\@ \ref{fig:plotrmin}). At first glance, 
this anticorrelation seems to be driven by the 
two stars with largest R$_{\rm min}$. However, even when 
restricting the sample to stars with 
R$_{\rm min} < 6$ kpc, the probability
of correlation is 99.8\%.
A similar anticorrelation is also present in the data
of Gratton et al. (\cite{G03b}, see their Fig.\ 6).
If R$_{\rm min}$ is representative
of the radius of formation of a star,
neglecting the phenomena of orbital diffusion, 
then the simplest interpretation is that the thick disk experienced a higher SFR
in the inner regions than in the outer regions.
This does not contrast with the notion that the
thick is very old and massive 
(Fuhrmann \cite{Fuhrmann98,Fuhrmann00,Fuhrmann04,Fuhrmann08}).
In fact the star formation is supposed to have continued
for about 2 Gyr (Fuhrmann \cite{Fuhrmann04}), which is
ample time for  the external regions  experiencing
a low SFR to develop a lower $\alpha$/Fe 
than the inner regions where the SFR was higher.
Note also that the most Be-poor, i.e. oldest, stars
are found at low  R$_{\rm min}$.
The emerging 
picture is that of a dissipative thick disk formed inside-out.  

Such an anticorrelation is not present in the log(Be/H) vs.\@ R$_{\rm min}$ 
plot, as shown in the upper panel of Figure \ref{fig:plotrmin}. It is
also not present in the [Be/Fe] vs.\@ R$_{\rm min}$ plot. This lack of
anticorrelation is an important result. The major objection to the use of Be
as a time indicator is the possibility that the abundance of this
element is dominated by local events rather than by the relativistic
component. The log(Be/H) vs.\@ R$_{\rm min}$  diagram does not show
evidence of a general decrease in Be with the distance from the
galactic center, as could be expected if the local flux of GCR was the
dominant factor. 

The range in Be covered by the thick-disk stars of our sample 
is relatively small. Moreover, most of the Be-poor stars are present
at a relatively small R$_{\rm min}$. This agrees with the notion of a
rather fast formation for the thick disk (Fuhrmann
\cite{Fuhrmann04,Fuhrmann08}). As a reference time scale, the reader
may consider that the thick-disk models shown in the upper panel of
Fig.\@ \ref{fig:alfafebe} have a star formation period of 1 Gyr.


\subsection{The Be chronometer in the halo}

For a long time the debate on the
halo formation was centered on the opposite
views of ``monolithic collapse'' (Eggen et al. \cite{ELS})
and ``fragment assembly'' (Searle \& Zinn \cite{SZ78}).
It is now  understood that these represent rather idealized
extreme cases that do not happen in reality.
Major insights into halo assembly  have come
from simulations of galaxy formation in a cosmological context
(see e.g. Font et al. \cite{font} and references
therein). In these simulations
in the $\Lambda$CDM scenario, the galaxy formation
proceeds hierarchically, nevertheless they predict
the existence of metallicity gradients and other
features typical of the ``monolithic collapse''.
In light of these simulations, we should expect
to find, in a halo sample, stars that have been accreted
from many dwarf satellite galaxies, as well as stars 
formed on short time scales
in the halo (i.e. a ``monolithic collapse''-like component).

Even before considering our data one can expect  that 
the ``Be chronometer'' was running at a different
pace in the different components out of which  
the halo was assembled.
This in fact appears quite obvious 
from Fig.\@ \ref{fig:alfafebe}. 
The stars with high Be abundances and high [$\alpha$/Fe]
cannot be reproduced by our model. 
It is tempting to interpret these stars as ``accreted'' from 
satellites in which the Be production was more efficient than 
in the Milky Way. However, these stars do not
display any peculiar kinematic property.

The good news is that the halo 
stars do not fill uniformly the log(Be/H) -- [$\alpha$/Fe]
plane, but position around only two, rather well-defined
sequences.
This means that there is some prospect for also using the
``Be chronometer'' for the halo stars.
Hopefully future models will be able to explain the two
sequences and calibrate the Be abundance versus time.
By use of a diagram like  Fig.\@ \ref{fig:alfafebe}, 
one should be able to decide which sequence any given
star belongs to.

 The reality of the separation seen in Fig.\@ \ref{fig:alfafebe} 
is dependent on the error of the abundances. A statistical test shows 
that the halo stars in the top right of panel b in this figure have 
probabilities between 40\% to 9\% of following the model curve. This 
suggests that the separation seen in the figure is real. These probabilities 
are likely to improve, and more stringent constraints might be derived, when 
oxygen, which is a pure SN II product, is determined in a homogeneous way for all the 
sample stars (Smiljanic et al., in preparation).

In Figs.\ \ref{fig:plotv} and \ref{fig:plotrmin} we plot both 
[$\alpha$/Fe] and log(Be/H) as a function of V, the spatial velocity 
in the direction of the rotation of Galaxy, and R$_{\rm min}$, 
the perigalactic distance, respectively.
At the end of the '90s, several halo stars with   
a low [$\alpha$/Fe] were identified
(Carney et al. \cite{carney97}, Nissen \& Schuster \cite{NS97})
and were considered as  ``accreted'' objects. 
Inspection of Fig. \ref{fig:alfafebe} shows that such stars
are however predicted by our evolutionary model.
To easily identify the ``low $\alpha$'' stars
(which we define as stars with [$\alpha$/Fe] $\leq$ 0.25 and log(Be/H)
$\leq -$ 11.4) in the figures, we plot them as open circles.  
These stars have very similar kinematic properties. They tend to 
have V close to zero and R$_{\rm min}$ $\leq$ 1 kpc, although the 
statistics are too few to allow a robust claim. They seem to be 
a group of non-rotating stars going very close to the Galactic 
center, a behavior that might be expected to be shown by 
accreted stars that sink to the Galactic center by dynamical friction. 
They also show a remarkably narrow range of Be abundances, 
suggesting a narrow range in age for this component. These 
characteristics might hint at a possible common origin for these stars.

A robust interpretation of these components 
in light of our evolutionary model does not seem possible. 
Unlike what is assumed by the model, our results suggest that the halo 
is not a single uniform population where a clear age-metallicity 
relation can be defined. The splitting of the halo into the 
two components identified in this work might 
be related to the accretion of external components or simply 
to variations in the star formation in different and initially 
independent regions of the early halo.

\section{Summary}\label{sec:con}

We have presented the largest sample, to date,
of Be abundances in stars. We confirm the existence
of a linear increase in Be with [Fe/H] and [$\alpha$/H],
as has been found by all the previous investigations.
However, thanks to the large dimension of our sample
and the homogeneous analysis, we 
have a marginal detection of an excess
scatter in the log(Be/H) --  [Fe/H] relation,
above what may be expected by observational errors.
A similar scatter is not obvious in the
log(Be/H) --  [$\alpha$/H] relation, although we identify
a pair of stars with similar parameters and similar [$\alpha$/H]
but with a 0.4 dex difference in Be abundance.
Our interpretation of this scatter is that it arises
from the different evolutionary  time scales of 
different Galactic components (halo, thick disk). Beryllium
is a good indicator of time, anywhere, but metal enrichment
proceeds at a different pace in different environments.

We tested the use of Be as a chronometer by comparing our observations
to model predictions. The result appears to be fairly
satisfactory for the thick disk, which we confirm to 
be a homogeneous population. 
There appears to be no trend in Be abundance
with perigalactic distance. We take this as evidence
that the Be production is dominated by the
relativistic component of the cosmic rays and largely
insensitive to local events. 
Our models of Be evolution,
however, fail to reproduce the observations for a subset
of halo stars. More sophisticated models that explicitly
consider the merger history of the halo should be
envisaged.
In a plot of log(Be/H) versus [$\alpha$/Fe] (Fig. \ref{fig:alfafebe} ), the halo
stars do not fill the plane randomly, but align 
in two rather clearly defined sequences. Only one of these
two sequences can be explained by our models. 
However, most stars of this same component show 
very similar kinematics and a narrow range in Be (age). 
They seem to be a group of non-rotating stars going 
very close to the Galactic center so might have a common 
origin.

For the halo stars we do not identify any
significant correlation between chemical
and kinematic properties. For the thick disk, instead,
we find a significant anticorrelation of [$\alpha$/Fe]
with perigalactic distance. This anticorrelation might be 
interpreted as evidence that the SFR was lower in the outer regions of
the thick disk, pointing towards an inside-out formation.

Even though not all our observations can be readily
interpreted in terms of existing models, we believe that
our study has highlighted the usefulness of Be abundance
in studying the different Galactic populations.

\begin{acknowledgements}

This work was developed during the visit of R.S. to ESO made possible
by a CAPES fellowship (1521/06-3) and support from the ESO DGDF. R.S. 
acknowledges an FAPESP PhD fellowship (04/13667-4).  
P.B. acknowledges support from EU contract MEXT-CT-2004-014265 (CIFIST). 
D.G. acknowledges support from EC grant MRTN-CT-2006-035890-Constellation. 
We thank Francesca Primas for making her line list available and 
for her much appreciated help in analyzing the spectra.

\end{acknowledgements}

\Online

\setcounter{table}{0}
\begin{longtable}{ccccccccccc}
\caption{\label{tab:data} Data of the sample stars.
  } \\ 
\hline\hline
HIP & HD & Other &  V  & $\pi$ & $\sigma_{\pi}$ & M$_{\rm V}$ & BC  &
  M$_{\rm bol}$ & log (L$_{\star}$/L$_{\odot}$) &  binary? \\ 
   &     & name  & mag. & mas & mas &  mag. &   mag. & mag. & 
       &          \\
\hline  
\endfirsthead
\caption{continued.} \\
\hline\hline
HIP & HD & Other &  V  & $\pi$ &  $\sigma_{\pi}$ & M$_{\rm V}$ & BC  &
M$_{\rm bol}$ &  log (L$_{\star}$/L$_{\odot}$) &  binary? \\
   &     & name  & mag. & mas & mas & mag. &   mag. & mag. & 
        &        \\
\hline
\endhead
\hline
\multicolumn{11}{l}{(1) Dommanget \& Nys (\cite{DN94}), (2) Latham et al. (\cite{La02}), 
(3) Fouts (\cite{Fo87}), (4) Nordstr\"om et al. (\cite{No04}), (5) This work,} \\
\multicolumn{11}{l}{(6) Lu et al. (\cite{LU87}), (7) Lindgren \& Ardeberg (\cite{LA96}), (8) Gliese \& Jahreiss (\cite{GJ88}), 
(9) Goldberg et al.\@ (\cite{Gol02}),} \\
\multicolumn{11}{l}{(10) Carney et al.\@ (\cite{Car01}), (11) Stryker et al.\@ (\cite{St85}), (12) Patience et 
al.\@ (\cite{Pat02}), (13) McAlister et al. (\cite{Mc87}),} \\
\multicolumn{3}{l}{(14) MacConnell et al. (\cite{MC97})}
\endfoot
171 & 224930 & 85 Peg A & 5.75 & 80.63 & 3.03 & 5.28 & $-$0.27 & 5.02 & $-$0.11 & ABCD$^{1}$ \\
3026 & 3567 & -- & 9.26 & 9.57 & 1.38 & 4.16 & $-$0.16 & 4.00 & +0.30 & no$^{2}$ \\
7459 & -- & CD-61 0282 & 10.10 & 11.63 & 1.19 & 5.43 & $-$0.16 & 5.27 & $-$0.21 & -- \\
10140 & -- & G 74-5 & 8.77 & 17.66 & 1.29 & 5.00 & $-$0.22 & 4.78 & $-$0.01 &
 ABCD$^{1}$ \\
10449 & -- & G 159-50 & 9.09 & 16.17 & 1.34 & 5.13 & $-$0.20 & 4.93 & $-$0.07 &
 no$^{2}$ \\
11952 & 16031 & -- & 9.77 & 8.67 & 1.81 & 4.46 & $-$0.18 & 4.28 & +0.19 & Susp.$^{3}$ \\
13366 & 17820 & -- & 8.38 & 15.38 & 1.39 & 4.31 & $-$0.16 & 4.15 & +0.24 & no$^{4}$ \\
14086 & 18907 & $\epsilon$ For & 5.89 & 32.94 & 0.72 & 3.48 & $-$0.33 & 3.15 & +0.64 & yes$^{5}$ \\
14594 & 19445 & -- & 8.06 & 25.85 & 1.14 & 5.12 & $-$0.23 & 4.89 & $-$0.06 & no$^{6}$ \\
17001 & -- & CD-24 1782 & 9.92 & 4.42 & 1.75 & -- & -- & -- & -- & -- \\
17147 & 22879 & G 80-15 & 6.70 & 41.07 & 0.86 & 4.77 & $-$0.17 & 4.60 & +0.06 & no$^{2}$ \\
18802 & 25704 & -- & 8.10 & 19.02 & 0.87 & 4.50 & $-$0.15 & 4.35 & +0.16 & AB$^{1}$ \\
19007 & 25673 & -- & 9.56 & 24.23 & 1.53 & 6.48 & $-$0.26 & 6.23 & $-$0.59 & -- \\
19814 & -- & G 82-5 & 10.60 & 24.27 & 23.10 & -- & $-$0.25 & -- & -- & no$^{2}$ \\
21609 & 29907 & -- & 9.85 & 17.00 & 0.98 & 6.00 & $-$0.28 & 5.72 & $-$0.39 & AB$^{7}$ \\
22632 & 31128 & -- & 9.13 & 15.55 & 1.20 & 5.09 & $-$0.21 & 4.88 & $-$0.05 & no$^{4}$ \\
24030 & 241253 & G 84-37 & 9.72 & 10.29 & 1.66 & 4.78 & $-$0.19 & 4.60 & +0.06 & no$^{2}$ \\
24316 & 34328 & -- & 9.44 & 14.55 & 1.01 & 5.25 & $-$0.21 & 5.04 & $-$0.12 & -- \\
31188 & 46341 & -- & 8.62 & 16.86 & 0.98 & 4.75 & $-$0.16 & 4.59 & +0.06 & -- \\
31639 & -- & CD-25 3416 & 9.65 & 17.59 & 1.34 & 5.88 & $-$0.25 & 5.63 & $-$0.35 & -- \\
33221 & -- & CD-33 3337 & 9.03 & 9.11 & 1.01 & 3.83 & $-$0.17 & 3.66 & +0.44 & -- \\
33582 & 51754 & -- & 9.05 & 14.63 & 1.39 & 4.88 & $-$0.16 & 4.72 & +0.01 & no$^{2}$ \\
34285 & -- & CD-57 1633 & 9.54 & 10.68 & 0.91 & 4.68 & $-$0.16 & 4.52 & +0.09 & no$^{4}$ \\
36491 & 59374 & G 88-31 & 8.50 & 20.00 & 1.66 & 5.01 & $-$0.17 & 4.84 & $-$0.03 & no$^{2}$ \\
36640 & 59984 & HR 2883 & 5.90 & 33.40 & 0.93 & 3.52 & -- & -- & -- & AB$^{1}$ \\
36818 & -- & CD-45 3283 & 10.43 & 15.32 & 1.38 & 6.36 & $-$0.19 & 6.17 & $-$0.57 & -- \\
36849 & 60319 & G 88-40 & 8.95 & 12.15 & 1.24 & 4.37 & $-$0.17 & 4.20 & +0.22 & no$^{2}$ \\
37853 & 63077 & HR 3018 & 5.36 & 65.79 & 0.56 & 4.46 & $-$0.15 & 4.31 & +0.18 & wide$^{8}$ \\
38625 & 64606 & G 112-54 & 7.44 & 52.01 & 1.85 & 6.02 & $-$0.25 & 5.77 & $-$0.41 & yes$^{2}$ \\
42592 & 74000 & -- & 9.66 & 7.26 & 1.32 & 3.96 & $-$0.21 & 3.76 & +0.40 & no$^{2}$ \\
44075 & 76932 & HR 3578 & 5.86 & 46.90 & 0.97 & 4.22 & $-$0.14 & 4.08 & +0.27 & yes$^{4}$ \\
44124 & -- & G 114-26 & 9.69 & 12.37 & 1.72 & 5.15 & -- & -- & -- & yes$^{9}$ \\
45554 & -- & G 46-31 & 10.86 & 3.79 & 2.16 & 3.75 & $-$0.15 & 3.60 & +0.46 & yes$^{2}$ \\
48152 & 84937 & -- & 8.31 & 12.44 & 1.06 & 3.78 & $-$0.19 & 3.59 & +0.46 & no$^{10}$ \\
50139 & 88725 & -- & 7.74 & 27.67 & 1.01 & 4.95 & $-$0.19 & 4.76 & 0.00 & no$^{2}$ \\
52771 & -- & BD+29 2091 & 10.24 & 10.55 & 1.75 & 5.36 & $-$0.21 & 5.15 & $-$0.16 & Susp.$^{11}$ \\
53070 & 94028 & G 58-25 & 8.23 & 19.23 & 1.13 & 4.65 & $-$0.18 & 4.47 & +0.11 & no$^{2}$ \\
55022 & 97916 & -- & 9.21 & 7.69 & 1.23 & 3.64 & $-$0.10 & 3.54 & +0.48 & yes$^{10}$ \\
57265 & -- & G 121-12 & 10.37 & 6.14 & 1.82 & 4.31 & $-$0.19 & 4.12 & +0.25 & no$^{2}$ \\
58145 & -- & BD-21 3420 & 10.15 & 5.43 & 1.44 & 3.82 & $-$0.16 & 3.66 & 0.43 & -- \\
58962 & 105004 & -- & 10.21 & 2.68 & 4.49 & -- & $-$0.17 & -- & -- & ABC$^{1}$ \\
59490 & 106038 & G 12-21 & 10.18 & 9.16 & 1.50 & 4.99 & $-$0.16 & 4.83 & $-$0.03 & no$^{2}$ \\
59750 & 106516 & HR 4657 & 6.11 & 44.34 & 1.01 & 4.34 & $-$0.11 & 4.23 & +0.21 & AB$^{1}$ \\
60632 & 108177 & G 13-35 & 9.67 & & & 4.87 & $-$0.17 & 4.70 & +0.02 & no$^{2}$ \\
62882 & 111980 & -- & 8.38 & 12.48 & 1.38 & 3.86 & -- & -- & -- & AB$^{1}$ \\
63559 & 113083 & -- & 8.05 & 18.51 & 1.12 & 4.39 & $-$0.18 & 4.21 & +0.22 & yes$^{4}$ \\
63918 & 113679 & -- & 9.70 & 6.82 & 1.32 & 3.87 & $-$0.16 & 3.71 & +0.42 & no$^{4}$ \\
64426 & 114762 & -- & 7.31 & 24.65 & 1.44 & 4.27 & $-$0.16 & 4.11 & +0.26 & yes$^{12}$ \\
66665 & -- & BD+13 2698 & 9.37 & 7.44 & 1.70 & 3.73 & $-$0.20 & 3.53 & +0.49 & no$^{2}$ \\
67655 & 120559 & --  & 7.97 & 40.02 & 1.00 & 5.98 & $-$0.26 & 5.72 & $-$0.39 & -- \\
67863 & 121004 & -- & 9.04 & 16.73 & 1.35 & 5.16 & $-$0.16 & 5.00 & $-$0.10 & no$^{4}$ \\
70681 & 126681 & -- & 9.33 & 19.16 & 1.44 & 5.74 & $-$0.22 & 5.52 & $-$0.31 & no$^{4}$\\
71458 & 128279 & -- & 7.97 & 5.96 & 1.32 & -- & -- & -- & -- & no$^{4}$ \\
72461 & -- & BD+26 2606 & 9.74 & 10.28 & 1.42 & 4.80 & $-$0.21 & 4.59 & +0.06 & yes$^{11}$ \\
74067 & 134088 & -- & 8.00 & 28.29 & 1.04 & 5.26 & $-$0.20 & 5.06 & $-$0.12 & no$^{4}$ \\
74079 & 134169 & -- & 7.68 & 16.80 & 1.11 & 3.81 & $-$0.16 & 3.65 & +0.44 & yes$^{4}$ \\
77946 & 142575 & -- & 8.61 & 6.56 & 1.23 & 2.69 & $-$0.11 & 2.58 & +0.87 & -- \\
80003 & -- & G 168-42 & 11.51 & 9.12 & 3.01 & 6.31 & $-$0.21 & 6.10 & $-$0.54 & yes$^{2}$ \\
80837 & 148816 & G 17-21 & 7.28 & 24.34 & 0.90 & 4.21 & $-$0.17 & 4.04 & +0.28 & no$^{2}$ \\
81170 & 149414 & -- & 9.63 & 20.71 & 1.50 & 6.21 & $-$0.26 & 5.95 & $-$0.48 & ABCDE$^{1}$ \\
85963 & 159307 & -- & 7.41 & 13.40 & 0.99 & 3.05 & $-$0.13 & 2.92 & +0.73  & yes$^{4}$  \\
87693 & -- & BD+20 3603 & 9.71 & 6.47 & 7.85 & 3.76 & $-$0.19 & 3.58 & +0.47 & yes$^{4}$ \\
88010 & 163810 & -- & 9.63 & 11.88 & 2.21 & 5.00 & $-$0.28 & 4.72 & +0.01 & AB$^{1}$ \\
92781 & 175179 & -- & 9.08 & 11.55 & 1.81 & 4.45 & $-$0.16 & 4.29 & +0.18 & no$^{2}$ \\
94449 & 179626 & -- & 9.17 & 7.52 & 1.36 & 3.55 & $-$0.21 & 3.34 & +0.56 & no$^{2}$ \\
98020 & 188510 & -- & 8.83 & 25.32 & 1.17 & 5.85 & $-$0.27 & 5.58 & $-$0.33 & yes$^{4}$ \\
98532 & 189558 & -- & 7.72 & 14.76 & 1.10 & 3.57 & -- & -- & -- & AB$^{4}$ \\
100568 & 193901 & -- & 8.65 & 22.88 & 1.24 & 5.45 & $-$0.19 & 5.26 & $-$0.20 & no$^{2}$ \\
100792 & 194598 & G 24-15 & 8.36 & 17.94 & 1.24 & 4.63 & $-$0.17 & 4.46 & +0.12 & no$^{2}$ \\ 
101346 & 195633 & -- & 8.53 & 8.63 & 1.16 & 3.21 & $-$0.13 & 3.08 & +0.67 & no$^{4}$ \\
103498 & 199289 & -- & 8.29 & 18.94 & 1.03 & 4.68 & $-$0.17 & 4.51 & +0.10 & no$^{4}$ \\ 
104660 & 201889 & -- & 8.04 & 17.95 & 1.44 & 4.31 & $-$0.20 & 4.11 & +0.26 & AB$^{1}$ \\
105858 & 203608 & HR 8181 & 4.22 & 108.50 & 0.59 & 4.40 & $-$0.13 & 4.27 & +0.19 & no$^{4}$ \\ 
105888 & 204155 & G 25-29 & 9.03 & 13.02 & 1.11 & 4.07 & $-$0.16 & 3.91 & +0.33 & no$^{2}$ \\
106447 &  -- & G 26-12 & 12.15 & $-$1.57 & 4.54 & -- & $-$0.21 & -- & -- & no$^{2}$ \\
107975 & 207978 & 15 Peg & 5.54 & 36.15 & 0.69 & 3.32 & $-$0.11 & 3.21 & +0.62 & no$^{13}$ \\
108490 & 208906 & -- & 6.95 & 34.12 & 0.70 & 4.62 & $-$0.15 & 4.47 & +0.11 & AB$^{1}$ \\
109067 & -- & G 18-28 & 9.55 & 21.52 & 1.59 & 6.21 & $-$0.26 & 5.96 & $-$0.48 & AB$^{2}$ \\
109558 & -- & G 126-62 & 9.47 & 8.43 & 1.42 & 4.10 & $-$0.18 & 3.92 & +0.33 & AB$^{2}$ \\
109646 & 210752 & -- & 7.40 & 26.57 & 0.85 & 4.44 & $-$0.14 & 4.30 & +0.18 & no$^{4}$ \\
112229 & 215257 & G 27-44 & 7.40 & 23.66 & 0.97 & 4.14 & $-$0.14 & 4.14 & +0.25 & no$^{2}$ \\
114271 & 218502 & -- & 8.50 & 14.33 & 1.20 & 4.28 & $-$0.11 & 4.17 & +0.26 & -- \\
114962 & 219617 & G 273-1 & 8.17 & 12.04 & 2.41 & 3.57 & $-$0.20 & 3.37 & +0.55 & ABCD$^{1}$ \\
115167 &  -- & G 29-23 & 10.21 & 3.26 & 2.20 & 2.78 & $-$0.20 & 2.58 & +0.87 & AB$^{1}$ \\
117041 & 222766 & -- & 10.15 & 8.46 & 1.76 & 4.79 & $-$0.26 & 4.53 & +0.09 & no$^{6}$ \\
 -- & -- & G05-19  & 11.12 &  -- &  --  &  --  & $-$0.16 & -- & -- & no$^{2}$ \\
 -- & -- & G05-40 & 10.79 & -- & -- & -- & $-$0.17 & -- & -- & no$^{2}$ \\
 -- & -- & G66-51 & 10.63 & -- & -- & -- & $-$0.26 & -- & -- & no$^{2}$ \\
 -- & -- & G166-37 & 12.66 & 5.20$^{14}$ & 0.70 & 6.24 & $-$ 0.27 & 5.97 & $-$0.49 & no$^{2}$ \\
 -- & -- & G170-21 & 12.51 & -- & -- & -- & $-$0.21 & -- & -- & no$^{2}$ \\
\hline
\end{longtable}
\setcounter{table}{1}
\begin{longtable}{ccccccc}
\caption{\label{tab:log} Log book of the observations. 
} \\ 
\hline\hline
HIP & Date of obs. &  Type$^{1}$  & Exp. Time & R & $\lambda_{\rm c}$  & S/N$^{2}$  \\ 
    &              &        &  (s)      &   &     (\AA)         & (final)  \\
\hline  
\endfirsthead
\caption{continued.} \\
\hline\hline
HIP & Date of obs. &  Type  & Exp. Time & R & $\lambda_{\rm c}$  & S/N  \\ 
    &              &        &  (s)      &   &     (\AA)         & (final)   \\
\hline
\endhead
\hline
\multicolumn{7}{l}{(1) Whether the spectra are from new observations, 
the ESO archive data, or the UVES POP library.} \\
\multicolumn{7}{l}{(2) The S/N per pixel of the final combined spectrum 
in the Be region.} \\
\endfoot
171   & 13.oct.2005 & New UVES & 2$\times$150  & 35\,000 & 3460  & 100 \\
      & 13.oct.2005 & New UVES & 2$\times$30   &    & 5800  &    \\
      & 13.oct.2005 & New UVES & 2$\times$30   &    & 8600  &    \\
3026  & 08.oct.2001 & Archive  & 4200          & 35\,000 & 3460  & 130 \\
7459  & 22.sep.2005 & New UVES & 2$\times$1800 & 35\,000 & 3460  &  60 \\
      & 22.sep.2005 & New UVES & 2$\times$800  &    & 5800  &    \\
      & 22.sep.2005 & New UVES & 2$\times$800  &    & 8600  &    \\
10140 & 14.oct.2005 & New UVES & 2$\times$900  & 35\,000 & 3460  &  70 \\
      & 14.oct.2005 & New UVES & 2$\times$350  &    & 5800  &    \\
      & 14.oct.2005 & New UVES & 2$\times$350  &    & 8600  &    \\
10449 & 14.oct.2005 & New UVES & 2$\times$900  & 35\,000 & 3460  & 90 \\
      & 14.oct.2005 & New UVES & 2$\times$350  &    & 5800  &    \\
      & 14.oct.2005 & New UVES & 2$\times$350  &    & 8600  &    \\
11952 & 13.oct.2005 & New UVES & 1500          & 35\,000 & 3460  & 75 \\
      & 13.oct.2005 & New UVES & 2$\times$650  &    & 8600  &    \\
13366 & 13.oct.2005 & New UVES & 450 and 900   & 35\,000 & 3460  & 90 \\
      & 13.oct.2005 & New UVES & 2$\times$150  &    & 5800  &    \\
      & 13.oct.2005 & New UVES & 2$\times$700  &    & 8600  &    \\
14086 & 23.sep.2005 & New UVES & 2$\times$150  & 35\,000 & 3460  & 70 \\
      & 23.sep.2005 & New UVES & 2$\times$30   &    & 5800  &    \\
      & 23.sep.2005 & New UVES & 2$\times$30   &    & 8600  &    \\
14594 & 26.nov.2001 & Archive  & 1800          & 40\,000 & 3460  & 150 \\
17001 & 13.dec.2001 & Archive  & 4$\times$1175 & 40\,000 & 3460  & 110 \\
17147 & 26.nov.2001 & Archive  &  480          & 40\,000 & 3460  & 140 \\
18802 & 28.nov.2001 & Archive  &  1500         & 35\,000 & 3460  & 100 \\
19007 & 16.oct.2005 & New UVES & 2$\times$1200 & 35\,000 & 3460  &  50 \\
      & 16.oct.2005 & New UVES & 2$\times$500  &    & 5800  &    \\
      & 16.oct.2005 & New UVES & 2$\times$500  &    & 8600  &    \\
19814 & 16.oct.2005 & New UVES & 2$\times$2400 & 35\,000 & 3460  &  55 \\
      & 16.oct.2005 & New UVES & 2$\times$1100 &    & 5800  &    \\
      & 16.oct.2005 & New UVES & 2$\times$1100 &    & 8600  &    \\
21609 & 23.sep.2005 & New UVES & 2$\times$1800 & 35\,000 & 3460  &  60 \\
      & 23.sep.2005 & New UVES & 2$\times$800  &    & 5800  &    \\
      & 23.sep.2005 & New UVES & 2$\times$800  &    & 8600  &    \\
22632 & 21.sep.2005 & New UVES & 2$\times$900  & 35\,000 & 3460  &  60 \\
      & 21.sep.2005 & New UVES & 2$\times$350  &    & 5800  &    \\
      & 21.sep.2005 & New UVES & 2$\times$350  &    & 8600  &    \\
24030 & 08.oct.2005 & New UVES & 2$\times$1500 & 35\,000 & 3460  &  75 \\
      & 08.oct.2005 & New UVES & 2$\times$650  &    & 5800  &    \\
      & 08.oct.2005 & New UVES & 2$\times$650  &    & 8600  &    \\
24316 & 22.sep.2005 & New UVES & 2$\times$1200 & 35\,000 & 3460  & 70 \\
      & 22.sep.2005 & New UVES & 2$\times$500  &    & 5800  &    \\
      & 22.sep.2005 & New UVES & 2$\times$500  &    & 8600  &    \\
31188 & 26.nov.2005 & New UVES & 2$\times$700  & 35\,000 & 3460  &  80 \\
      & 26.nov.2005 & New UVES & 2$\times$300  &    & 5800  &    \\
      & 26.nov.2005 & New UVES & 2$\times$300  &    & 8600  &    \\
31639 & 24.sep.2005 & New UVES & 2$\times$1200 & 35\,000 & 3460  &  25 \\
      & 24.sep.2005 & New UVES & 2$\times$500  &    & 5800  &    \\
      & 24.sep.2005 & New UVES & 2$\times$500  &    & 8600  &    \\
33221 & 01.dec.2001 & Archive  & 4200          & 35\,000 & 3460  & 145 \\
      & 26.dec.2001 & Archive  & 4200          &    & 3460  &    \\
      & 27.dec.2001 & Archive  & 4200          &    & 3460  &    \\
33582 & 10.nov.2005 & New UVES & 2$\times$900  & 35\,000 & 3460  &  65 \\
      & 10.nov.2005 & New UVES & 2$\times$350  &    & 5800  &    \\
      & 10.nov.2005 & New UVES & 2$\times$350  &    & 8600  &    \\
34285 & 10.apr.2000 & Archive  & 4500          & 40\,000 & 3460  & 100 \\
      & 11.apr.2000 & Archive  & 3600          & 50\,000 & 3460  &    \\
36491 & 24.dec.2005 & New UVES & 2$\times$500  & 35\,000 & 3460  &  40 \\
      & 24.dec.2005 & New UVES & 2$\times$250  &    & 5800  &    \\
      & 24.dec.2005 & New UVES & 2$\times$250  &    & 8600  &    \\
36640 & 08.oct.2005 & New UVES & 2$\times$150  & 35\,000 & 3460  & 130 \\
      & 08.oct.2005 & New UVES & 2$\times$30   &    & 5800  &    \\
      & 08.oct.2005 & New UVES & 2$\times$30   &    & 8600  &    \\
36818 & 25.nov.2001 & Archive  & 5400          & 35\,000 & 3460  &  75 \\
36849 & 25.nov.2001 & Archive  & 2$\times$2100 & 35\,000 & 3460  & 130 \\
37853 & 08.apr.2001 & UVES POP & 2$\times$60   & 80\,000 & 3460  &  80 \\
38625 & 20.dec.2005 & New UVES & 4$\times$360  & 35\,000 & 3460  &  60 \\
      & 20.dec.2005 & New UVES & 6$\times$120  &    & 5800  &    \\
      & 20.dec.2005 & New UVES & 2$\times$120  &    & 8600  &    \\
42592 & 20.dec.2005 & New UVES & 2$\times$900  & 35\,000 & 3460  &  90 \\
      & 20.dec.2005 & New UVES & 2$\times$350  &    & 5800  &    \\
      & 20.dec.2005 & New UVES & 2$\times$360  &    & 8600  &    \\
44075 & 14.apr.2001 & UVES POP & 200           & 80\,000 & 3460  &  60 \\
44124 & 01.feb.2005 & Archive  & 3$\times$1066 & 40\,000 & 3460  &  105 \\
      & 01.feb.2005 & Archive  & 3$\times$800  &    & 5800  &    \\
45554 & 24.dec.2005 & New UVES & 2$\times$3000 & 35\,000 & 3460  &  45 \\
      & 24.dec.2005 & New UVES & 2$\times$1400 &    & 5800  &    \\
      & 24.dec.2005 & New UVES & 2$\times$1400 &    & 8600  &    \\
48152 & 29.nov.2002 & UVES POP & 3$\times$400  & 80\,000 & 3460  & 115 \\
50139 & 05.jan.2006 & New UVES & 2$\times$400  & 35\,000 & 3460  &  90 \\
      & 05.jan.2006 & New UVES & 2$\times$130  &    & 5800  &    \\
      & 05.jan.2006 & New UVES & 2$\times$130  &    & 8600  &    \\
52771 & 20.mar.2005 & Archive  & 3$\times$1066 & 40\,000 & 3460  & 40 \\
53070 & 05.jan.2006 & New UVES & 600           & 35\,000 & 3460  & 60 \\
      & 05.jan.2006 & New UVES & 2$\times$200  &    & 8600  &    \\ 
55022 & 10.apr.2000 & Archive  & 3$\times$1800 & 40\,000 & 3460  & 90 \\
57265 & 14.jan.2006 & New UVES & 2$\times$2100 & 35\,000 & 3460  & 50 \\
      & 14.jan.2006 & New UVES & 2$\times$900  &    & 5800  &    \\
      & 14.jan.2006 & New UVES & 2$\times$900  &    & 8600  &    \\
58145 & 14.jan.2006 & New UVES & 2$\times$1800 & 35\,000 & 3460  & 85 \\
      & 14.jan.2006 & New UVES & 2$\times$800  &    & 5800  &    \\
      & 14.jan.2006 & New UVES & 2$\times$800  &    & 8600  &    \\
58962 & 12.jan.2002 & Archive  & 2$\times$4800 & 35\,000 & 3460  & 70 \\
59490 & 12.apr.2000 & Archive  & 2$\times$3600 & 40\,000 & 3460  & 70 \\
59750 & 30.jan.2006 & New UVES & 2$\times$150  & 35\,000 & 3460  & 100 \\
      & 30.jan.2006 & New UVES & 2$\times$30   &    & 5800  &    \\
      & 30.jan.2006 & New UVES & 2$\times$30   &    & 8600  &    \\
60632 & 28.mar.2004 & Archive  & 3$\times$1066 & 40\,000 & 3460  & 70 \\
62882 & 10.apr.2000 & Archive  & 2$\times$1800 & 40\,000 & 3460  & 155 \\
63559 & 19.jan.2006 & New UVES & 2$\times$300  & 35\,000 & 3460  & 75 \\
      & 19.jan.2006 & New UVES & 1$\times$200  &    & 5800  &    \\
      & 19.jan.2006 & New UVES & 2$\times$200  &    & 8600  &    \\
63918 & 10.apr.2000 & Archive  & 2$\times$3600 & 40\,000 & 3460  & 100 \\
64426 & 08.jun.2006 & New UVES & 3$\times$450  & 35\,000 & 3460  & 145 \\
      & 08.jun.2006 & New UVES & 4$\times$150  &    & 5800  &    \\
      & 08.jun.2006 & New UVES & 2$\times$150  &    & 8600  &    \\
66665 & 15.jun.2006 & New UVES & 2$\times$1000 & 35\,000 & 3460  &  50 \\
      & 15.jun.2006 & New UVES & 2$\times$400  &    & 5800  &    \\
      & 15.jun.2006 & New UVES & 2$\times$400  &    & 8600  &    \\
67655 & 21.feb.2002 & Archive  & 2$\times$900  & 35\,000 & 3460  & 100 \\
67863 & 11.apr.2001 & Archive  & 3600          & 40\,000 & 3460  & 145 \\
70681 & 09.apr.2000 & Archive  & 2$\times$2700 & 50\,000 & 3460  & 100 \\
71458 & 04.feb.2002 & Archive  & 5$\times$250  & 40\,000 & 3460  & 170 \\
72461 & 25.jun.2006 & New UVES & 2$\times$1800 & 40\,000 & 3460  &  60 \\
      & 25.jun.2006 & New UVES & 2$\times$800  &    & 5800  &    \\
      & 25.jun.2006 & New UVES & 2$\times$800  &    & 8600  &    \\
74067 & 12.jun.2006 & New UVES & 2$\times$500  & 35\,000 & 3460  & 100 \\
      & 12.jun.2006 & New UVES & 2$\times$200  &    & 5800  &    \\
      & 12.jun.2006 & New UVES & 2$\times$200  &    & 8600  &    \\
74079 & 12.jun.2006 & New UVES & 2$\times$400  & 35\,000 & 3460  &  85 \\
      & 12.jun.2006 & New UVES & 2$\times$150  &    & 5800  &    \\
      & 12.jun.2006 & New UVES & 2$\times$150  &    & 8600  &    \\
77946 & 11.jun.2006 & New UVES & 2$\times$800  & 35\,000 & 3460  &  65 \\
      & 11.jun.2006 & New UVES & 2$\times$300  &    & 5800  &    \\
      & 11.jun.2006 & New UVES & 2$\times$300  &    & 8600  &    \\
80003 & 12.jun.2006 & New UVES & 2$\times$2400 & 35\,000 & 3460  &  30 \\
      & 12.jun.2006 & New UVES & 2$\times$900  &    & 5800  &    \\
      & 12.jun.2006 & New UVES & 2$\times$900  &    & 8600  &    \\
80837 & 04.jun.2006 & New UVES & 4$\times$300  & 35\,000 & 3460  & 110 \\
      & 04.jun.2006 & New UVES & 4$\times$100  &    & 5800  &    \\
      & 04.jun.2006 & New UVES & 4$\times$100  &    & 8600  &    \\
81170 & 05.aug.2003 & Archive  & 3$\times$1333 & 40\,000 & 3460  &  45 \\
85963 & 04.jun.2006 & New UVES & 2$\times$400  & 35\,000 & 3460  &  70 \\
      & 04.jun.2006 & New UVES & 2$\times$150  &    & 5800  &    \\
      & 04.jun.2006 & New UVES & 2$\times$150  &    & 8600  &    \\
87693 & 10.apr.2000 & Archive  & 2$\times$3600 & 40\,000 & 3460  &  70 \\
88010 & 07.aug.2003 & Archive  & 3$\times$1333 & 40\,000 & 3460  &  70 \\
92781 & 10.apr.2000 & Archive  & 2$\times$1800 & 40\,000 & 3460  & 100 \\
94449 & 12.jun.2006 & New UVES & 2$\times$1000 & 35\,000 & 3460  &  90 \\
      & 12.jun.2006 & New UVES & 2$\times$400  &    & 5800  &    \\
      & 12.jun.2006 & New UVES & 2$\times$400  &    & 8600  &    \\
98020 & 11.apr.2001 & Archive  & 2700          & 50\,000 & 3460  & 110 \\
98532 & 09.apr.2000 & Archive  & 2$\times$900  & 50\,000 & 3460  & 140 \\
100568 & 12.jun.2006 & New UVES & 2$\times$800  & 35\,000 & 3460  & 120 \\
       & 12.jun.2006 & New UVES & 2$\times$300  &    & 5800  &    \\
       & 12.jun.2006 & New UVES & 2$\times$300  &    & 8600  &    \\
100792 & 25.jun.2006 & New UVES & 2$\times$700  & 35\,000 & 3460  &  50 \\
       & 25.jun.2006 & New UVES & 2$\times$300  &    & 5800  &    \\
       & 25.jun.2006 & New UVES & 2$\times$300  &    & 8600  &    \\
101346 & 10.apr.2000 & Archive  & 2$\times$1500 & 40\,000 & 3460  & 85 \\
103498 & 04.jun.2006 & New UVES & 2$\times$600  & 35\,000 & 3460  & 80 \\
       & 04.jun.2006 & New UVES & 2$\times$200  &    & 5800  &    \\
       & 04.jun.2006 & New UVES & 2$\times$200  &    & 8600  &    \\
104660 & 30.jun.2006 & New UVES & 2$\times$500  & 35\,000 & 3460  &  40 \\
       & 30.jun.2006 & New UVES & 2$\times$200  &    & 5800  &    \\
       & 30.jun.2006 & New UVES & 2$\times$200  &    & 8600  &    \\
105858 & 07.dec.2002 & UVES POP & 3$\times$80   & 80\,000 & 3460  & 400 \\
105888 & 18.oct.2005 & Archive  & 800           & 40\,000 & 3460  &  65 \\
106447 & 25.jun.2006 & New UVES & 2$\times$3600 & 35\,000 & 3460  &  45 \\
       & 25.jun.2006 & New UVES & 6$\times$1100 &    & 5800  &    \\
       & 30.jun.2006 & New UVES & 3600          & 35\,000 & 3460  &    \\
       & 30.jun.2006 & New UVES & 3$\times$1100 &    & 8600  &    \\
107975 & 30.jun.2006 & New UVES & 2$\times$150  & 35\,000 & 3460  &  55 \\
       & 30.jun.2006 & New UVES & 2$\times$10   &    & 5800  &    \\
       & 30.jun.2006 & New UVES & 2$\times$10   &    & 8600  &    \\
108490 & 12.jun.2006 & New UVES & 300           & 35\,000 & 3460  &  90 \\
       & 12.jun.2006 & New UVES & 2$\times$30   &    & 8600  &    \\
       & 09.jul.2006 & New UVES & 300           & 35\,000 & 3460  &    \\
       & 09.jul.2006 & New UVES & 2$\times$30   &    & 5800  &    \\
109067 & 13.oct.2005 & New UVES & 1500          & 35\,000 & 3460  &  50 \\
       & 13.oct.2005 & New UVES & 2$\times$650  &    & 5800  &    \\
109558 & 29.may.2006 & New UVES & 1800          & 35\,000 & 3460  &  55 \\
       & 29.may.2006 & New UVES & 2$\times$600  &    & 5800  &    \\
109646 & 30.jun.2006 & New UVES & 2$\times$400  & 35\,000 & 3460  & 105 \\
       & 30.jun.2006 & New UVES & 2$\times$150  &    & 5800  &    \\
       & 30.jun.2006 & New UVES & 2$\times$150  &    & 8600  &    \\
112229 & 30.jun.2006 & New UVES & 400           & 35\,000 & 3460  & 45 \\
       & 30.jun.2006 & New UVES & 2$\times$150  &    & 8600  &    \\
114271 & 17.oct.2005 & New UVES & 1200          & 35\,000 & 3460  & 120 \\
       & 17.oct.2005 & New UVES & 2$\times$500  &    & 8600  &    \\
114962 & 29.may.2006 & New UVES & 2$\times$600  & 35\,000 & 3460  & 100 \\
       & 29.may.2006 & New UVES & 2$\times$200  &    & 5800  &    \\
       & 29.may.2006 & New UVES & 2$\times$200  &    & 8600  &    \\
115167 & 11.aug.2003 & Archive  & 3$\times$1066 & 40\,000 & 3460  &  65 \\
117041 & 13.oct.2005 & New UVES & 1800          & 35\,000 & 3460  &  35 \\
       & 13.oct.2005 & New UVES & 2$\times$800  &    & 5800  &    \\
G05-19  & 15.oct.2005 & New UVES & 2700          & 35\,000 & 3460  & 55 \\
        & 15.oct.2005 & New UVES & 3000          & 35\,000 & 3460  &    \\
        & 15.oct.2005 & New UVES & 2$\times$1300  &    & 5800  &    \\
        & 15.oct.2005 & New UVES & 2$\times$1400  &    & 8600  &    \\
G05-40  & 08.oct.2001 & Archive  & 2$\times$4800  & 35\,000 & 3460  & 90 \\
        & 22.nov.2001 & Archive  & 2$\times$4800  & 35\,000 & 3460  &    \\
G66-51  & 02.jun.2006 & New UVES & 2$\times$2400  & 35\,000 & 3460  & 30 \\
        & 02.jun.2006 & New UVES & 2$\times$900  &    & 5800  &    \\
        & 02.jun.2006 & New UVES & 2$\times$900  &    & 8600  &    \\
G166-37 & 02.jun.2006 & New UVES & 3600          & 35\,000 & 3460  & 25 \\
        & 02.jun.2006 & New UVES & 3$\times$1100 &    & 5800  &    \\
        & 12.jun.2006 & New UVES & 2$\times$3600 & 35\,000 & 3460  &    \\
        & 12.jun.2006 & New UVES & 3$\times$1100 &    & 5800  &    \\
        & 12.jun.2006 & New UVES & 3$\times$1100 &    & 8600  &    \\
G170-21 & 01.jun.2006 & New UVES & 3600          & 35\,000 & 3460  &  35 \\
        & 01.jun.2006 & New UVES & 3$\times$1100 &    & 5800  &    \\
        & 02.jun.2006 & New UVES & 3600          & 35\,000 & 3460  &    \\
        & 02.jun.2006 & New UVES & 3$\times$1100 &    & 5800  &    \\
        & 02.jun.2006 & New UVES & 2$\times$1100 &    & 8600  &    \\
        & 12.jun.2006 & New UVES & 3600          & 35\,000 & 3460  &    \\
        & 12.jun.2006 & New UVES & 3$\times$1100 &    & 8600  &    \\
\hline
\end{longtable}
\setcounter{table}{2}
\begin{longtable}{ccccccccccc}
\caption{\label{tab:par} Atmospheric parameters, as adopted from the
  literature, beryllium abundances as derived in this work, and lithium
  abundances, adopted from the literature when available or derived in
  this work when not. Details on the parameters and the abundances are
  given in the text. The reference papers of the atmospheric parameters
  and lithium abundances are also listed.} \\ 
\hline\hline
Star &  T$_{\rm eff}$ & log g &  log g & $\xi$      & [Fe/H] &
  Ref.   & log(Be/H) & log(Be/H) & A(Li) & Ref. \\
   &     Kelvin     & Liter. & Paral. & km s$^{-1}$ &
  &        & 3131 &  average &  &   \\
\hline  
\endfirsthead
\caption{continued.} \\
\hline\hline
Star &  T$_{\rm eff}$ & log g &  log g & $\xi$      & [Fe/H] &
  Ref. & log(Be/H) & log(Be/H) & A(Li) & Ref.   \\
   &     Kelvin     &  Liter. & Paral.  & km s$^{-1}$ &
  &       & 3131  &  average  &    &      \\
\hline
\endhead
\hline
\multicolumn{11}{l}{(1) Takeda \& Kawanomoto (\cite{TK05}), (2) Charbonnel \& Primas 
(\cite{CP05}), (3) Chen et al.\ (\cite{Ch01}), (4) This work,} \\
\multicolumn{11}{l}{(5) Boesgaard et al. (\cite{Bo05}), (6) Fulbright (\cite{F00}), 
(7) Favata et al.\ (\cite{Fav96}), (8) Ryan \& Deliyannis (\cite{RyD95}),} \\
\multicolumn{11}{l}{(9) Romano et al.\ (\cite{Ro99}), (10) Spite et al.\ (\cite{Sp94}), 
(11) Gratton et al.\ (\cite{Gr00}).} \\
\endfoot
HIP 171 & 5275 & 4.10 & 4.30 & 1.05 & $-$0.90 & F00 & $-$11.55 & $-$11.51 & $\leq$0.80 & (1) \\
HIP 3026 & 5950 & 3.90 & 4.10 & 1.40 & $-$1.20 & F00 & $-$12.13 & $-$12.11 & 2.42 & (2) \\
HIP 7459 & 5909 & 4.46 & 4.59 & 1.23 & $-$1.15 & NS97 & $-$11.90 & $-$11.88 & 2.12 & (3) \\
HIP 10140 & 5425 & 4.10 & 4.25 & 0.85 & $-$1.00 & F00 & $-$11.68 & $-$11.64 & 1.54 & (2) \\
HIP 10449 & 5640 & 4.40 & 4.38 & 1.00 & $-$0.80 & F00 & $-$11.31 & $-$11.31 & 1.00 & (4) \\
HIP 11952 & 6100 & 4.20 & 4.25 & 0.95 & $-$1.60 & F00 & $-$12.28 & $-$12.26 & 2.16 & (2) \\
HIP 13366 & 5700 & 4.20 & 4.09 & 0.95 & $-$0.70 & F00 & $-$11.07 & $-$11.04 & 1.28 & (3) \\
HIP 14086 & 5075 & 3.60 & 3.48 & 1.10 & $-$0.60 & F00 & -- & -- & -- & -- \\
HIP 14594 & 5825 & 4.20 & 4.42 & 1.10 & $-$2.00 & F00 & $-$12.63 & $-$12.58 & 2.18 & (2) \\
HIP 17001 & 5360 & 3.00 & -- & 1.20 & $-$2.35 & GS88 & $\leq$$-$13.83 & -- & 1.17 & (2) \\
HIP 17147 & 5800 & 4.30 & 4.29 & 1.10 & $-$0.80 & F00 & $-$11.20 & $-$11.20 & 1.45 & (3) \\
HIP 18802 & 5886 & 4.33 & 4.22 & 1.38 & $-$0.85 & NS97 & $-$11.41 & $-$11.41 & 1.89 & (2) \\
HIP 19007 & 5150 & 4.50 & 4.74 & 1.20 & $-$0.50 & F00 & -- & -- & -- & -- \\ 
HIP 19814 & 5378 & 4.43 & -- & 0.84 & $-$0.69 & NS97 & -- & $\leq$$-$11.64 & 0.98 & (5) \\
HIP 21609 & 5200 & 3.80 & 4.55 & 1.55 & $-$1.60 & F00 & -- & -- & 1.43 & (2) \\
HIP 22632 & 5825 & 4.30 & 4.41 & 1.35 & $-$1.40 & F00 & $-$12.17 & $-$12.14 & 2.17 & (2) \\
HIP 24030 & 5700 & 4.20 & 4.26 & 1.11 & $-$1.20 & Pr00 & $-$11.55 & $-$11.53 & 2.08 & (3) \\
HIP 24316 & 5725 & 4.40 & 4.45 & 1.30 & $-$1.50 & F00 & $-$11.98 & $-$11.96 & 2.00 & (2) \\
HIP 31188 & 5750 & 4.10 & 4.28 & 1.65 & $-$0.70 & F00 & $-$11.46 & $-$11.46 & 1.95 & (4) \\
HIP 31639 & 5300 & 4.30 & 4.55 & 0.60 & $-$0.50 & F00 & $-$11.27 & $-$11.15 & $\leq$0.50 & (4) \\
HIP 33221 & 6097 & 4.09 & 4.00 & 1.86 & $-$1.30 & NS97 & $-$11.67 & $-$11.67 & 2.32 & (3) \\
HIP 33582 & 5725 & 4.30 & 4.32 & 1.25 & $-$0.50 & F00 & $-$11.09 & $-$11.09 & 1.10 & (4) \\
HIP 34285 & 5933 & 4.26 & 4.30 & 1.50 & $-$0.90 & NS97 & $-$11.85 & $-$11.85 & 2.22 & (3) \\
HIP 36491 & 5800 & 4.40 & 4.39 & 1.10 & $-$0.80 & F00 & $-$11.40 & $-$11.40 & 1.70 & (6) \\
HIP 36640 & 5830 & 3.64 & -- & 0.70 & $-$0.70 & GS91 & $-$11.44 & $-$11.42 & 2.36 & (3) \\
HIP 36818 & 5672 & 4.57 & 4.88 & 0.90 & $-$0.83 & NS97 & -- & $\leq$$-$12.58 & $\leq$0.70 & (3) \\
HIP 36849 & 5850 & 4.10 & 4.15 & 1.10 & $-$0.70 & F00 & $-$11.35 & $-$11.33 & 1.96 & (3) \\
HIP 37853 & 5822 & 4.42 & 4.18 & 1.21 & $-$0.78 & Ed93 & $-$11.16 & $-$11.19 & 1.37 & (3) \\
HIP 38625 & 5200 & 4.40 & 4.57 & 0.30 & $-$0.70 & F00 & $-$11.42 & -- & $\leq$0.16 & (7) \\
HIP 42592 & 6025 & 4.10 & 4.02 & 1.20 & $-$2.00 & F00 & $-$12.58 & $-$12.58 & 2.12 & (2) \\
HIP 44075 & 5900 & 4.20 & 4.11 & 1.25 & $-$0.70 & F00 & $-$11.12 & $-$11.12 & 2.06 & (3) \\
HIP 44124 & 5750 & 3.60 & -- & 0.10 & $-$1.80 & F00 & $-$12.92 & $-$12.86 & 1.96 & (4) \\
HIP 45554 & 6021 & 4.44 & 3.96 & 1.34 & $-$0.75 & NS97 & $-$11.45 & $-$11.45 & 2.09 & (3) \\
HIP 48152 & 6375 & 4.10 & 4.06 & 2.25 & $-$2.00 & F00 & $-$12.67 & $-$12.67 & 2.25 & (2) \\
HIP 50139 & 5600 & 4.30 & 4.30 & 0.35 & $-$0.70 & F00 & $-$11.28 & $-$11.23 & $\leq$0.40 & (4) \\
HIP 52771 & 5700 & 4.50 & 4.48 & 1.30 & $-$1.80 & F00 & -- & $-$12.55 & 2.16 & (2) \\
HIP 53070 & 5900 & 4.20 & 4.27 & 1.45 & $-$1.40 & F00 & $-$11.80 & $-$11.80 & 2.21 & (2) \\
HIP 55022 & 6450 & 4.20 & 4.05 & 1.80 & $-$0.80 & F00 & -- & $\leq$$-$12.75 & $\leq$1.31 & (2) \\
HIP 57265 & 5875 & 4.00 & 4.13 & 1.50 & $-$1.00 & F00 & $-$11.90 & $-$11.90 & 2.55 & (5) \\
HIP 58145 & 5946 & 4.41 & 3.96 & 1.32 & $-$1.04 & NS97 & $-$11.31 & $-$11.31 & 1.95 & (3) \\
HIP 58962 & 5831 & 4.36 & -- & 1.30 & $-$0.80 & NS97 & $-$11.68 & $-$11.64 & 1.95 & (3) \\
HIP 59490 & 6046 & 4.46 & 4.46 & 1.34 & $-$1.26 & NS97 & $-$10.60 & $-$10.58 & 2.51 & (3) \\
HIP 59750 & 6200 & 4.40 & 4.26 & 1.10 & $-$0.60 & F00 & -- & $\leq$$-$12.50 & $\leq$1.10 & (3) \\
HIP 60632 & 6200 & 4.40 & 4.45 & 1.35 & $-$1.50 & F00 & $-$12.10 & $-$12.10 & 2.21 & (2) \\
HIP 62882 & 5600 & 3.70 & -- & 0.04 & $-$1.10 & F00 & $-$11.20 & $-$11.20 & 2.19 & (8) \\
HIP 63559 & 5865 & 4.41 & 4.16 & 1.26 & $-$0.93 & NS97 & -- & -- & 2.14 & (2) \\
HIP 63918 & 5720 & 4.14 & 3.91 & 1.49 & $-$0.65 & NS97 & $-$10.95 & $-$10.95 & 1.99 & (3) \\
HIP 64426 & 5800 & 4.10 & 4.10 & 1.25 & $-$0.70 & F00 & $-$11.31 & $-$11.31 & 2.01 & (3) \\
HIP 66665 & 5500 & 3.80 & 3.77 & 1.05 & $-$0.80 & F00 & $-$11.61 & $-$11.63 & 0.90 & (4) \\
HIP 67655 & 5396 & 4.38 & 4.62 & 0.92 & $-$0.93 & NS97 & $-$11.69 & $-$11.50 & $\leq$1.25 & (2) \\
HIP 67863 & 5686 & 4.40 & 4.42 & 1.13 & $-$0.70 & NS97 & $-$10.95 & $-$10.95 & $\leq$1.34 & (2) \\
HIP 70681 & 5450 & 4.50 & 4.55 & 0.80 & $-$1.10 & F00 & $-$11.12 & $-$11.12 & 1.54 & (3) \\
HIP 71458 & 5480 & 3.10 & -- & 1.98 & $-$2.10 & McW95 & $\leq$$-$13.90 & -- & 1.29 & (2) \\
HIP 72461 & 5875 & 4.10 & 4.31 & 0.40 & $-$2.30 & F00 & -- & -- & 2.22 & (2) \\
HIP 74067 & 5575 & 4.30 & 4.41 & 1.10 & $-$0.80 & F00 & $-$11.36 & $-$11.36 & 0.95 & (4) \\
HIP 74079 & 5825 & 4.00 & 3.92 & 1.30 & $-$0.70 & F00 & $-$11.28 & $-$11.28 & 2.24 & (2) \\
HIP 77946 & 6550 & 3.60 & 3.70 & 1.65 & $-$0.90 & F00 & $\leq$$-$12.95 & $\leq$$-$12.85 & $\leq$1.45 & (6) \\
HIP 80003 & 5486 & 4.80 & 4.80 & 0.00 & $-$0.87 & SB02 & $-$11.62 & $-$11.62 & 1.25 & (5) \\
HIP 80837 & 5800 & 4.10 & 4.07 & 1.15 & $-$0.70 & F00 & $-$11.22 & $-$11.20 & 1.53 & (9) \\
HIP 81170 & 5175 & 4.70 & 4.64 & 0.30 & $-$1.10 & F00 & $-$11.53 & $-$11.42 & $\leq$0.30 & (10) \\
HIP 85963 & 6227 & 3.94 & 3.75 & 2.16 & $-$0.71 & Ed93 & $-$11.78  & $-$11.78  & $\leq$0.70  & (4) \\
HIP 87693 & 6175 & 4.00 & 3.99 & 0.90 & $-$2.00 & F00 & $-$12.77 & $-$12.77 & 2.20 & (2) \\
HIP 88010 & 5200 & 4.00 & 4.15 & 0.70 & $-$1.40 & F00 & $-$12.40 & $-$12.34 & 1.67 & (2) \\
HIP 92781 & 5650 & 4.20 & 4.12 & 0.95 & $-$0.60 & F00 & $-$11.14 & $-$11.14 & -- & -- \\
HIP 94449 & 5625 & 3.70 & 3.74 & 1.15 & $-$1.20 & F00 & $-$11.50 & $-$11.50 & 1.81 & (6) \\
HIP 98020 & 5325 & 4.60 & 4.54 & 1.10 & $-$1.60 & F00 & -- & -- & 1.61 & (2) \\
HIP 98532 & 5550 & 3.60 & -- & 1.30 & $-$1.10 & F00 & $-$11.43 & $-$11.43 & 2.15 & (3) \\
HIP 100568 & 5650 & 4.40 & 4.51 & 1.10 & $-$1.00 & F00 & $-$11.98 & $-$11.92 & 1.92 & (2) \\
HIP 100792 & 5875 & 4.20 & 4.26 & 1.40 & $-$1.10 & F00 & $-$11.97 & $-$11.93 & 2.02 & (2) \\
HIP 101346 & 6000 & 3.90 & 3.74 & 1.40 & $-$0.50 & F00 & $-$11.52 & $-$11.52 & 2.15 & (9) \\
HIP 103498 & 5894 & 4.38 & 4.29 & 1.32 & $-$1.03 & Ed93 & $-$11.36 & $-$11.39 & 2.00 & (11) \\
HIP 104660 & 5500 & 3.90 & 4.01 & 1.15 & $-$0.80 & F00 & $-$11.41 & $-$11.44 & 1.10 & (2) \\
HIP 105858 & 6139 & 4.34 & 4.26 & 1.57 & $-$0.67 & Ed93 & $-$11.32 & $-$11.32 & 2.39 & (3) \\
HIP 105888 & 5700 & 4.00 & 3.99 & 1.00 & $-$0.60 & F00 & $-$11.13 & $-$11.13 & 1.37 & (6) \\
HIP 106447 & 6089 & 4.04 & -- & 1.50 & $-$2.48 & SB02 & $-$12.88 & -- & 2.23 & (5) \\
HIP 107975 & 6275 & 3.90 & 3.87 & 1.50 & $-$0.50 & F00 & $-$12.25 & $-$12.25 & $\leq$1.10 & (9) \\ 
HIP 108490 & 6009 & 4.41 & 4.30 & 1.37 & $-$0.72 & Ed93 & $-$11.34 & $-$11.30 & 2.31 & (3) \\
HIP 109067 & 5300 & 4.30 & 4.68 & 0.85 & $-$0.80 & F00 & $-$11.39 & $-$11.37 & $\leq$0.10 & (4) \\
HIP 109558 & 6025 & 4.00 & 4.09 & 1.10 & $-$1.50 & F00 & $-$12.30 & $-$12.34 & 2.10 & (2) \\
HIP 109646 & 5910 & 4.25 & 4.21 & 1.50 & $-$0.64 & Ed93 & $-$11.38 & $-$11.36 & 2.23 & (3) \\
HIP 112229 & 5983 & 4.37 & 4.16 & 1.40 & $-$0.65 & Ed93 & $-$11.21 & $-$11.25 & 2.31 & (3) \\
HIP 114271 & 6200 & 4.10 & 4.21 & 0.90 & $-$1.70 & F00 & $-$12.44 & $-$12.39 & 2.32 & (2) \\
HIP 114962 & 5825 & 4.30 & 3.81 & 1.40 & $-$1.40 & F00 & $-$12.40 & $-$12.34 & 2.25 & (2) \\
HIP 115167 & 6100 & 3.80 & 3.57 & 1.15 & $-$1.50 & F00 & $-$12.45 & $-$12.45 & 2.17 & (2) \\
HIP 117041 & 5300 & 4.20 & 4.11 & 0.90 & $-$0.80 & F00 & $-$11.34 & $-$11.34 & $\leq$0.10 & (4) \\
G05-19 & 5942 & 4.24 & -- & 0.89 & $-$1.10 & SB02 & $-$11.98 & $-$12.01 & 2.26 & (5) \\
G05-40 & 5863 & 4.24 & -- & 1.48 & $-$0.83 & NS97 & $-$11.15 & $-$11.15 & 1.90 & (3) \\
G66-51 & 5255 & 4.48 & -- & 0.90 & $-$1.00 & Pr00 & $-$11.47 & $-$11.37 & $\leq$0.30 & (4) \\
G166-37 & 5300 & 4.80 & 4.69 & 0.60 & $-$1.20 & F00 & -- & -- & 1.28 & (5) \\
G170-21 & 5664 & 4.65 & -- & 0.00 & $-$1.45 & SB02 & $-$11.65 & $-$11.58 & 1.91 & (5) \\
\hline
\end{longtable}
\setcounter{table}{5}
\begin{longtable}{ccccccccccc}
\caption{\label{tab:prob} The probabilities of each star of belonging 
to the thin disk, the thick disk, and the halo as calculated by Venn (\cite{Venn04}), 
the kinematic classification$^{1}$ according to Gratton et al. (\cite{G03,G03b}), the average 
[$\alpha$/Fe], and the orbital parameters: minimum radii (R$_{\rm rmin}$), 
maximum radii (R$_{\rm max}$), maximum distance from the plane (Z$_{\rm max}$), 
eccentricity, and rotational velocity.
} \\ 
\hline\hline
Star & Thin disk & Thick disk &  Halo  &  Gratton  & [$\alpha$/Fe]  &  R min. & R max. & Z max &  ecc. & V$_{\rm rot}$ \\
     &  prob.    &  prob.     &  prob. &  clas.$^{1}$    &                &         &        &       &       &           \\
\hline  
\endfirsthead
\caption{continued.} \\
\hline\hline
Star & Thin disk & Thick disk &  Halo  &  Gratton  & [$\alpha$/Fe]  &  R min. & R max. & Z max &  ecc. & V$_{\rm rot}$ \\
     &  prob.    &  prob.     &  prob. &  clas.$^{1}$    &                &         &        &       &       &           \\
\hline
\endhead
\hline
\multicolumn{11}{l}{(1) The following code is adopted: 0 for the accretion component , 1 for 
the dissipative component, and 2 for thin disk stars.} \\
\multicolumn{11}{l}{(2) The orbital parameters for these stars were calculated in this work.}
\endfoot
HIP 171 & 0.20 & 0.80 & 0.00 & 0 & 0.41 & 4.58 & 8.50 & 0.31 & 0.30 & 152.63 \\
HIP 3026 & 0.00 & 0.00 & 1.00 & 1 & 0.30 & 0.19 & 10.72 & 7.62 & 0.97 & $-$10.61 \\
HIP 7459 & 0.00 & 0.00 & 1.00 & 1 & 0.17 & 0.67 & 16.87 & 8.34 & 0.92 & $-$38.62 \\
HIP 10140 & 0.00 & 0.90 & 0.10 & 0 & 0.29 & 4.29 & 8.97 & 0.61 & 0.35 & 147.98 \\
HIP 10449 & 0.00 & 0.00 & 1.00 & 1 & 0.33 & 0.40 & 12.82 & 6.64 & 0.94 & 22.69 \\
HIP 11952$^{2}$ & 0.00 & 0.90 & 0.10 & 0 & 0.43 & 3.70 & 8.67 & 0.42 & 0.40 & 122.0 \\
HIP 13366 & 0.00 & 0.70 & 0.30 & 0 & 0.31 & 3.35 & 8.85 & 1.65 & 0.45 & 120.06 \\
HIP 14086 & 0.10 & 0.90 & 0.00 & 0 & 0.34 & 5.27 & 9.02 & 1.05 & 0.26 & 167.76 \\
HIP 14594 & 0.00 & 0.10 & 0.90 & 0 & 0.41 & 2.34 & 12.24 & 1.52 & 0.68 & 102.15 \\
HIP 17001 & 0.00 & 0.00 & 1.00 & -- & 0.16 & -- & -- & -- & -- & -- \\
HIP 17147 & 0.00 & 0.90 & 0.10 & 0 & 0.33 & 3.74 & 9.93 & 0.52 & 0.45 & 140.82 \\
HIP 18802 & 0.30 & 0.70 & 0.00 & 0 & 0.18 & 4.30 & 11.06 & 0.05 & 0.44 & 162.33 \\
HIP 19007 & 0.90 & 0.10 & 0.00 & 2 & 0.21 & 7.95 & 10.97 & 0.11 & 0.16 & 236.72 \\
HIP 19814$^{2}$ & 0.00 & 0.00 & 1.00 & 1 & 0.05 & 0.58 & 31.41 & 10.39 & 0.96 & 28.00 \\
HIP 21609 & 0.00 & 0.00 & 1.00 & 1 & 0.45 & 1.51 & 47.65 & 2.24 & 0.94 & 89.89 \\
HIP 22632 & 0.00 & 0.90 & 0.10 & 0 & 0.36 & 3.48 & 8.87 & 0.24 & 0.44 & 128.43 \\
HIP 24030 & 0.00 & 0.80 & 0.20 & 0 & 0.28 & 4.06 & 8.66 & 2.49 & 0.36 & 131.64 \\
HIP 24316 & 0.00 & 0.00 & 1.00 & 1 & 0.37 & 2.65 & 15.78 & 4.71 & 0.71 & $-$120.15 \\
HIP 31188 & 0.00 & 0.90 & 0.10 & 0 & 0.13 & 6.98 & 8.68 & 1.42 & 0.11 & 191.26 \\
HIP 31639 & 0.20 & 0.70 & 0.00 & 0 & 0.27 & 5.71 & 8.55 & 0.55 & 0.20 & 173.93 \\
HIP 33221 & 0.00 & 0.70 & 0.30 & 0 & 0.28 & 6.67 & 8.88 & 3.43 & 0.14 & 175.82 \\
HIP 33582 & 0.00 & 0.00 & 1.00 & 0 & 0.37 & 1.74 & 14.68 & 0.18 & 0.79 & 89.37 \\
HIP 34285 & 0.00 & 0.00 & 1.00 & 1 & 0.05 & 0.63 & 25.25 & 2.14 & 0.95 & $-$43.26 \\
HIP 36491 & 0.00 & 0.80 & 0.20 & 0 & 0.30 & 2.60 & 8.71 & 0.04 & 0.54 & 103.85 \\
HIP 36640 & 0.00 & 0.00 & 1.00 & -- & 0.06 & -- & -- & -- & -- & -- \\
HIP 36818 & 0.00 & 0.00 & 1.00 & 1 & 0.04 & 0.66 & 14.63 & 1.73 & 0.91 & $-$37.34 \\
HIP 36849 & 0.00 & 0.90 & 0.10 & 0 & 0.26 & 3.69 & 9.27 & 0.78 & 0.43 & 134.34 \\
HIP 37853$^{2}$ & 0.00 & 0.00 & 1.00 & 0 & 0.28 & 3.96 & 11.00 & 0.38 & 0.47 & 141.60 \\
HIP 38625 & 0.20 & 0.80 & 0.00 & 0 & 0.28 & 4.92 & 9.39 & 0.08 & 0.31 & 166.78 \\
HIP 42592 & 0.00 & 0.00 & 1.00 & 1 & 0.30 & 2.92 & 21.49 & 3.08 & 0.76 & $-$137.69 \\
HIP 44075 & 0.00 & 0.80 & 0.20 & 0 & 0.34 & 3.95 & 8.75 & 1.74 & 0.38 & 134.02 \\
HIP 44124 & 0.00 & 0.00 & 1.00 & 0 & 0.29 & 0.76 & 13.22 & 0.17 & 0.89 & 46.74 \\
HIP 45554$^{2}$ & 0.00 & 0.00 & 1.00 & 1 & 0.08 & 1.02 & 8.65 & 2.90 & 0.79 & $-$0.62 \\
HIP 48152 & 0.00 & 0.00 & 1.00 & 1 & 0.38 & 0.19 & 15.58 & 9.08 & 0.98 & $-$12.26 \\
HIP 50139 & 0.50 & 0.40 & 0.00 & 0 & 0.25 & 5.80 & 10.44 & 0.22 & 0.29 & 193.21 \\
HIP 52771 & 0.00 & 0.00 & 1.00 & 1 & 0.45 & 2.89 & 12.65 & 1.75 & 0.63 & $-$117.71 \\
HIP 53070 & 0.00 & 0.60 & 0.40 & 0 & 0.44 & 2.01 & 8.58 & 0.18 & 0.62 & 85.88 \\
HIP 55022 & 0.00 & 0.50 & 0.50 & 1 & 0.38 & 6.60 & 16.23 & 2.46 & 0.42 & 235.35 \\
HIP 57265$^{2}$ & 0.00 & 0.00 & 1.00 & 1 & 0.21 & 0.13 & 43.31 & 21.72 & 0.99 & $-$6.00 \\
HIP 58145$^{2}$ & 0.00 & 0.90 & 0.10 & 0 & 0.24 & 4.87 & 8.55 & 1.37 & 0.27 & 143.00 \\
HIP 58962$^{2}$ & 0.00 & 0.00 & 1.00 & 1 & 0.06 & 0.56 & 8.54 & 2.75 & 0.88 & 21.00 \\
HIP 59490 & 0.00 & 0.00 & 1.00 & 1 & 0.25 & 0.82 & 8.61 & 0.55 & 0.83 & $-$44.47 \\
HIP 59750 & 0.00 & 0.90 & 0.10 & 0 & 0.31 & 4.36 & 9.23 & 0.89 & 0.36 & 150.82 \\
HIP 60632 & 0.00 & 0.00 & 1.00 & 1 & 0.41 & 0.21 & 10.92 & 6.77 & 0.96 & 11.03 \\
HIP 62882 & 0.00 & 0.00 & 1.00 & 1 & 0.37 & 0.33 & 21.76 & 10.23 & 0.97 & 22.14 \\
HIP 63559 & 0.00 & 0.00 & 1.00 & 1 & 0.11 & 0.38 & 9.00 & 5.19 & 0.92 & $-$18.24 \\
HIP 63918 & 0.00 & 0.00 & 1.00 & 1 & 0.21 & 1.67 & 9.70 & 0.07 & 0.71 & $-$80.00 \\
HIP 64426 & 0.00 & 0.90 & 0.10 & 0 & 0.29 & 4.58 & 9.52 & 1.35 & 0.35 & 155.72 \\
HIP 66665$^{2}$ & 0.00 & 0.20 & 0.80 & 0 & 0.36 & 3.28 & 12.57 & 2.19 & 0.59 & 123.00 \\
HIP 67655 & 0.30 & 0.70 & 0.00 & 0 & 0.22 & 5.82 & 8.75 & 0.38 & 0.19 & 177.72 \\
HIP 67863 & 0.00 & 0.00 & 1.00 & 1 & 0.22 & 0.59 & 9.01 & 6.24 & 0.88 & $-$27.24 \\
HIP 70681 & 0.00 & 0.90 & 0.10 & 0 & 0.31 & 6.14 & 8.50 & 1.35 & 0.16 & 178.04 \\
HIP 71458 & 0.00 & 0.00 & 1.00 & -- & 0.35 & -- & -- & -- & -- & -- \\
HIP 72461 & 0.00 & 0.60 & 0.40 & 0 & 0.44 & 2.83 & 10.82 & 0.46 & 0.59 & 119.11 \\
HIP 74067 & 0.00 & 0.90 & 0.10 & 0 & 0.30 & 4.95 & 8.53 & 1.09 & 0.27 & 157.5 \\
HIP 74079 & 0.90 & 0.10 & 0.00 & 0 & 0.27 & 7.99 & 9.41 & 0.25 & 0.08 & 222.97 \\
HIP 77946 & 0.00 & 0.00 & 1.00 & 0 & 0.42 & 2.20 & 10.66 & 4.03 & 0.66 & 94.05 \\
HIP 80003$^{2}$ & 0.00 & 0.00 & 1.00 & 1 & 0.21 & 3.19 & 25.24 & 4.32 & 0.78 & 146.00 \\
HIP 80837 & 0.00 & 0.00 & 1.00 & 1 & 0.32 & 0.70 & 9.70 & 5.24 & 0.87 & $-$39.51 \\
HIP 81170 & 0.00 & 0.00 & 1.00 & 0 & 0.36 & 1.51 & 9.28 & 6.03 & 0.72 & 52.21 \\
HIP 85963$^{2}$ & 0.80 & 0.20 & 0.00 & 2 & 0.15 & 6.93 & 8.55 & 0.05 & 0.11 & 185.90 \\
HIP 87693$^{2}$ & 0.00 & 0.00 & 1.00 & 1 & 0.44 & 4.42 & 8.89 & 0.45 & 0.34 & $-$164.00 \\
HIP 88010 & 0.00 & 0.00 & 1.00 & 1 & 0.31 & 0.94 & 21.21 & 2.37 & 0.92 & $-$56.94 \\
HIP 92781 & 0.00 & 0.20 & 0.80 & 0 & 0.29 & 1.73 & 9.83 & 0.41 & 0.70 & 80.00 \\
HIP 94449 & 0.00 & 0.00 & 1.00 & 1 & 0.38 & 1.92 & 11.42 & 1.41 & 0.71 & $-$90.38 \\
HIP 98020 & 0.00 & 0.20 & 0.80 & 0 & 0.27 & 2.67 & 11.29 & 1.81 & 0.62 & 111.38 \\
HIP 98532 & 0.00 & 0.50 & 0.50 & 0 & 0.39 & 2.34 & 9.31 & 1.00 & 0.60 & 97.36 \\
HIP 100568 & 0.00 & 0.00 & 1.00 & 1 & 0.20 & 0.45 & 10.59 & 1.87 & 0.92 & $-$21.64 \\
HIP 100792 & 0.00 & 0.00 & 1.00 & 1 & 0.24 & 0.97 & 8.88 & 0.34 & 0.80 & $-$50.74 \\
HIP 101346 & 0.60 & 0.40 & 0.00 & 0 & 0.16 & 5.76 & 9.53 & 0.06 & 0.25 & 187.90 \\
HIP 103498$^{2}$ & 0.10 & 0.80 & 0.00 & 0 & 0.30 & 4.94 & 8.75 & 0.33 & 0.28 & 151.00 \\
HIP 104660 & 0.00 & 0.80 & 0.10 & 0 & 0.42 & 3.64 & 10.73 & 0.48 & 0.49 & 143.58 \\
HIP 105858 & 0.90 & 0.10 & 0.00 & 0 & 0.09 & 8.43 & 12.69 & 0.12 & 0.20 & 249.80 \\
HIP 105888$^{2}$ & 0.00 & 0.70 & 0.30 & 0 & 0.35 & 2.47 & 8.54 & 0.56 & 0.55 & 99.52 \\
HIP 106447$^{2}$ & 0.00 & 0.00 & 1.00 & 1 & 0.42 & 0.89 & 23.80 & 2.15 & 0.93 & 44.00 \\
HIP 107975 & 0.90 & 0.10 & 0.00 & 2 & 0.18 & 8.31 & 10.73 & 0.01 & 0.13 & 241.03 \\
HIP 108490$^{2}$ & 0.80 & 0.20 & 0.00 & 0 & 0.15 & 7.01 & 10.61 & 0.16 & 0.20 & 207.20 \\
HIP 109067 & 0.00 & 0.00 & 1.00 & 1 & 0.31 & 0.15 & 8.57 & 6.14 & 0.97 & 4.57 \\
HIP 109558 & 0.00 & 0.00 & 1.00 & 1 & 0.41 & 0.94 & 22.67 & 0.32 & 0.92 & $-$55.85 \\
HIP 109646$^{2}$ & 0.20 & 0.70 & 0.00 & 0 & 0.10 & 8.39 & 11.73 & 1.36 & 0.17 & 235.40 \\
HIP 112229$^{2}$ & 0.50 & 0.40 & 0.00 & 0 & 0.12 & 7.56 & 11.14 & 0.76 & 0.19 & 219.20 \\
HIP 114271 & 0.00 & 0.90 & 0.10 & 0 & 0.37 & 3.18 & 8.50 & 0.06 & 0.46 & 119.82 \\
HIP 114962$^{2}$ & 0.00 & 0.00 & 1.00 & 1 & 0.30 & 1.49 & 59.03 & 3.13 & 0.95 & $-$109.00 \\
HIP 115167$^{2}$ & 0.00 & 0.00 & 1.00 & 1 & 0.38 & 4.19 & 38.98 & 3.17 & 0.81 & $-$231.00 \\ 
HIP 117041$^{2}$ & 0.00 & 0.00 & 1.00 & 1 & 0.38 & 1.22 & 22.74 & 18.60 & 0.90 & 42.00 \\ 
G05-19$^{2}$ & 0.00 & 0.00 & 1.00 & 1 & 0.20 & 0.81 & 17.35 & 0.21 & 0.91 & 39.00 \\
G05-40$^{2}$ & 0.00 & 0.00 & 1.00 & 1 & 0.23 & 0.44 & 10.45 & 5.86 & 0.92 & 15.00 \\
G66-51$^{2}$ & 0.00 & 0.80 & 0.20 & 1 & 0.29 & 4.00 & 10.21 & 1.06 & 0.44 & 136.00 \\
G166-37$^{2}$ & 0.00 & 0.00 & 1.00 & 1 & 0.16 & 1.75 & 105.74 & 64.65 & 0.97 & 82.00 \\
G170-21$^{2}$ & 0.00 & 0.00 & 1.00 & 0 & 0.36 & 1.72 & 8.48 & 0.65 & 0.66 & $-$88.00 \\ 
\hline
\end{longtable}

\appendix

\section{Detailed comparison}\label{app:comp}

In this section we present a detailed comparison of our Be abundance results with previous 
results from the literature on a star-by-star basis. In most of the cases where differences 
in the abundances are seen, differences in the log g values can also be found. In these 
cases we are confident our results are robust since our log g values show excellent 
agreement with the ones derived using Hipparcos parallaxes.

\subsubsection*{HIP 171 (HD 224930)}

 The Be  abundance of star HIP 171 was previously determined
 by Stephens et al.\@ (\cite{St97}), log(Be/H) = $-$11.02. The value we found, 
log(Be/H) = $-$11.55, is somewhat different. This difference is most probably
 a result of the different log g adopted; log g = 4.62 by Stephens et al.\@
 (\cite{St97}) and log g = 4.10 by us.

\subsubsection*{HIP 11952 (HD 16031)}

 The Be abundance of HIP 11952 was previously determined by
 Gilmore et al.\@ (\cite{GGEN92}). Adopting log g = 3.90 and [Fe/H] =
 $-$1.96 they found log(Be/H) = $-$12.37. Adopting log g = 4.20 and
 [Fe/H] = $-$1.60, we found log(Be/H) = $-$12.28, which shows a good
 agreement within the uncertainties, in spite of the different
 atmospheric parameters.

\subsubsection*{HIP 14594 (HD 19445)}

 The Be abundance of star HIP 14594 was previously determined
 by four papers. Rebolo et al.\@ (\cite{RABM88}) found an upper limit of
 log(Be/H) $<$ $-$11.70, Ryan et al.\@ (\cite{Ry90}) found an
 upper limit of log(Be/H) $<$ $-$12.30, Boesgaard \& King (\cite{BK93})
 found log(Be/H) = $-$12.14, and Boesgaard et al.\@ (\cite{BDKR99})
 found a range of values from log(Be/H)= $-$12.45 to $-$12.55. The
 value we derived, log(Be/H) = $-$12.63, is in good agreement with the
 lower limit of the range derived by Boesgaard et al.\@
 (\cite{BDKR99}).

\subsubsection*{HIP 17001 (CD$-$24 1782)}

 The Be abundance of star HIP 17001 was previously
 determined by Garc\'{\i}a P\'erez \& Primas (\cite{GPP06}) who found
 log(Be/H) = $-$13.45 in LTE and log(Be/H) = $-$13.54 in NLTE. We
 derived an upper limit of log(Be/H) $<$ $-$13.83. The difference in
 the values is mainly due to the different log g adopted, 3.00 by us
 and 3.46 by Garc\'{\i}a P\'erez \& Primas (\cite{GPP06}).

\subsubsection*{HIP 17147 (HD 22879)}

 The Be abundance of HIP 17147 was previously determined
 by Beckman et al.\@ (\cite{Be89}), log(Be/H) = $-$11.25, This value agrees well with 
the one found in this work, log(Be/H) = $-$11.20.

\subsubsection*{HIP 18802 (HD 25704)}

 The Be abundance of star HIP 18802 was previously
 determined by Molaro et al. \cite{MBCP97}, log(Be/H) = $-$11.61. In
 this work we found log(Be/H) = $-$11.41. The log g values, although similar, 
log g = 4.20 by Molaro et al. \cite{MBCP97} and log g = 4.33
 by us, account for a difference of $\sim$ 0.08 dex in
 the abundance, what would bring them to an agreement within the
 uncertainties.

\subsubsection*{HIP 24316 (HD 34328)}

 The Be abundance of HIP 24316 was previously determined by
 Gilmore et al.\@ (\cite{GGEN92}), log(Be/H) = $-$11.90. This value agrees
 well with the value found in this work, log(Be/H) = $-$11.98.

\subsubsection*{HIP 37853 (HR 3018)}

 The Be abundance of HIP 37853 was previously determined by
 Gilmore et al.\@ (\cite{GGEN92}), log(Be/H) = $-$11.20.  This value 
 agrees well with the one found in this
 work, log(Be/H) = $-$11.16.

\subsubsection*{HIP 42592 (HD 74000)}

 The Be abundance of star HIP 42592 was previously determined
 by Ryan et al.\@ (\cite{Ry90}), log(Be/H) $<$ $-$12.20 and by Boesgaard 
et al.\@ (\cite{BDKR99}), which found a range of values from log(Be/H) =
 $-$12.03 to $-$12.47. The value found in this work, log(Be/H) = $-$12.58, 
agrees with the lower limit of the range determined by Boesgaard et al.\@ (\cite{BDKR99}), within
 the uncertainties. The small difference can be removed if we
 adopt the same log g used by Boesgaard et al., 0.16 dex higher than ours (increasing
 our abundance by $\sim$ 0.08 dex).

\subsubsection*{HIP 44075 (HD 76932)}

 The Be abundance of star HIP 44075 was determined by a number
 of papers in literature. Molaro \& Beckman (\cite{MB84}) found an
 upper limit of log(Be/H) $<$ $-$11.52, Beckman et al.\@ (\cite{Be89})
 found log(Be/H) = $-$11.69, both Gilmore et al.\@ (\cite{GGEN92}) and
 Ryan et al.\@ (\cite{Ry92}) found log(Be/H) = $-$11.30, Boesgaard \&
 King (\cite{BK93}) found log(Be/H) = $-$11.04, Garcia Lopez et al.\@
 (\cite{GL95b}) found log(Be/H) = $-$11.36 (in NLTE), Thorburn \&
 Hobbs (\cite{TH96}) found log(Be/H) = $-$11.45, Molaro et al.\@
 (\cite{MBCP97}) found $-$11.21, and Boesgaard et al.\@
 (\cite{BDKR99}) found a range of values from log(Be/H)= $-$11.17 to
 $-$11.24. The value we found, log(Be/H) = $-$11.12, is in the upper
 range of the values listed above and in good agreement with Boesgaard
 et al.\@ (\cite{BDKR99}).

\subsubsection*{HIP 48152 (HD 84937)}

 For star HIP 48152, Ryan et al.\@ (\cite{Ry92}) found log(Be/H) $<$ $-$12.85, Boesgaard \&
 King (\cite{BK93}) found log(Be/H) = $-$12.85, Thorburn \& Hobbs
 (\cite{TH96}) found log(Be/H) $<$ $-$12.95, and Boesgaard et al.\@
 (\cite{BDKR99}), from log(Be/H)= $-$12.83 to $-$12.94. The value we
 derived is log(Be/H) = $-$12.67. Reducing our log g 
by 0.20 dex to match the one adopted by Boesgaard et al.\@ (\cite{BDKR99}), 
for example, would reduce our abundance by $\sim$ 0.11 dex, resulting 
in an agreement within the uncertainties.

\subsubsection*{HIP 53070 (HD 94028)}

 For HIP 53070, Boesgaard \& King (\cite{BK93}) found log(Be/H) = $-$11.56, Garcia Lopez et al.\@
 (\cite{GL95b}) found log(Be/H) = $-$11.66 (in NLTE), Thorburn \&
 Hobbs (\cite{TH96}) found log(Be/H) = $-$11.65, and Boesgaard et al.\@
 (\cite{BDKR99}) found a range of values from log(Be/H)= $-$11.51 to
 $-$11.55. The value we derived is log(Be/H) = $-$11.80. Once again, a change 
in the adopted log g would bring the abundances to a better agreement.

\subsubsection*{HIP 55022 (HD 97916)}

 For HIP 55022, Boesgaard (\cite{Boe07}) found log(Be/H) $<$ $-$13.30 while 
 we derived log(Be/H) $<$ = $-$12.75.

\subsubsection*{HIP 59750 (HD 106516)}

 For HIP 59750, Molaro et al.\@ (\cite{MBCP97}) found log(Be/H) $<$ $-$12.76 
and Stephens et al.\@ (\cite{St97}) found log(Be/H) $<$ $-$12.61. The upper-limit 
we derived is log(Be/H) $<$ $-$12.50, agrees with the previous results within the 
uncertainties.

\subsubsection*{HIP 64426 (HD 114762)}

 For HIP 64426, Stephens et al.\@ (\cite{St97}) found log(Be/H) = $-$11.05, Boesgaard \&
 King (\cite{BK93}) found log(Be/H) = $-$11.14, Santos et al.\@
 (\cite{San02}) found log(Be/H) = $-$11.03, and Santos et al.\@
 (\cite{San04}) found log(Be/H) = $-$11.18. The value we derived is
 log(Be/H) = $-$11.31. The difference is again related to the 
different choice of log g values.

\subsubsection*{HIP 71458 (HD 128279)}

 For HIP 71458, Molaro et al. (\cite{MBCP97}) found log(Be/H) = $-$12.75 and
 by Garc\'{\i}a P\'erez \& Primas (\cite{GPP06}) found
 log(Be/H) $<$ $-$14.01, in LTE, and log(Be/H) = $-$13.94 in NLTE. We
 derived an upper limit of log(Be/H) $<$ $-$13.90.

\subsubsection*{HIP 74079 (HD 134169)}

 The Be abundance of HIP 74079 was previously determined
 by five papers. Gilmore et al.\@ (\cite{GGEN92}) and  Ryan et al.\@
 (\cite{Ry92}) both found log(Be/H) = $-$11.35, Boesgaard \&
 King (\cite{BK93}) found log(Be/H) = $-$11.29, Garcia Lopez et al.\@
 (\cite{GL95b}) found log(Be/H) = $-$11.23 (in NLTE), and Boesgaard et al.\@
 (\cite{BDKR99}) found a range of values from log(Be/H) = $-$11.32 to
 log(Be/H) = $-$11.40. All these values agree with ours, log(Be/H) =
 $-$11.28, within the uncertainties.

\subsubsection*{HIP 80837 (HD 148816)}

 The Be abundance of HIP 80837 was previously determined
 by Stephens et al.\@ (\cite{St97}), log(Be/H) = $-$11.07 and by Boesgaard \&
 King (\cite{BK93}), log(Be/H) = $-$11.10. Both values are consistent with ours, 
log(Be/H) = $-$11.22, within the uncertainties.

\subsubsection*{HIP 87693 (BD+20 3603)}

 The Be abundance of star HIP 87693 was previously determined
 by Boesgaard et al.\@ (\cite{BDKR99}) who found a range of values from
 log(Be/H) = $-$12.40 to log(Be/H) = $-$12.62. Our value, log(Be/H) =
 $-$12.77, is again consistent, within the uncertainties, with the
 lower limit of the range found by Boesgaard et al.\@
 (\cite{BDKR99}). The difference in the adopted log g values,
 4.33 or 4.27 by  Boesgaard et al.\@ and 4.00 by us, seems to be the
 main reason for the different abundances.

\subsubsection*{HIP 98532 (HD 189558)}

 The Be abundance of star HIP 98532 was previously determined
 by Rebolo et al.\@ (\cite{RABM88}) who found log(Be/H) = $-$11.70 and
 by Boesgaard \& King (\cite{BK93}) who found log(Be/H) = $-$10.99.
 Both papers claim higher uncertainties for this star when compared
 with the other stars of the sample. We determined a value of
 log(Be/H) = $-$11.43, which is intermediate between the two previous
 determinations.

\subsubsection*{HIP 100792 (HD 194598)}

 The Be abundance of star HIP 100792 was previously determined
 by three papers. Rebolo et al.\@ (\cite{RABM88}) found log(Be/H) =
 $-$11.70, Thorburn \& Hobbs (\cite{TH96}) found log(Be/H) = $-$11.95,
 and Boesgaard et al.\@ (\cite{BDKR99}) found a range of values from
 log(Be/H)= $-$11.73 to $-$11.88. The value determined in this work is
 log(Be/H) = $-$11.97. Our result agrees very well with the one from
 Thorburn \& Hobbs (\cite{TH96}) although they adopt a smaller log g,
 4.00, when compared to ours, log g = 4.20. Within the uncertainties,
 our result agrees with the lower limit of the range of values found
 by Boesgaard et al.\@ (\cite{BDKR99}).

\subsubsection*{HIP 101346 (HD 195633)}

 The Be abundance of star HIP 101346 was previously determined
 in three papers. Boesgaard \& King (\cite{BK93}) found log(Be/H) =
 $-$11.21, Stephens et al.\@ (\cite{St97}) found log(Be/H) = $-$11.29,
 and Boesgaard \& Novicki (\cite{BN06}) found log(Be/H)= $-$11.34. Our
 value is somewhat smaller than these, log(Be/H) = $-$11.52. The
 parameter controlling most of this difference seems to be the
 metallicity. We adopt [Fe/H] = $-$0.50, while Boesgaard \& King
 (\cite{BK93}) adopt [Fe/H] = $-$1.07, Stephens et al.\@ (\cite{St97})
 [Fe/H] = $-$1.00, and Boesgaard \& Novicki (\cite{BN06}) [Fe/H] =
 $-$0.88. Adopting the value of Boesgaard \& Novicki (\cite{BN06}),
 for example, would increase our result by 0.08 dex, bringing the
 values to agree within the uncertainties.

\subsubsection*{HIP 104660 (HD 201889)}

 The Be abundance of star HIP 104660 was previously determined
 by Boesgaard \& King (\cite{BK93}), log(Be/H) = $-$11.43, and by
 Boesgaard et al.\@ (\cite{BDKR99}) which found a range of values from  
 log(Be/H)= $-$11.30 to $-$11.38. Our value, log(Be/H) = $-$11.41, is
 in excellent agreement with these.

\subsubsection*{HIP 105858 (HR 8181)}

 The Be abundance of HIP 105858 was previously determined by
 Gilmore et al.\@ (\cite{GGEN92}). A value of log(Be/H) = $-$11.40 was
 found, in excellent agreement with ours, log(Be/H) = $-$11.32.

\subsubsection*{HIP 107975 (HD 207978)}

 The Be abundance of star HIP 107975 was previously determined
 by Stephens et al.\@ (\cite{St97}), which found an upper limit of
 log(Be/H) $<$ $-$12.38. Our analysis provides a detection mainly
 based on line 3131 \AA, log(Be/H) = $-$12.25.

\subsubsection*{HIP 108490 (HD 208906)}

 The Be abundance of star HIP 108490 was previously determined
 in three papers. Boesgaard \& King (\cite{BK93}) found log(Be/H) =
 $-$11.10, Stephens et al.\@ (\cite{St97}) found log(Be/H) = $-$11.19,
 and Boesgaard et al.\@ (\cite{B04}) found log(Be/H)= $-$11.30. This
 last value is in excellent agreement with our own,
 log(Be/H) = $-$11.34. The difference with Boesgaard \& King (\cite{BK93}) 
can be explained by the smaller log g
 adopted by them, log g = 4.25, when compared with the one adopted in
 this work, log g = 4.41 (and also by Boesgaard et al. 2004).

\subsubsection*{HIP 109558 (BD+17 4708)}

 The Be abundance of star HIP 109558 was previously determined
 by Boesgaard et al.\@ (\cite{BDKR99}), who found a range of values
 from log(Be/H)= $-$12.28 to $-$12.42. Our result, log(Be/H) =
 $-$12.30, is in excellent agreement with this range.

\subsubsection*{HIP 114271 (HD 218502)}

 The Be abundance of HIP 114271 was previously determined
 by Molaro et al.\@ (\cite{MBCP97}), log(Be/H)=
 $-$12.56. Our result, log(Be/H) = $-$12.44, agrees with it within
 the uncertainties.

\subsubsection*{HIP 114962 (HD 219617)}

 For HIP 114962, Rebolo et al.\@ (\cite{RABM88}) found log(Be/H) $<$ $-$11.60, Molaro et
 al.\@ (\cite{MBCP97}) found log(Be/H) = $-$12.56, and Boesgaard et al.\@
 (\cite{BDKR99}) found a range from log(Be/H) = $-$12.09 to $-$12.15. Our value, 
log(Be/H) = $-$12.40, is closer to the value found by
 Molaro et al.\@ (\cite{MBCP97}) than to the one by Boesgaard et al.\@
 (\cite{BDKR99}) although our parameters are very different from the
 former and similar to the latter.

\section{F00 subsample}\label{ap:f00}

\begin{figure}
\begin{centering}
\includegraphics[width=7cm]{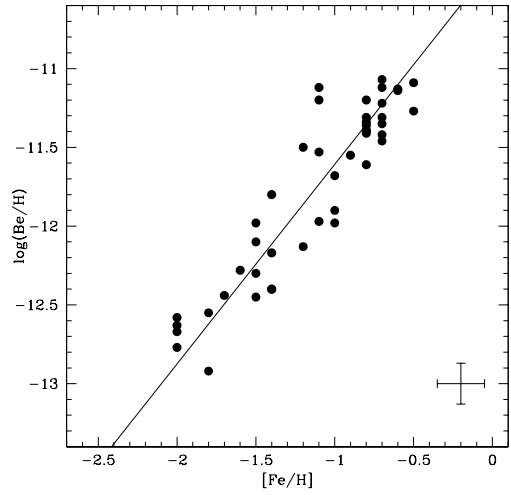}
\caption{Diagram of [Fe/H] vs.\@ log(Be/H) for the stars analyzed by Fulbright (\cite{F00}). Only detections 
are shown. An example of error bar is shown in the lower right corner.}
\label{fig:logbefehf00}
\end{centering}
\end{figure}

\begin{figure}
\begin{centering}
\includegraphics[width=7cm]{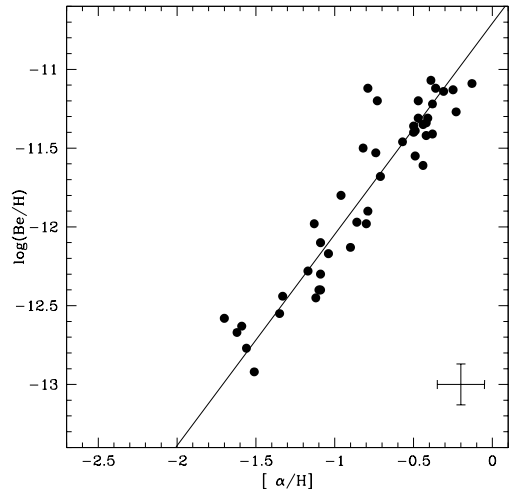}
\caption{Diagram of [$\alpha$/H] vs.\@ log(Be/H) for the stars analyzed by Fulbright (\cite{F00}). Only detections 
are shown. An example of error bar is shown in the lower right corner.}
\label{fig:logbealfahf00}
\end{centering}
\end{figure}

In this appendix we show plots of the relations between log(Be/H) and [Fe/H] (Fig.\@ \ref{fig:logbefehf00}) 
and between log(Be/H) and [$\alpha$/H] (Fig.\@ \ref{fig:logbealfahf00}) for the stars of the F00 subsample. 
The fits are statistically identical to the ones obtained with the whole clean sample.


\section{\label{acc-diss}
 Accretion and Dissipative Components}

In this appendix we divide our sample
in an ``accretion'' and a ``dissipative'' 
component according to the prescriptions
of Gratton et al. (\cite{G03,G03b}).

\begin{figure}
\begin{centering}
\includegraphics[width=7cm]{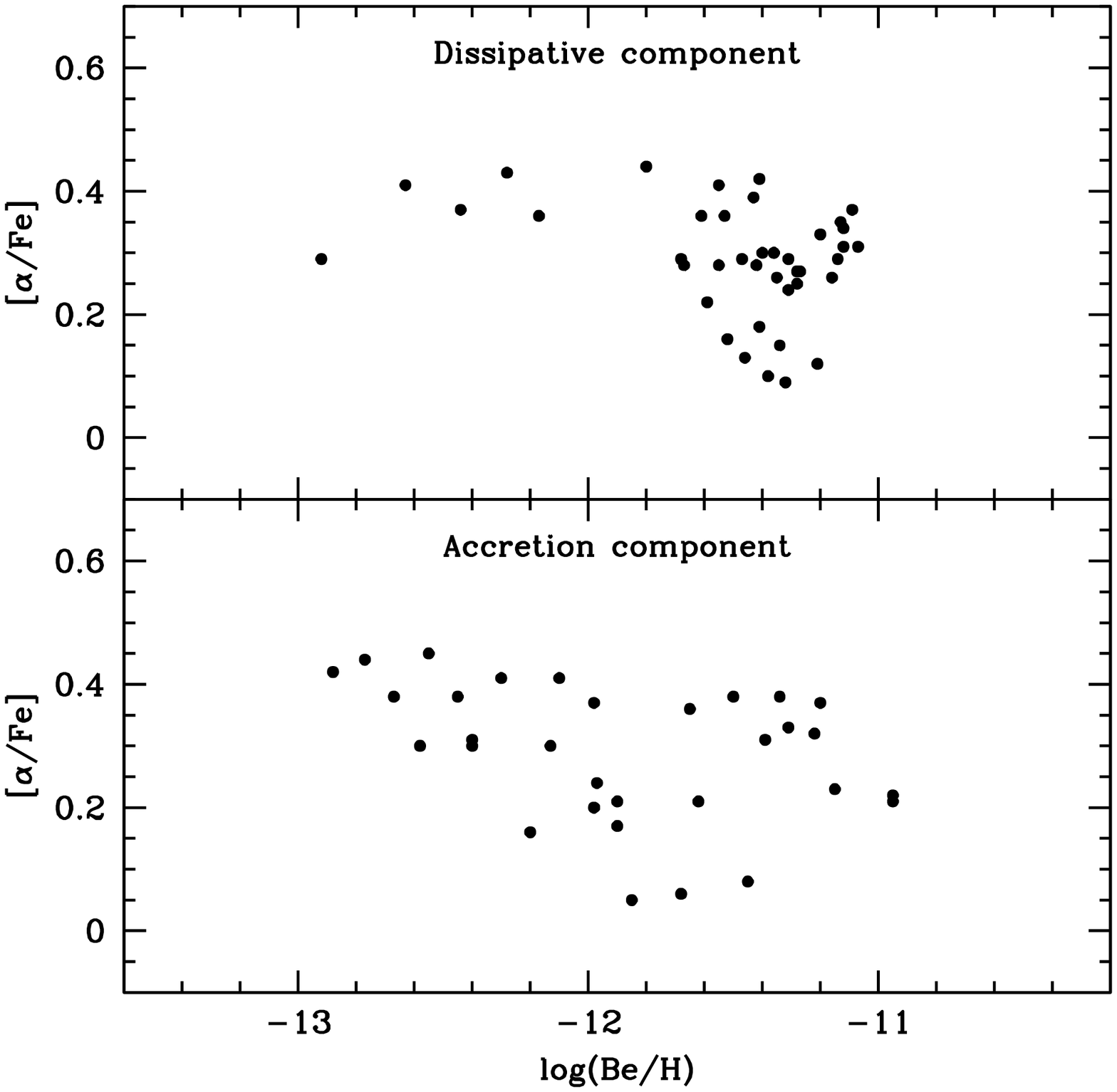}
\caption{Diagram of [$\alpha$/Fe] vs.\@ log(Be/H). The dissipative component stars are shown in the 
upper panel while the accretion component stars are shown in the lower panel. The dissipative stars 
diagram is characterized by a large scatter while the accretion stars diagram clearly divides in two sequences.}
\label{fig:alfafebegra}
\end{centering}
\end{figure}
\begin{figure}
\begin{centering}
\includegraphics[width=7cm]{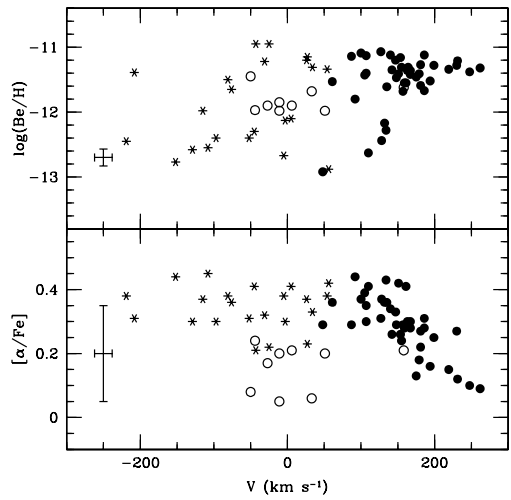}
\caption{Diagram of \@ log(Be/H) vs.\@ the V, the component of the space velocity of the star in the 
direction of the disk rotation. The dissipative component stars are shown as filled circles, accretion 
stars are shown as starred symbols, and the subgroup of accretion stars as open circles. A typical 
error of $\pm$ 12 Km s$^{-1}$ in V was adopted (see Gratton et al.\@ \cite{G03b}).}
\label{fig:plotvgra}
\end{centering}
\end{figure}
\begin{figure}
\begin{centering}
\includegraphics[width=7cm]{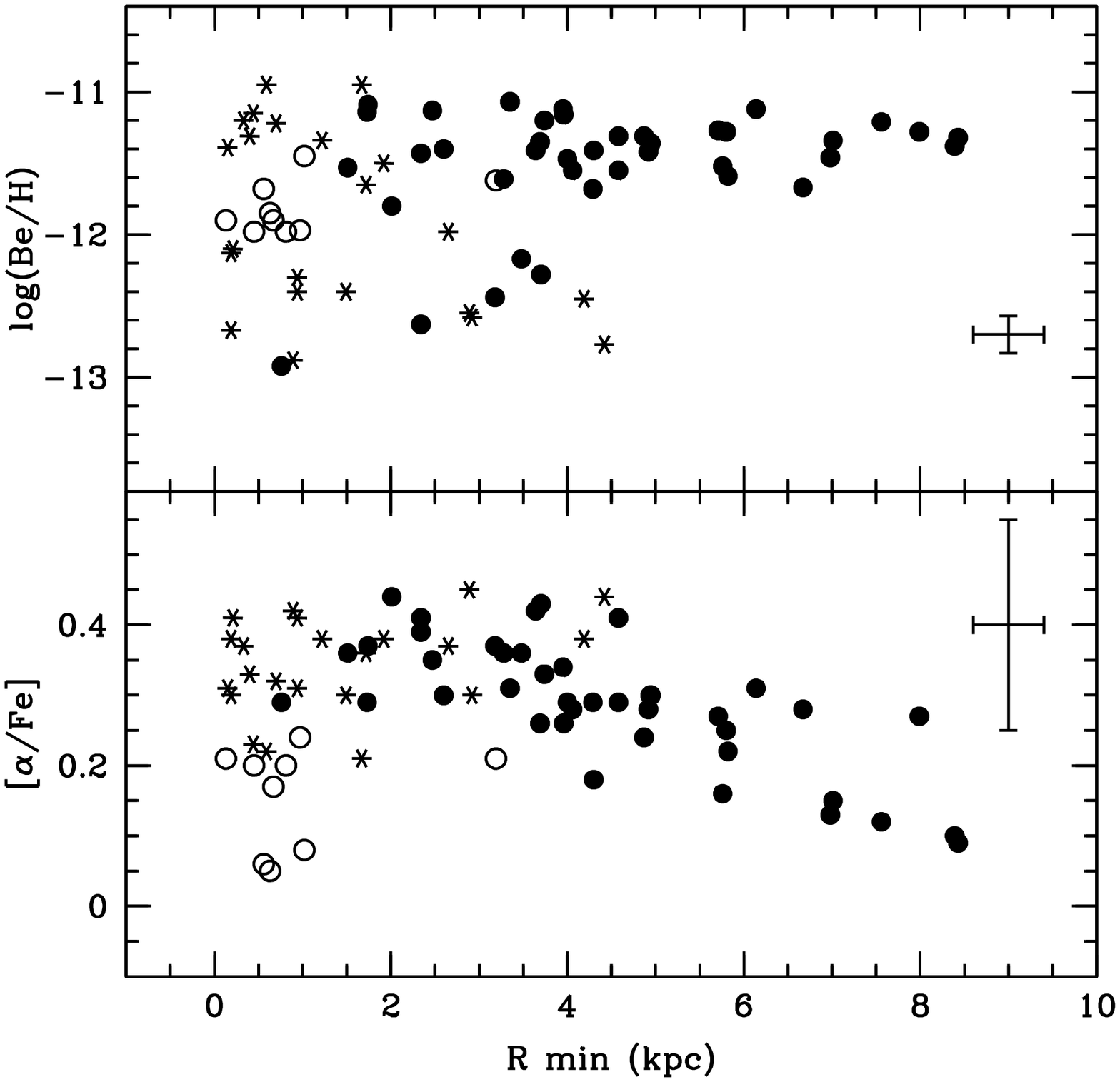}
\caption{Diagram of log(Be/H) vs.\@ R min, the perigalactic distance of stellar orbit. The dissipative  
stars are shown as filled circles, accretion stars are shown as starred symbols, and the subgroup of 
accretion stars as open circles. A typical error of $\pm$ 0.40 Kpc in R$_{\rm min}$ was adopted 
(see Gratton et al.\@ \cite{G03b}).}
\label{fig:plotrmingra}
\end{centering}
\end{figure}
\begin{figure}
\begin{centering}
\includegraphics[width=7cm]{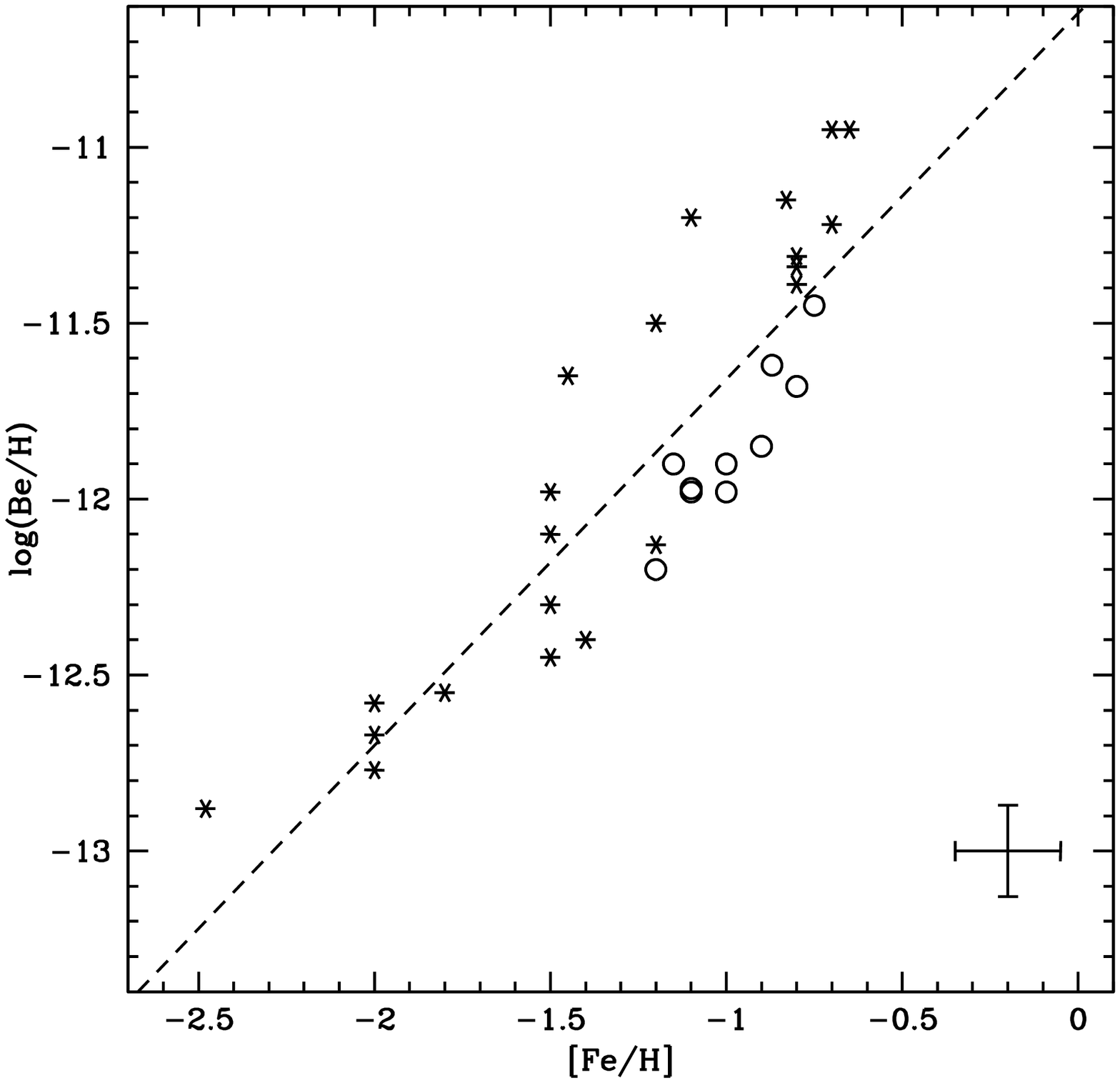}
\caption{Diagram of log(Be/H) vs.\@ [Fe/H] where only the accretion stars are shown. The subgroup of stars 
with low alpha is shown as open circles, the remaining accretion stars are shown as starred symbols. The linear fit 
for all the accretion stars is shown to guide the eye.}
\label{fig:logbefehacc}
\end{centering}
\end{figure}

The purpose is to test whether the same characteristics of our sample would be obtained with this different 
classification. The division of the 
stars according to this classification is also listed in Table \ref{tab:prob}. Some of the stars 
classified as thin disk stars by Venn et al.\@ (\cite{Venn04}) are classified as stars of the 
dissipative component by the criteria of Gratton et al.\@ (\cite{G03,G03b}). In the clean sample 
we have 34 stars classified as accretion component and 39 as dissipative component. In figures 
\ref{fig:alfafebegra}, \ref{fig:plotvgra}, \ref{fig:plotrmingra}, and \ref{fig:logbefehacc}, we show 
how these components behave in each of the plots presented before for the halo and thick disk components. 
Therefore, these should be directly compared with figures \ref{fig:alfafebe}, \ref{fig:plotv}, 
\ref{fig:plotrmin}. As it can be seen, the overlap is really  high and the 
conclusions obtained are the same.

\end{document}